\documentclass[a4paper,11pt]{article} 

\pdfoutput=1

\usepackage[x11names, svgnames, rgb, table]{xcolor}
\usepackage[T1]{fontenc}
\usepackage{microtype}
\usepackage[english]{babel}
\usepackage{tikz}
\usepackage{slashed}
\usepackage{tabularx}
\usepackage{wrapfig}
\usepackage{afterpage}
\usepackage{physics}
\usepackage{comment}
\usepackage{emptypage}
\usepackage{listings}
\usepackage{pdfpages}
\usepackage{enumerate}
\usepackage{appendix}
\usepackage{bbold}
\usepackage{bm}
\usepackage{mathrsfs}
\usepackage{setspace}
\usepackage{pifont}
\usepackage{xfrac}
\usepackage{cancel}
\usepackage{mathtools}
\usepackage{multirow}
\usepackage{booktabs}
\usepackage{makecell}
\usepackage[export]{adjustbox}
\usepackage{amsthm}
\usepackage{lmodern}
\usepackage{colortbl}

\definecolor{hdrcolor}{gray}{0.9}

\usetikzlibrary{snakes,arrows,shapes}
\usepackage{jheppub} 

\hypersetup{
    pdfencoding=unicode,
	colorlinks=true,
	urlcolor=RoyalBlue,
	linkcolor=Maroon,
	citecolor=ForestGreen,
    pdftitle={Unitarity bounds and sum rules in the SMEFT: a systematic analysis},
	pdfauthor={Luigi C. Bresciani, Paride Paradisi, Andrea Sainaghi},
	pdfdisplaydoctitle=true,
	pdfstartview=FitH
}

\makeatletter\g@addto@macro\bfseries{\boldmath}\makeatother

\newcommand{\Pij}[3]{\delta_{#1 #3}\delta_{#3 #2}}

\allowdisplaybreaks

\preprint{IRMP-CP3-26-39}

\title{Unitarity bounds and sum rules in the SMEFT: a systematic analysis}

\author[a,b,c]{Luigi~C.~Bresciani,}
\author[a,b]{Paride~Paradisi,}
\author[a,b,d]{Andrea~Sainaghi}
\affiliation[a]{Dipartimento di Fisica e Astronomia ``G.~Galilei'', Università degli Studi di Padova,\\Via F.~Marzolo 8,
I-35131 Padova, Italy}
\affiliation[b]{Istituto Nazionale di Fisica Nucleare, Sezione di Padova,\\Via F.~Marzolo 8,
I-35131 Padova, Italy}
\affiliation[c]{Centre for Cosmology, Particle Physics and Phenomenology, Université Catholique de Louvain,\\Chemin du Cyclotron 2, B-1348 Louvain-la-Neuve, Belgium}
\affiliation[d]{Physik-Institut, Universit\"at Z\"urich,\\Winterthurerstrasse 190, CH-8057 Z\"urich, Switzerland}

\emailAdd{luigi.bresciani@uclouvain.be}
\emailAdd{paride.paradisi@unipd.it}
\emailAdd{andrea.sainaghi@phd.unipd.it}

\abstract{We present a systematic analysis of perturbative unitarity bounds and spinning sum rules in the baryon-number-conserving dimension-six Standard Model Effective Field Theory (SMEFT), with particular emphasis on flavor structure. Employing an on-shell partial-wave formalism for generic $N \to M$ scattering amplitudes, we perform a coupled-channel analysis of the complete operator basis and derive bounds on individual Wilson coefficients by marginalizing over the remaining interactions. We formulate spinning sum rules for four-fermion operators in the Warsaw basis and translate their flavor-space definiteness conditions into constraints on Wilson coefficients, under specified assumptions about the ultraviolet dynamics. We then extend both analyses to the three-generation four-fermion sector within Minimal Flavor Violation, retaining leading up-type Yukawa insertions, and an exact $\mathrm{U}(2)^5$ flavor symmetry. The resulting constraints reveal correlations among flavor structures that are absent in the single-flavor limit. Through explicit matching to representative ultraviolet completions, we illustrate how spinning sum rules can distinguish scalar- from vector-mediated interactions, while also identifying the limitations of this interpretation. These results provide a general framework for assessing the consistency of new-physics scenarios within the SMEFT, relying on minimal theoretical assumptions.}

\begin{document}
\maketitle
\pagestyle{myplain}
\flushbottom

\section{Introduction}
\label{sec:introduction}

The remarkable success of the Standard Model (SM), culminating in the discovery of a Higgs boson with properties consistent with its predictions, 
has so far not been accompanied by direct evidence for new particles at the electroweak scale. 
If new degrees of freedom are sufficiently heavier than the energies probed experimentally, their low-energy effects can be systematically described within an effective field theory (EFT). 
The Standard Model Effective Field Theory (SMEFT) provides a model-independent framework for this purpose, extending the SM Lagrangian by 
higher-dimensional operators constructed from SM fields and invariant under the SM gauge symmetries. 
The corresponding Wilson coefficients encode the effects of heavy dynamics, while 
the operators are suppressed by powers of a characteristic new-physics scale $\Lambda$~\cite{Buchmuller:1985jz,Grzadkowski:2010es,Brivio:2017vri,Isidori:2023pyp,Aebischer:2025qhh}. 
In the absence of compelling evidence for a particular ultraviolet (UV) completion, the SMEFT has become a standard framework for interpreting electroweak, flavor, Higgs, and collider measurements in searches for physics beyond the SM.

The generality of the SMEFT also presents a major challenge. 
Already at dimension six, the number of independent Wilson coefficients becomes very large once the full flavor structure is retained~\cite{Faroughy:2020ina,Greljo:2022cah}. 
Experimental constraints must therefore be interpreted in a multidimensional parameter space, 
where correlations among operators, approximate flat directions, and assumptions about flavor can substantially affect the results. 
In this context, theoretical constraints derived from general properties of scattering amplitudes provide valuable complementary information. 
Two particularly useful approaches are perturbative unitarity and analyticity-based dispersion relations~\cite{Lee:1977yc,Lee:1977eg,Falkowski:2019tft,Gu:2020thj,Remmen:2020uze,Remmen:2022orj,Azatov:2021ygj,Low:2009di,Altmannshofer:2023bfk,Altmannshofer:2025lun}. 
While both are rooted in fundamental properties of the scattering matrix, they provide distinct yet complementary constraints on the effective theory.

Perturbative unitarity provides a quantitative criterion for assessing the validity of the EFT description at a given energy. 
Higher-dimensional operators typically generate scattering amplitudes that grow with the center-of-mass energy, eventually challenging the perturbative expansion. 
Partial-wave unitarity translates this behavior into bounds on combinations of Wilson coefficients~\cite{Gounaris:1994cm,Corbett:2014ora,Corbett:2017qgl,Cao:2023qks,Eboli:2026ccq,DiLuzio:2017chi,DiLuzio:2016sur,Mahmud:2024iyn,Allwicher:2021jkr,Cohen:2021ucp,Almeida:2020ylr,Abu-Ajamieh:2020yqi,Anders:1993dk,Brivio:2021fog,Falkowski:2019tft,Chang:2019vez,Alonso:2025ksv,Bresciani:2025ojh,Dobado:2026kll,Bresciani:2025toe,Bresciani:2026acy,Degrande:2025uil,Singh:2026xak}.
These bounds should not be interpreted as absolute exclusions of EFT parameter space: rather, their saturation indicates the energy scale at which the perturbative description ceases to be reliable and additional dynamics must become relevant. 
This issue is particularly important for collider measurements probing energies potentially comparable to the scale of new physics~\cite{Cohen:2021gdw}.

While traditional partial-wave analyses are generally formulated in terms of $2 \to 2$ scattering, a complete treatment of higher-dimensional interactions 
may require processes with larger numbers of external particles. 
In particular, some SMEFT operators do not generate the relevant energy-growing contributions in $2 \to 2$ amplitudes, making higher-multiplicity scattering essential. 
The on-shell formalism developed in Ref.~\cite{Bresciani:2025toe} extends partial-wave unitarity to generic \(N\to M\) amplitudes, allowing scattering processes with different particle multiplicities to be treated within a common framework; see also Ref.~\cite{Degrande:2025uil}.  
Moreover, the strongest bounds are obtained by imposing unitarity on the eigenvalues of the full coupled-channel partial-wave matrices, which include all states with the same conserved quantum numbers.
The resulting constraints generally involve several Wilson coefficients simultaneously, rather than restricting each operator independently.

Previous studies have investigated partial-wave unitarity in various sectors of the dimension-six SMEFT~\cite{Gounaris:1994cm,Corbett:2014ora,Corbett:2017qgl,Cao:2023qks}, focusing predominantly on bosonic interactions and selected operators containing fermions. 
In Ref.~\cite{Bresciani:2026acy}, we performed a complete coupled-channel reassessment of perturbative unitarity in the baryon-number-conserving dimension-six SMEFT, 
extending the analysis to the full four-fermion basis, gluonic operators, \(\psi^2\varphi^3\) interactions, and the purely scalar operator $(\varphi^\dag \varphi)^3$. 
An important aspect of this approach is the treatment of correlations among Wilson coefficients: bounds on individual coefficients are obtained by marginalizing over the remaining interactions, yielding conservative projections of the multidimensional unitarity region rather than conventional single-operator limits. 
In the present work, we provide a detailed account of this analysis, including explicit examples of the coupled-channel construction and the role of different angular-momentum and particle-multiplicity sectors.

Perturbative unitarity primarily constrains the magnitude of effective interactions, 
but generally provides limited information about the quantum numbers of the heavy states responsible for them. 
Additional insights can be obtained from dispersion relations derived from the analyticity, crossing symmetry, and unitarity of the scattering matrix. 
In particular, spinning sum rules relate suitable combinations of dimension-six four-fermion Wilson coefficients to the angular momentum of intermediate states in the UV~\cite{Gu:2020thj,Remmen:2020uze,Remmen:2022orj,Azatov:2021ygj,Altmannshofer:2023bfk,Altmannshofer:2025lun}. 
Under appropriate assumptions, these relations lead to sign constraints that can distinguish scalar- from vector-dominated heavy dynamics. 
Therefore, they offer a complementary perspective on the possible microscopic origin of effective four-fermion interactions.

The interpretation of spinning sum rules, however, requires care. 
Unlike perturbative unitarity bounds, their sign constraints are not universal consistency conditions on arbitrary SMEFT coefficients. 
In the form considered here, they rely on assumptions about the UV dynamics, including a tree-level origin of the relevant interactions, 
dominance by states of definite spin, and the absence of boundary terms in the dispersion relations~\cite{Remmen:2020uze,Remmen:2022orj,Azatov:2021ygj}. 
These conditions need not hold in general. 
Loop-induced interactions, contributions from states of different spins, or non-vanishing boundary terms associated with \(t\)-channel vector exchange can modify or invalidate the resulting inequalities. 
Consequently, a violation of a scalar- or vector-dominated sum-rule condition does not necessarily imply an inconsistent UV completion. 
Rather, the applicability and limitations of these relations must be assessed in conjunction with the underlying dynamics. 
In this work, we formulate the spinning sum rules in the Warsaw basis and investigate their interpretation through representative UV completions.

A central objective of our analysis is to extend these theoretical constraints beyond the single-flavor limit. 
This is particularly relevant for four-fermion operators, whose Wilson coefficients carry four flavor indices and whose number of independent components grows rapidly when all three SM generations are included~\cite{Faroughy:2020ina,Greljo:2022cah}. 
At the same time, flavor non-universality is a characteristic feature of many well-motivated new-physics scenarios, especially those involving enhanced couplings to third-generation fermions~\cite{Greljo:2018tuh,Fuentes-Martin:2020bnh,Allwicher:2023shc}. 
A completely general treatment of the flavor indices is therefore difficult to implement and interpret, whereas the single-flavor limit misses important correlations among generations.

We address this problem by considering two theoretically motivated flavor hypotheses. 
The first is Minimal Flavor Violation (MFV), based on the $\mathrm{U}(3)^5$ flavor symmetry of the SM gauge interactions, broken only by the Yukawa matrices, treated as spurions~\cite{DAmbrosio:2002vsn,Cirigliano:2005ck}.
In our analysis, we retain the flavor-symmetric structures and the leading corrections involving the up-type Yukawa matrix, which capture the dominant effects associated with the top quark. 
The second hypothesis is an approximate $\mathrm{U}(2)^5$ flavor symmetry, under which the first two generations transform as doublets while the third generation is a singlet. 
This structure provides a natural framework for describing non-universal interactions involving third-generation fermions and is particularly relevant to a broad class of TeV-scale new-physics scenarios~\cite{Bordone:2017bld,Greljo:2018tuh,Allwicher:2020esa,Fuentes-Martin:2020bnh,Fuentes-Martin:2022xnb,Barbieri:2023qpf,Barbieri:2024zkh,Davighi:2023evx,Davighi:2023iks,Fuentes-Martin:2024fpx,Covone:2024elw,Lizana:2024jby,Beneito:2025fzf,FernandezNavarro:2023rhv,FernandezNavarro:2024hnv,FernandezNavarro:2025zmb,Isidori:2025rci,Greljo:2025mwj,Masiero:2026qon}. 
Both hypotheses reduce the number of independent four-fermion coefficients while retaining non-trivial information about their flavor structure.

Within these flavor frameworks, we derive perturbative unitarity bounds and spinning sum-rule constraints on the independent four-fermion Wilson coefficients. 
The resulting conditions involve correlated combinations of flavor structures and can reveal restrictions that are absent in the single-flavor limit. 
In particular, the allowed regions need not be described by independent intervals for each coefficient, and their geometry can reflect non-trivial relations among flavor components. 
We further investigate the interpretation of these constraints by matching representative UV completions onto the corresponding effective operators, illustrating both the information encoded in spinning sum rules and the circumstances under which their conventional spin interpretation can fail.

The paper is organized as follows. In Sec.~\ref{sec:unitarity}, we review the generalized partial-wave formalism and present the perturbative unitarity bounds for the complete baryon-number-conserving dimension-six SMEFT, including illustrative coupled-channel examples. 
In Sec.~\ref{sec:sum_rules}, we formulate spinning sum rules for four-fermion operators in the Warsaw basis, derive their implications in flavor space, and discuss their assumptions and limitations.
In Sec.~\ref{sec:flavorful}, we extend both analyses to the three-generation four-fermion sector under MFV and $\mathrm{U}(2)^5$ flavor symmetry, and examine representative UV completions. We summarize our findings in Sec.~\ref{sec:conclusion}. 
Mathematical tools used in the analysis are collected in App.~\ref{app:mathtools}, while the complete multiflavor results are presented in App.~\ref{app:flavor_results}.

\section{Partial-wave unitarity bounds}
\label{sec:unitarity}

In this section we derive perturbative partial-wave unitarity bounds for the
baryon-number-conserving dimension-six SMEFT.
We first review the on-shell formalism used to decompose generic $N\to M$
scattering amplitudes into states of definite total angular momentum.
We then illustrate its implementation through two representative examples.
The operator $O_{\varphi}=(\varphi^\dagger\varphi)^3$ highlights the
role of higher-multiplicity processes, since its dimension-six $2\to2$
contribution does not grow with energy, whereas
$O_{\varphi e}$ provides a simple example in which different
partial waves compete in setting the strongest constraint.
We then systematically apply the formalism to the complete
baryon-number-conserving dimension-six Warsaw basis, performing a fully
coupled-channel analysis over all relevant scattering processes and retaining
the correlations among Wilson coefficients that contribute to the same
partial-wave amplitudes.

\subsection{Review of the method}
\label{sec:unitarity_method}

We adopt the on-shell partial-wave formalism developed in
Ref.~\cite{Bresciani:2025toe}, which generalizes the usual $2\to2$
partial-wave decomposition to scattering amplitudes with arbitrary
multiplicity.
For a generic transition from an initial state $i$ to a final state $f$,
the amplitude can be regarded as a vector in the corresponding kinematic
space and decomposed onto a basis of states with definite total angular
momentum $J$,
\begin{equation}
    |{\cal A}_{i\to f}\rangle
    =
    \sum_J
    a^J_{i\to f}\,
    |{\cal B}^J_{i\to f}\rangle ,
\end{equation}
where $a^J_{i\to f}$ denotes the partial-wave coefficient.
The angular-momentum basis is chosen to satisfy
\begin{equation}
    \langle {\cal B}^{J}_{i\to f}
    |{\cal B}^{J'}_{i\to f}\rangle
    =
    \int d\Phi_i\,d\Phi_f\,
    {\cal B}^{J'}_{i\to f}
    \left({\cal B}^{J}_{i\to f}\right)^*
    =
    (2J+1)\,\delta^{JJ'} ,
\end{equation}
with $d\Phi_i$ and $d\Phi_f$ the Lorentz-invariant phase-space measures
associated with the initial and final states.
For massless $n$-particle states of distinguishable particles, the phase-space volume is normalized as
\begin{equation}
    \Phi_n
    =
    \frac{1}{8\pi (n-1)!(n-2)!}
    \left(\frac{\sqrt{s}}{4\pi}\right)^{2n-4}.
\end{equation}
With this normalization, unitarity of the $S$ matrix implies
\begin{equation}
    \left|{\rm Re}\,a^J_{i\to i}\right|\leq 1,
    \qquad
    0\leq {\rm Im}\,a^J_{i\to i}\leq 2,
    \qquad
    \left|a^J_{i\to f}\right|\leq 1
    \quad (i\neq f).
    \label{eq:unitarity_bounds}
\end{equation}
For $2\to2$ scattering, the basis elements
$|{\cal B}^J_{i\to f}\rangle$ reduce, up to normalization, to the familiar
Wigner $d$-functions~\cite{Jacob:1959at,Varshalovich:1988ifq}.
The advantage of the present formulation is that the same construction
continues to apply when either the initial or final state contains more than
two particles, without requiring a separate partial-wave parametrization for
each multiplicity.

For a fixed value of $J$, the strongest bounds are obtained by considering
simultaneously all states carrying the same conserved quantum numbers.
The corresponding partial-wave coefficients therefore assemble into a
coupled-channel matrix in the space of particle species, helicities, gauge
indices, flavor indices, and, when relevant, different particle
multiplicities.
Perturbative unitarity is then imposed on the eigenvalues of this matrix,
rather than separately on individual scattering amplitudes.
This step is essential whenever several processes communicate through the
same interaction, since coherent combinations of channels can strengthen the
resulting constraint.

In practice, the construction proceeds directly from on-shell amplitudes.
For each operator, we generate all scattering processes that display the
leading energy growth at a given order in the EFT expansion, determine the
allowed values of $J$ from their helicity structure, and project the
amplitudes onto the corresponding angular-momentum basis~\cite{Bresciani:2025toe,Jiang:2020rwz,Shu:2021qlr}.
All independent gauge configurations and, whenever retained, flavor
configurations are included in the coupled-channel matrix, together with the appropriate normalization factors for identical particles.

When more than one angular momentum contributes, the unitarity condition is imposed independently in each $J$ sector and the strongest resulting
constraint is retained.
As we shall see explicitly below, the lowest allowed angular momentum does
not necessarily provide the strongest bound once the full multiplicity of
internal quantum numbers is taken into account. 

Throughout this section we work in the high-energy limit, where particle masses can be neglected, and retain the leading energy-growing contributions induced by the dimension-six interactions.
Purely SM terms, which do not exhibit the energy growth relevant for the
unitarity bounds derived here, are therefore neglected unless explicitly
stated otherwise. Besides isolating the UV behavior of interest, this prescription avoids infrared singularities associated with possible massless $t$-channel exchanges in the SM amplitudes, whose partial-wave projection would require a separate infrared treatment; see, e.g., Refs.~\cite{Bellazzini:2021oaj,Hannesdottir:2019umk,Hannesdottir:2019opa,FuentesZamoro:2025exp,Plestid:2026pzn,Carney:2017oxp,Carney:2018ygh,Lippstreu:2023vvg,Lippstreu:2025jit,Caron-Huot:2022ugt,Caron-Huot:2022jli,Chang:2025cxc,Bellazzini:2025bay}.
An exception is provided by the $X^3$ operators, for which the role of the
SM contribution is discussed separately in
Sec.~\ref{sec:bosonic_unitarity}.

The resulting bounds should therefore be interpreted as perturbative
consistency conditions on the SMEFT at a given center-of-mass energy,
rather than as sharp exclusions of points in Wilson-coefficient space.
Their saturation identifies the scale at which the perturbative EFT
description ceases to be reliable and additional dynamics must become
relevant.

For later convenience, we label an external particle by $N_I$, where
$N=1,2,\ldots$ denotes its momentum label and $I$ collectively denotes its
gauge quantum numbers.
An antiparticle is indicated by placing a bar over the corresponding
momentum label.
This notation will be used in the explicit examples below.

\subsubsection{Example: \texorpdfstring{$O_\varphi$}{Ophi} operator}
\label{sec:example O_phi}

A particularly instructive application of the generalized $N\to M$
formalism is provided by the purely Higgs operator
\begin{equation}
    O_\varphi=(\varphi^\dagger\varphi)^3.
\end{equation}
Unlike the derivative dimension-six operators, its $2\to2$ contribution does
not display the energy growth relevant for the high-energy unitarity bounds.
The relevant unitarity constraint instead arises from processes involving
six external scalar fields, so that one must consider $3\to3$ and $2\to4$ scattering. This makes $O_\varphi$ a simple example in which the extension beyond the
standard $2\to2$ partial-wave decomposition is essential.

We first consider the $3\to3$ channels.
Up to charge conjugation, the independent amplitudes can be written as
\begin{align}
    \mathcal{A}(1_a,2_b,3_c\to 4_d,5_e,6_f)
    &=
    \mathcal{A}(\overline{1}_a,\overline{2}_b,\overline{3}_c
    \to \overline{4}_d,\overline{5}_e,\overline{6}_f)
    =
    36\,\frac{C_\varphi}{\Lambda^2}\,
    \delta_{a}^{(d}\delta_{b}^{e}\delta_{c}^{f)},
    \\
    \mathcal{A}(1_a,2_b,\overline{3}_c\to 4_d,5_e,\overline{6}_f)
    &=
    \mathcal{A}(\overline{1}_a,\overline{2}_b,3_c
    \to \overline{4}_d,\overline{5}_e,6_f)
    =
    36\,\frac{C_\varphi}{\Lambda^2}\,
    \delta_{a}^{(d}\delta_{b}^{e}\delta_{f}^{c)},
\end{align}
where $a,b,c,d,e,f=1,2$ are $\mathrm{SU}(2)_L$ indices and we have defined
the symmetrized tensor
\begin{equation}
    \delta_{a}^{(d}\delta_{b}^{e}\delta_{c}^{f)}
    =
    \frac{1}{6}\Big[
    \delta_{a}^{d}\delta_{b}^{e}\delta_{c}^{f}
    +\delta_{a}^{d}\delta_{b}^{f}\delta_{c}^{e}
    +\delta_{a}^{e}\delta_{b}^{d}\delta_{c}^{f}
    +\delta_{a}^{e}\delta_{b}^{f}\delta_{c}^{d}
    +\delta_{a}^{f}\delta_{b}^{d}\delta_{c}^{e}
    +\delta_{a}^{f}\delta_{b}^{e}\delta_{c}^{d}
    \Big].
\end{equation}
The corresponding $2\to4$ amplitudes, together with their crossed
$4\to2$ processes, are
\begin{align}
    \mathcal{A}(1_a,2_b\to 3_c,4_d,5_e,\overline{6}_f)
    &=
    \mathcal{A}(\overline{1}_a,\overline{2}_b
    \to \overline{3}_c,\overline{4}_d,\overline{5}_e,6_f)
    =
    36\,\frac{C_\varphi}{\Lambda^2}\,
    \delta_{a}^{(c}\delta_{b}^{d}\delta_{f}^{e)},
    \\
    \mathcal{A}(1_a,\overline{2}_b
    \to 3_c,4_d,\overline{5}_e,\overline{6}_f)
    &=
    36\,\frac{C_\varphi}{\Lambda^2}\,
    \delta_{a}^{(b}\delta_{e}^{c}\delta_{f}^{d)}.
\end{align}

Since all external particles are scalars and the interaction contains no
derivatives, these amplitudes contribute entirely to the $J=0$ partial wave.
The phase-space integrations are therefore particularly simple, and the
relevant normalized basis elements are
\begin{equation}
\label{eq:basis_J0_scalars}
    \mathcal{B}_{3\to3}^{0}
    =
    \frac{256\pi^3}{s},
    \qquad
    \mathcal{B}_{2\to4}^{0}
    =
    \mathcal{B}_{4\to2}^{0}
    =
    \frac{256\pi^3\sqrt{3}}{s}.
\end{equation}
In particular,
$\mathcal{B}_{2\to4}^{0}=1/\sqrt{\Phi_2\Phi_4}$, consistently with the
normalization introduced in Sec.~\ref{sec:unitarity_method}.
When a state
contains identical particles the
symmetry factor $S=\prod_k n_k!$ of its phase-space measure is accounted for by a factor $1/\sqrt{S_iS_f}$ in the
partial-wave coefficient.
The full $J=0$ scattering matrix decomposes into independent blocks
according to the quantum numbers and field content of the states connected
by $O_\varphi$.

As a first example, consider the
$\varphi\varphi\varphi^\dagger\to
\varphi\varphi\varphi^\dagger$ sector.
The corresponding partial-wave matrix is
\begin{equation}
    a^0_{\varphi\varphi\varphi^\dagger
    \to\varphi\varphi\varphi^\dagger}
    =
    \frac{s\,C_\varphi}{256\pi^3\Lambda^2}
    \begin{pmatrix}
        18 & 0 & 0 & 6\sqrt{2} & 0 & 0 \\
        0 & 6 & 0 & 0 & 0 & 0 \\
        0 & 0 & 12 & 0 & 0 & 6\sqrt{2} \\
        6\sqrt{2} & 0 & 0 & 12 & 0 & 0 \\
        0 & 0 & 0 & 0 & 6 & 0 \\
        0 & 0 & 6\sqrt{2} & 0 & 0 & 18
    \end{pmatrix}.
\end{equation}
The matrix is $6\times6$, corresponding to the six independent
$\mathrm{SU}(2)_L$ configurations
\begin{equation}
    (a,b,c)\in
    \{(1,1),(1,2),(2,2)\}\otimes\{1,2\},
\end{equation}
where $a$ and $b$ label the two $\varphi$ fields and $c$ labels $\varphi^\dagger$.
Imposing unitarity on the eigenvalue of largest modulus yields
\begin{equation}
    \frac{3s}{32\pi^3\Lambda^2}\,
    |C_\varphi|
    \leq 1.
\end{equation}

A qualitatively different block couples the two- and four-particle states
\begin{equation}
    i,f\in
    \left\{
    \varphi\varphi^\dagger,\,
    \varphi\varphi\varphi^\dagger\varphi^\dagger
    \right\}.
\end{equation}
Since $O_\varphi$ does not connect the two-particle sector to itself at the
energy-growing order considered here, nor the four-particle sector to itself
through the same six-field contact vertex, the partial-wave matrix takes the
off-diagonal form
\begin{equation}
    a^0_{i\to f}
    =
    \begin{pmatrix}
        \mathbb{O} &
        M_{\varphi\varphi^\dagger
        \to\varphi\varphi\varphi^\dagger\varphi^\dagger}
        \\
        M^\dagger_{\varphi\varphi^\dagger
        \to\varphi\varphi\varphi^\dagger\varphi^\dagger}
        &
        \mathbb{O}
    \end{pmatrix},
\end{equation}
with
\begin{equation}
M_{\varphi\varphi^\dagger
    \to\varphi\varphi\varphi^\dagger\varphi^\dagger}
    =
    \frac{s\,C_\varphi}{256\pi^3\Lambda^2}
    \begin{pmatrix}
        6\sqrt{3} & 0 & 0 & 0 & 4\sqrt{3} & 0 & 0 & 0 & 2\sqrt{3} \\
        0 & 2\sqrt{6} & 0 & 0 & 0 & 2\sqrt{6} & 0 & 0 & 0 \\
        0 & 0 & 0 & 2\sqrt{6} & 0 & 0 & 0 & 2\sqrt{6} & 0 \\
        2\sqrt{3} & 0 & 0 & 0 & 4\sqrt{3} & 0 & 0 & 0 & 6\sqrt{3}
    \end{pmatrix}.
\end{equation}
Here the $4\times9$ rectangular block follows from the independent gauge
configurations
\begin{equation}
    (a,b)\in\{1,2\}\otimes\{1,2\},
    \qquad
    (c,d,e,f)\in
    \{(1,1),(1,2),(2,2)\}
    \otimes
    \{(1,1),(1,2),(2,2)\}.
\end{equation}
The non-vanishing eigenvalues of the complete Hermitian matrix
are given by the singular values of $M$ with both signs (see App.~\ref{app:JW}).
The largest one leads to
\begin{equation}
    \frac{3\sqrt{2}\,s}{64\pi^3\Lambda^2}\,
    |C_\varphi|
    \leq 1.
\end{equation}

\begin{table}[tbp]
\renewcommand{\arraystretch}{1.5}
\centering
\begin{tabular}{c|cc|c}
\toprule
\rowcolor{hdrcolor}
\boldmath$\varphi^6$
& \textbf{Channel} & \boldmath$N\to M$
& \textbf{Unitarity bound} \\
\hline
\multirow{4}{*}{$C_\varphi$}
& $\varphi\varphi\varphi^\dagger\to\varphi\varphi\varphi^\dagger$ & $3\to3$
& $32\pi^3/3$ \\
& $\varphi\varphi\varphi\to\varphi\varphi\varphi$ & $3\to3$
& $128\pi^3/3$ \\
& $\varphi\varphi^\dagger\to\varphi\varphi\varphi^\dagger\varphi^\dagger$ & $2\to4$
& $32\pi^3\sqrt{2}/3$ \\
& $\varphi\varphi\to\varphi\varphi\varphi\varphi^\dagger$ & $2\to4$
& $64\pi^3/\sqrt{3}$ \\
\bottomrule
\end{tabular}
\caption{
Partial-wave unitarity bounds on the modulus of $C_\varphi$ obtained from
the independent $J=0$ scattering sectors generated by $O_\varphi$.
Channels related by CP conjugation, such as
$\varphi\varphi^\dagger\varphi^\dagger\to\varphi\varphi^\dagger\varphi^\dagger$,
yield identical bounds and are not listed separately.
The factors $s/\Lambda^2$ are left implicit.
}
\label{tab:phi6-bounds-example}
\end{table}

The remaining independent blocks can be analyzed in the same way.
Their bounds are collected in Table~\ref{tab:phi6-bounds-example}.
The strongest constraint is obtained from the
$\varphi\varphi\varphi^\dagger\to\varphi\varphi\varphi^\dagger$ sector,
\begin{equation}
    \frac{s}{\Lambda^2}|C_\varphi|
    \leq
    \frac{32\pi^3}{3}.
\end{equation}
This example illustrates two features of the generalized partial-wave
formalism. First, higher-multiplicity channels are indispensable for operators whose energy-growing contribution is not captured by $2\to2$ scattering. Second, channels with different initial- and final-state multiplicities can lead to parametrically comparable constraints once the normalization of the multiparticle phase space and the symmetry factors of identical states are properly included.

\subsubsection{Example: \texorpdfstring{$O_{\varphi e}$}{Ophi e} operator}
\label{sec:example OHe}

A second instructive example is provided by the operator\footnote{
We define $\varphi^\dagger
    \overset{\leftrightarrow}{D}_{\mu}
    \varphi\equiv \varphi^\dag (D_\mu \varphi)-(D_\mu\varphi)^\dag \varphi$
and $\varphi^\dagger
    \overset{\leftrightarrow}{D}{}_{\mu}^I
    \varphi\equiv \varphi^\dag \sigma^I(D_\mu \varphi)-(D_\mu\varphi)^\dag \sigma^I\varphi$.}
\begin{equation}
    O_{\varphi e}
    =
    \Big(
    \varphi^\dagger
    i\overset{\leftrightarrow}{D}_{\mu}
    \varphi
    \Big)
    \left(
    \overline e_p \gamma^\mu e_r
    \right).
\end{equation}
Unlike the purely scalar example discussed above, this interaction illustrates
the non-trivial interplay between partial waves with different total angular
momentum.
At the $2\to2$ level, the relevant amplitudes are
\begin{align}
    \mathcal{A}
    \Big(
    1_p^+,\overline 2_r^-
    \to
    3_a,\overline 4_b
    \Big)
    &=
    \frac{2}{\Lambda^2}
    [C_{\varphi e}]_{rp}\,
    \delta^{a}_{b}
    [23]\langle31\rangle ,
    \\
    \mathcal{A}
    \Big(
    1_p^+,2_a
    \to
    3_r^+,4_b
    \Big)
    &=
    \frac{2}{\Lambda^2}
    [C_{\varphi e}]_{rp}\,
    \delta_{a}^{b}
    [23]\langle21\rangle ,
\end{align}
where $p,r=1,2,3$ denote flavor indices and $a,b=1,2$ are
$\mathrm{SU}(2)_L$ indices.
For a right-handed fermion the physical helicity is $h=+1/2$.
The helicity differences therefore imply that the annihilation channel
$e\overline e\to\varphi\varphi^\dagger$ starts at $J=1$, whereas the
scattering channel $e\varphi\to e\varphi$ admits $J=1/2$.

To illustrate the relative strength of the two sectors, we work in the
single-flavor limit.
Projecting the annihilation amplitude onto the $J=1$ partial wave gives
\begin{equation}
    a^1_{i\to f}
    =
    \begin{pmatrix}
        \mathbb{O} & M \\
        M^\dagger & \mathbb{O}
    \end{pmatrix},
    \qquad
    M
    =
    -\frac{\sqrt{2}\,s}{24\pi\Lambda^2}
    C_{\varphi e}
    \begin{pmatrix}
        1 & 0 & 0 & 1
    \end{pmatrix},
\end{equation}
with
\begin{equation}
    i,f\in
    \left\{
    e\overline e,\,
    \varphi\varphi^\dagger
    \right\}.
\end{equation}
The rectangular block is $1\times4$, corresponding to the four independent
weak-isospin configurations
\begin{equation}
    (a,b)\in
    \{1,2\}\otimes\{1,2\}.
\end{equation}
The non-vanishing eigenvalues of the full Hermitian matrix are again given,
up to a sign, by the singular values of $M$ (see App.~\ref{app:JW}).
Since $M$ has rank one, its only non-vanishing singular value, $\sqrt{2}\times\sqrt{2}s|C_{\varphi e}|/(24\pi\Lambda^2)$, leads to
\begin{equation}
    \frac{s}{12\pi\Lambda^2}
    |C_{\varphi e}|
    \leq 1.
\end{equation}

The crossed channel $e\varphi\to e\varphi$ instead contributes already at
$J=1/2$.
Its partial-wave matrix is
\begin{equation}
    a^{1/2}_{e\varphi\to e\varphi}
    =
    \frac{s}{8\pi\Lambda^2}
    C_{\varphi e}
    \begin{pmatrix}
        1 & 0 \\
        0 & 1
    \end{pmatrix},
\end{equation}
and therefore yields
\begin{equation}
    \frac{s}{8\pi\Lambda^2}
    |C_{\varphi e}|
    \leq 1.
\end{equation}
For $O_{\varphi e}$, the $J=1/2$ channel is thus the most constraining one.

This conclusion should not, however, be interpreted as a general rule that
the smallest kinematically allowed angular momentum always gives the strongest
unitarity bound.
The strength of a coupled-channel constraint depends not only on the partial
wave itself, but also on the number and structure of the internal gauge
configurations that contribute coherently to the scattering matrix.
For operators involving fermions with non-trivial weak-isospin or color
quantum numbers, the enhancement associated with these additional degrees of
freedom can make a higher-$J$ sector more restrictive.
As discussed in Sec.~\ref{sec:fermionic_bilinears}, this occurs for several 
operators of the class $\psi^2\varphi^2D$, for which the $J=1$ bound can 
supersede the corresponding $J=1/2$ constraint.

\subsection{Results}
Using the framework described above, we systematically compute the
partial-wave unitarity bounds for the complete set of
baryon-number-conserving dimension-six SMEFT operators in the Warsaw basis \cite{Grzadkowski:2010es}, collected in Table~\ref{tab:WarsawBasis}.
For each operator class, we include all scattering channels displaying the
relevant high-energy behavior and account for all allowed helicity, gauge,
and, whenever applicable, flavor configurations. The resulting marginalized bounds are summarized in Table~\ref{tab:UniBounds}.
For operators involving fermions, the results presented in this section are
obtained in the single-flavor limit; the extension of the four-fermion
sector to three generations will be discussed in Sec.~\ref{sec:flavorful}.

\begin{table}[tbp]
\renewcommand{\arraystretch}{1.5}
    \centering
    \resizebox{\textwidth}{!}{
    \begin{tabular}{c | c|c | c|c| c|c| c|c| c}
    \toprule
    \rowcolor{hdrcolor}
    \multicolumn{2}{c|}{{\boldmath$X^3$}} & \multicolumn{2}{c|}{{\boldmath$\varphi^6$}/{\boldmath$\varphi^4D^2$}} & \multicolumn{2}{c|}{{\boldmath$\psi^2\varphi^3$}} & \multicolumn{2}{c|}{{\boldmath$(\overline{L}L)(\overline{L}L)$}} & \multicolumn{2}{c}{{\boldmath$(\overline{L}R)(\overline{R}L)$}/{\boldmath$(\overline{L}R)(\overline{L}R)$}} \\
    \hline
    $O_G$ & $f^{ABC} G_\mu^{A\nu} G_\nu^{B\rho} G_\rho^{C\mu}$ & $O_\varphi$ & $(\varphi^\dagger \varphi)^3$ & $^\ddagger O_{e\varphi}$ & $(\varphi^\dagger \varphi)(\overline{\ell}_p e_r \varphi)$ & $O_{\ell\ell}$ & $(\overline{\ell}_p \gamma_\mu \ell_r)(\overline{\ell}_s \gamma^\mu \ell_t)$ & $^\ddagger O_{\ell edq}$ & $(\overline{\ell}_p^j e_r)(\overline{d}_s q_{tj})$ \\
    
    $O_{\widetilde{G}}$ & $f^{ABC} \widetilde{G}_\mu^{A\nu} G_\nu^{B\rho} G_\rho^{C\mu}$ & $O_{\varphi\Box}$ & $(\varphi^\dagger \varphi)\Box(\varphi^\dagger \varphi)$ & $^\ddagger O_{u\varphi}$ & $(\varphi^\dagger \varphi)(\overline{q}_p u_r \widetilde{\varphi})$ & $O_{qq}^{(1)}$ & $(\overline{q}_p \gamma_\mu q_r)(\overline{q}_s \gamma^\mu q_t)$ & $^\ddagger O_{quqd}^{(1)}$ & $(\overline{q}_p^j u_r)\epsilon_{jk}(\overline{q}_s^k d_t)$ \\
    
    $O_W$ & $\epsilon^{IJK} W_\mu^{I\nu} W_\nu^{J\rho} W_\rho^{K\mu}$ & $O_{\varphi D}$ & $(\varphi^\dagger D^\mu \varphi)^* (\varphi^\dagger D_\mu \varphi)$ & $^\ddagger O_{d\varphi}$ & $(\varphi^\dagger \varphi)(\overline{q}_p d_r \varphi)$ & $O_{qq}^{(3)}$ & $(\overline{q}_p \gamma_\mu \sigma^I q_r)(\overline{q}_s \gamma^\mu \sigma^I q_t)$ & $^\ddagger O_{quqd}^{(8)}$ & $(\overline{q}_p^j T^A u_r)\epsilon_{jk}(\overline{q}_s^k T^A d_t)$ \\
    
    $O_{\widetilde{W}}$ & $\epsilon^{IJK} \widetilde{W}_\mu^{I\nu} W_\nu^{J\rho} W_\rho^{K\mu}$ & & & & & $O_{\ell q}^{(1)}$ & $(\overline{\ell}_p \gamma_\mu \ell_r)(\overline{q}_s \gamma^\mu q_t)$ & $^\ddagger O_{\ell equ}^{(1)}$ & $(\overline{\ell}_p^j e_r)\epsilon_{jk}(\overline{q}_s^k u_t)$ \\
    
    & & & & & & $O_{\ell q}^{(3)}$ & $(\overline{\ell}_p \gamma_\mu \sigma^I \ell_r)(\overline{q}_s \gamma^\mu \sigma^I q_t)$ & $^\ddagger O_{\ell equ}^{(3)}$ & $(\overline{\ell}_p^j \sigma_{\mu\nu} e_r)\epsilon_{jk}(\overline{q}_s^k \sigma^{\mu\nu} u_t)$ \\
    
    \midrule
    \rowcolor{hdrcolor}
    \multicolumn{2}{c|}{{\boldmath$X^2\varphi^2$}} & \multicolumn{2}{c|}{{\boldmath$\psi^2 X \varphi$}} & \multicolumn{2}{c|}{{\boldmath$\psi^2\varphi^2D$}} & \multicolumn{2}{c|}{{\boldmath$(\overline{R}R)(\overline{R}R)$}} & \multicolumn{2}{c}{{\boldmath$(\overline{L}L)(\overline{R}R)$}} \\
    \hline
    $O_{\varphi G}$ & $(\varphi^\dagger \varphi) G_{\mu\nu}^A G^{A\mu\nu}$ & $^\ddagger O_{eW}$ & $(\overline{\ell}_p \sigma^{\mu\nu} e_r)\sigma^I \varphi W_{\mu\nu}^I$ & $O_{\varphi\ell}^{(1)}$ & $(\varphi^\dagger i\overset{\leftrightarrow}{D}{}_\mu \varphi)(\overline{\ell}_p \gamma^\mu \ell_r)$ & $O_{ee}$ & $(\overline{e}_p \gamma_\mu e_r)(\overline{e}_s \gamma^\mu e_t)$ & $O_{\ell e}$ & $(\overline{\ell}_p \gamma_\mu \ell_r)(\overline{e}_s \gamma^\mu e_t)$ \\
    
    $O_{\varphi \widetilde{G}}$ & $(\varphi^\dagger \varphi) \widetilde{G}_{\mu\nu}^A G^{A\mu\nu}$ & $^\ddagger O_{eB}$ & $(\overline{\ell}_p \sigma^{\mu\nu} e_r)\varphi B_{\mu\nu}$ & $O_{\varphi\ell}^{(3)}$ & $(\varphi^\dagger i\overset{\leftrightarrow}{D}{}_\mu^I \varphi)(\overline{\ell}_p \sigma^I \gamma^\mu \ell_r)$ & $O_{uu}$ & $(\overline{u}_p \gamma_\mu u_r)(\overline{u}_s \gamma^\mu u_t)$ & $O_{\ell u}$ & $(\overline{\ell}_p \gamma_\mu \ell_r)(\overline{u}_s \gamma^\mu u_t)$ \\
    
    $O_{\varphi W}$ & $(\varphi^\dagger \varphi) W_{\mu\nu}^I W^{I\mu\nu}$ & $^\ddagger O_{uG}$ & $(\overline{q}_p \sigma^{\mu\nu} T^A u_r)\widetilde{\varphi} G_{\mu\nu}^A$ & $O_{\varphi e}$ & $(\varphi^\dagger i\overset{\leftrightarrow}{D}{}_\mu \varphi)(\overline{e}_p \gamma^\mu e_r)$ & $O_{dd}$ & $(\overline{d}_p \gamma_\mu d_r)(\overline{d}_s \gamma^\mu d_t)$ & $O_{\ell d}$ & $(\overline{\ell}_p \gamma_\mu \ell_r)(\overline{d}_s \gamma^\mu d_t)$ \\
    
    $O_{\varphi \widetilde{W}}$ & $(\varphi^\dagger \varphi) \widetilde{W}_{\mu\nu}^I W^{I\mu\nu}$ & $^\ddagger O_{uW}$ & $(\overline{q}_p \sigma^{\mu\nu} u_r)\sigma^I \widetilde{\varphi} W_{\mu\nu}^I$ & $O_{\varphi q}^{(1)}$ & $(\varphi^\dagger i\overset{\leftrightarrow}{D}{}_\mu \varphi)(\overline{q}_p \gamma^\mu q_r)$ & $O_{eu}$ & $(\overline{e}_p \gamma_\mu e_r)(\overline{u}_s \gamma^\mu u_t)$ & $O_{qe}$ & $(\overline{q}_p \gamma_\mu q_r)(\overline{e}_s \gamma^\mu e_t)$ \\
    
    $O_{\varphi B}$ & $(\varphi^\dagger \varphi) B_{\mu\nu} B^{\mu\nu}$ & $^\ddagger O_{uB}$ & $(\overline{q}_p \sigma^{\mu\nu} u_r)\widetilde{\varphi} B_{\mu\nu}$ & $O_{\varphi q}^{(3)}$ & $(\varphi^\dagger i\overset{\leftrightarrow}{D}{}_\mu^I \varphi)(\overline{q}_p \sigma^I \gamma^\mu q_r)$ & $O_{ed}$ & $(\overline{e}_p \gamma_\mu e_r)(\overline{d}_s \gamma^\mu d_t)$ & $O_{qu}^{(1)}$ & $(\overline{q}_p \gamma_\mu q_r)(\overline{u}_s \gamma^\mu u_t)$ \\
    
    $O_{\varphi \widetilde{B}}$ & $(\varphi^\dagger \varphi) \widetilde{B}_{\mu\nu} B^{\mu\nu}$ & $^\ddagger O_{dG}$ & $(\overline{q}_p \sigma^{\mu\nu} T^A d_r)\varphi G_{\mu\nu}^A$ & $O_{\varphi u}$ & $(\varphi^\dagger i\overset{\leftrightarrow}{D}{}_\mu \varphi)(\overline{u}_p \gamma^\mu u_r)$ & $O_{ud}^{(1)}$ & $(\overline{u}_p \gamma_\mu u_r)(\overline{d}_s \gamma^\mu d_t)$ & $O_{qu}^{(8)}$ & $(\overline{q}_p \gamma_\mu T^A q_r)(\overline{u}_s \gamma^\mu T^A u_t)$ \\
    
    $O_{\varphi WB}$ & $(\varphi^\dagger \sigma^I \varphi) W_{\mu\nu}^I B^{\mu\nu}$ & $^\ddagger O_{dW}$ & $(\overline{q}_p \sigma^{\mu\nu} d_r)\sigma^I \varphi W_{\mu\nu}^I$ & $O_{\varphi d}$ & $(\varphi^\dagger i\overset{\leftrightarrow}{D}{}_\mu \varphi)(\overline{d}_p \gamma^\mu d_r)$ & $O_{ud}^{(8)}$ & $(\overline{u}_p \gamma_\mu T^A u_r)(\overline{d}_s \gamma^\mu T^A d_t)$ & $O_{qd}^{(1)}$ & $(\overline{q}_p \gamma_\mu q_r)(\overline{d}_s \gamma^\mu d_t)$ \\
    
    $O_{\varphi \widetilde{W}B}$ & $(\varphi^\dagger \sigma^I \varphi) \widetilde{W}_{\mu\nu}^I B^{\mu\nu}$ & $^\ddagger O_{dB}$ & $(\overline{q}_p \sigma^{\mu\nu} d_r)\varphi B_{\mu\nu}$ & $^\ddagger O_{\varphi ud}$ & $(\widetilde{\varphi}^\dagger i D_\mu \varphi)(\overline{u}_p \gamma^\mu d_r)$ & & & $O_{qd}^{(8)}$ & $(\overline{q}_p \gamma_\mu T^A q_r)(\overline{d}_s \gamma^\mu T^A d_t)$ \\
    \bottomrule
    \end{tabular}
    }
    \caption{Dimension-six operators of the SMEFT in the Warsaw basis \cite{Grzadkowski:2010es} that conserve baryon number. Non-Hermitian operators are denoted with $^\ddagger O$ and the corresponding Hermitian conjugates must be included in the Lagrangian. Here $p,r,s,t$ are flavor indices, $j,k$ are $\mathrm{SU}(2)_L$ indices, $\epsilon_{jk}$ is the antisymmetric tensor with $\epsilon_{12}=+1$, $\sigma^I$ are the Pauli matrices, $T^A$ are the $\mathrm{SU}(3)_C$ generators, $\widetilde\varphi=i\sigma^2\varphi^*$, and $\widetilde X_{\mu\nu}=\frac{1}{2}\epsilon_{\mu\nu\rho\sigma}X^{\rho\sigma}$ with $\epsilon_{0123}=+1$.}
    \label{tab:WarsawBasis}
\end{table}

\begin{table}[tbp]
\renewcommand{\arraystretch}{1.5}
    \centering
    \resizebox{\columnwidth}{!}{
    \begin{tabular}{c|c|c|c|c|c|c|c|c|c}
    \toprule
    \rowcolor{hdrcolor}
    \multicolumn{2}{c|}{{\boldmath$X^3$}} & \multicolumn{2}{c|}{{\boldmath$\varphi^6$}/{\boldmath$\varphi^4D^2$}} & \multicolumn{2}{c|}{{\boldmath$   \psi^2\varphi^3$}} & \multicolumn{2}{c|}{{\boldmath$(\overline{L}L)(\overline{L}L)$}} & \multicolumn{2}{c}{{\boldmath$(\overline{L}R)(\overline{R}L)$}/{\boldmath$(\overline{L}R)(\overline{L}R)$}}\\
    \hline
    $C_G$ & $4\pi/(9g_s)$ & $C_\varphi$ & $32\pi^3/3$ & $C_{e\varphi}$ & $32\pi^2/\sqrt{3}$ & $C_{\ell\ell}$ & $2\pi$ & $C_{\ell e dq}$ & $8\pi/\sqrt{3}$ \\
    
    $C_{\widetilde G}$ & $4\pi/(9g_s)$ & $C_{\varphi\Box}$ & $8\pi/3$ & $C_{u\varphi}$ & $32\pi^2/3$ & $C_{qq}^{(1)}$ & $6\pi/5$ & $C_{quqd}^{(1)}$ & $16\pi(1+\sqrt{2})/9$\\ 
        
    $C_W$ & $2\pi/(3g)$ &$C_{\varphi D}$ & $32\pi/3$ & $C_{d\varphi}$ & $32\pi^2/3$ & $C_{qq}^{(3)}$ & $4\pi/3$ & $C_{quqd}^{(8)}$ & $4\pi(2+7\sqrt{2})/3$\\
    
    $C_{\widetilde W}$ & $2\pi/(3g)$ & & & & & $C_{\ell q}^{(1)}$ & $2\pi\sqrt{3}$ & $C_{\ell equ}^{(1)}$ & $8\pi/\sqrt{3}$ \\

    &&&&&&  $C_{\ell q}^{(3)}$ & $2\pi$ & $C_{\ell e qu}^{(3)}$ & $(2+\sqrt{2})\pi/3$   \\
    
         \midrule
         \rowcolor{hdrcolor}
         \multicolumn{2}{c|}{{\boldmath$X^2\varphi^2$}} & \multicolumn{2}{c|}{{\boldmath$\psi^2 X \varphi$}} & \multicolumn{2}{c|}{{\boldmath$\psi^2\varphi^2D$}} & \multicolumn{2}{c|}{{\boldmath$(\overline{R}R)(\overline{R}R)$}} & \multicolumn{2}{c}{{\boldmath$(\overline{L}L)(\overline{R}R)$}}\\
         \hline
         $C_{\varphi G}$ & $\pi$ & $C_{eW}$ & $4\pi\sqrt{2/3}$ & $C_{\varphi \ell}^{(1)}$ & $8\pi$ & $C_{ee}$ & $2\pi$ & $C_{\ell e}$ & $4\pi$ \\
         
         $C_{\varphi \widetilde G}$ & $\pi$ & $C_{eB}$ & $4\pi$ & $C_{\varphi \ell}^{(3)}$ & $8\pi/3$ & $C_{uu}$ & $3\pi/2$ & $C_{\ell u}$ & $4\pi$\\
         
         $C_{\varphi W}$ & $2\pi\sqrt{2/3}$ & $C_{uG}$ & $2\pi\sqrt{3}$ & $C_{\varphi e}$ & $8\pi$ & $C_{dd}$ & $3\pi/2$ & $C_{\ell d}$ & $4\pi$\\
         
         $C_{\varphi \widetilde W}$ & $2\pi\sqrt{2/3}$ & $C_{uW}$ & $4\pi\sqrt{2/3}$ & $C_{\varphi q}^{(1)}$ & $2\pi\sqrt{6}$ & $C_{eu}$ & $4\pi$ & $C_{q e}$ & $4\pi$ \\
         
        $C_{\varphi B}$ & $2\pi\sqrt{2}$ & $C_{uB}$ & $4\pi$ & $C_{\varphi q}^{(3)}$ & $8\pi/3$ & $C_{ed}$ & $4\pi$ & $C_{q u}^{(1)}$ & $2\pi\sqrt{2}$\\
        
        $C_{\varphi \widetilde B}$ & $2\pi\sqrt{2}$ & $C_{dG}$ & $2\pi\sqrt{3}$ & $C_{\varphi u}$ & $4\pi \sqrt{3}$ & $C_{ud}^{(1)}$ & $4\pi$ & $C_{q u}^{(8)}$ & $3\pi(1+1/\sqrt{2})$\\
        
        $C_{\varphi W B}$ & $4\pi$ & $C_{dW}$ & $4\pi\sqrt{2/3}$ & $C_{\varphi d}$ & $4\pi\sqrt{3}$ & $C_{ud}^{(8)}$ & $8\pi$ & $C_{qd}^{(1)}$ & $2\pi\sqrt{2}$\\
        
        $C_{\varphi \widetilde W B}$ & $4\pi$ & $C_{dB}$ & $4\pi$ & $C_{\varphi ud}$ & $8\pi$ & & & $C_{q d}^{(8)}$ & $3\pi(1+1/\sqrt{2})$ \\
        \bottomrule
    \end{tabular}
    }
\caption{Unitarity bounds on the modulus of the baryon-number-conserving dimension-six SMEFT Wilson coefficients in the Warsaw basis~\cite{Grzadkowski:2010es} (Table~\ref{tab:WarsawBasis}), obtained by marginalizing over all other coefficients. The factors $s/\Lambda^2$ are left implicit, i.e., the reported values constrain the combination $s|C|/\Lambda^2$. Fermionic operators are considered in the single-flavor limit.}
\label{tab:UniBounds}
\end{table}

An important feature of our analysis is that the bound on a given Wilson
coefficient is not obtained by switching off all the remaining ones.
Instead, coefficients contributing to the same partial-wave amplitudes are
treated simultaneously, and the limit on each coefficient is obtained by
marginalizing over the others.
Since the partial-wave amplitudes depend linearly on the Wilson coefficients,
the unitarity conditions define a convex allowed region in coefficient space,
whose projection onto a given coefficient axis yields the corresponding
marginalized bound.
Since the complete set of scattering channels is included, this region is
bounded in all independent directions of coefficient space.
The bounds reported in Table~\ref{tab:UniBounds} should therefore be
interpreted as conservative one-dimensional projections of the full
coupled-channel unitarity region, rather than as single-operator bounds.

This distinction becomes particularly relevant whenever several Wilson
coefficients contribute to the same scattering amplitudes.
This occurs, for instance, for the pair
$(O_{\varphi\Box},O_{\varphi D})$, for operators involving a field-strength
tensor and its dual, and for fermionic operators that differ only in the singlet or non-singlet
$\mathrm{SU}(2)_L$ or $\mathrm{SU}(3)_C$ contraction of their bilinears, such as $(O_{\ell q}^{(1)},O_{\ell q}^{(3)})$ or
$(O_{qu}^{(1)},O_{qu}^{(8)})$.
Keeping these correlations is essential for obtaining bounds that are
independent of arbitrary one-operator-at-a-time assumptions.

To fully exploit the coupled-channel construction, for each operator class
we consider all possible initial and final states in the allowed helicity
and gauge configurations.
At dimension six, the numerically leading constraints frequently arise
from the lowest-multiplicity amplitudes.
Nevertheless, the general $N\to M$ formalism of
Ref.~\cite{Bresciani:2025toe} is essential for treating the complete basis
in a uniform way.
In particular, higher-multiplicity amplitudes are required for
$O_\varphi$ and for the $\psi^2\varphi^3$ class, as discussed in
Secs.~\ref{sec:example O_phi} and~\ref{sec:fermionic_bilinears},
respectively.

Compared with previous SMEFT analyses
\cite{Corbett:2014ora,Corbett:2017qgl,Cao:2023qks}, our calculation covers
the complete set of four-fermion interactions and all operators involving
gluons.
The fully coupled-channel treatment can also strengthen the bounds whenever
several internal gauge configurations contribute coherently to the same
partial-wave matrix. In the following we discuss separately the purely bosonic operators, operators containing a fermionic bilinear, and four-fermion interactions.

\subsubsection{Purely bosonic operators}
\label{sec:bosonic_unitarity}

We begin with the purely bosonic sector.
For the operator classes whose leading high-energy behavior is captured by
$2\to2$ scattering, a representative set of channels contributing to the
$J=0$ partial wave is
\begin{equation}
    \varphi_i\varphi_j^{(\dagger)}
    \to
    \varphi_k\varphi_\ell^{(\dagger)},
    \qquad
    X^I X^J
    \to
    \varphi_i\varphi_j^\dagger,
    \qquad
    X^I X^J
    \to
    X^K X^L ,
\end{equation}
where $i,j,k,\ell$ denote $\mathrm{SU}(2)_L$ fundamental or
antifundamental indices, while $I,J,K,L$ denote adjoint gauge
indices and $X=B,W,G$.
All states with identical conserved quantum numbers are included
simultaneously in the corresponding coupled-channel matrices.

For illustration, with the $X^3$ coefficients set to zero, the weak-isospin- and color-singlet and the
weak-isospin-triplet $J=0$ sectors that contain the state $\varphi\varphi^\dagger$ lead, respectively, to the constraints
\begin{gather}
    \frac{s^2}{\Lambda^4}\sum_{X=B,W,G}n_X\abs{C_{\varphi X}+iC_{\varphi\widetilde X}}^2
    \leq
    \pi\left(8\pi-\frac{3s}{\Lambda^2}\abs{C_{\varphi\Box}}\right),
    \label{eq:bosonic_unitarity_1}
    \\
    \frac{s^2}{\Lambda^4}\abs{C_{\varphi WB}+iC_{\varphi\widetilde W B}}^2
    \leq
    2\pi\left(8\pi-\frac{s}{\Lambda^2}\abs{C_{\varphi D}-C_{\varphi\Box}}\right),
    \label{eq:bosonic_unitarity_2}
\end{gather}
where $n_B=1$, $n_W=3$, and $n_G=8$.
With the $X^3$ coefficients switched on, the singlet matrix also contains the entries $X^\pm X^\pm\to X^\mp X^\mp$, and Eq.~\eqref{eq:bosonic_unitarity_1}
generalizes to
\begin{equation}
    \frac{s^2}{\Lambda^4}\sum_{X=B,W,G} n_X\,
    \frac{\abs{\mathcal C_{\varphi X}}^2
    \mp \rho_X\,\mathrm{Re}\big(\mathcal C_{\varphi X}^2\,\mathcal C_X^*\big)}
    {1-\rho_X^2\abs{\mathcal C_X}^2}
    \leq
    \pi\left(8\pi\pm \frac{3s}{\Lambda^2}\,C_{\varphi\Box}\right),
    \label{eq:bosonic_unitarity_X3}
\end{equation}
where $\mathcal C_{\varphi X}\equiv C_{\varphi X}+iC_{\varphi\widetilde X}$,
$\mathcal C_{X}\equiv C_{X}+iC_{\widetilde X}$, $\rho_B=0$,
$\rho_W=3gs/(2\pi\Lambda^2)$, and $\rho_G=9g_ss/(4\pi\Lambda^2)$, with
$\rho_W\abs{\mathcal C_W}< 1$ and $\rho_G\abs{\mathcal C_G}< 1$.
Marginalizing these and the remaining independent coupled-channel
conditions over the other Wilson coefficients leads to the limits reported
in Table~\ref{tab:UniBounds}.
The corresponding $J=1,2$ sectors have also been included in the analysis,
but, for the purely bosonic operators, they yield weaker constraints than
those arising from $J=0$.

The operators $O_{\varphi\Box}$ and $O_{\varphi D}$ provide a simple
illustration of the importance of retaining correlations among Wilson
coefficients.
Both contribute to common Higgs scattering amplitudes, and unitarity
therefore constrains a two-dimensional region in the
$(C_{\varphi\Box},C_{\varphi D})$ plane rather than two independent
intervals.
The corresponding allowed region is shown in
Fig.~\ref{fig:uniCHbox,D} (left) and the unitarity conditions are listed in Table~\ref{tab:phi2D_channels}.
The marginalized limits quoted in Table~\ref{tab:UniBounds} are obtained
by projecting this region onto the two coefficient axes.
Consequently, imposing $C_{\varphi\Box}=0$ from the outset would lead to
an artificially stronger constraint on $C_{\varphi D}$.

\begin{figure}[tbp]
\centering
    \includegraphics[width=0.4\textwidth]{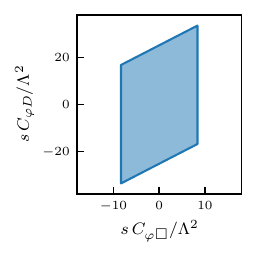}
    \hspace{0.05\textwidth}
    \includegraphics[width=0.4\textwidth]{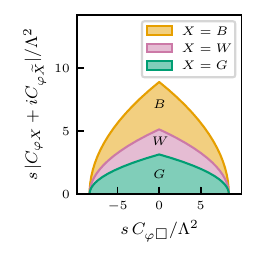}
    \caption{
    Parameter space allowed by partial-wave unitarity in the planes
    $(C_{\varphi\Box},C_{\varphi D})$ (left) and
    $(C_{\varphi\Box},|C_{\varphi X}+iC_{\varphi\widetilde X}|)$ for $X=B,W,G$ (right).
    The left region is bounded by the weak-isospin singlet and triplet
    conditions of the $J=0$ channel $\varphi\varphi^\dagger\to\varphi\varphi^\dagger$
    listed in Table~\ref{tab:phi2D_channels}.
    The right regions follow from the $J=0$ weak-isospin- and color-singlet condition of
    Eq.~\eqref{eq:bosonic_unitarity_1}, for one $X$ at a time and with all other
    Wilson coefficients set to zero.
    In both panels, the extent of each region along each axis gives the marginalized
    bounds of Table~\ref{tab:UniBounds}.}
    \label{fig:uniCHbox,D}
\end{figure}

\begin{table}[tbp]
\renewcommand{\arraystretch}{1.5}
\centering
\begin{tabular}{c|cc|cc}
\toprule
\rowcolor{hdrcolor}
\boldmath$\varphi^4D^2$
& \textbf{Channel} & \boldmath$J$
& \multicolumn{2}{c}{\textbf{Unitarity conditions}} \\
\hline
\multirow{2}{*}{$C_{\varphi\Box},\,C_{\varphi D}$}
& $\varphi\varphi^\dagger\to\varphi\varphi^\dagger$ & $0$
& $|C_{\varphi\Box}|\leq8\pi/3$
& $|C_{\varphi\Box}-C_{\varphi D}|\leq8\pi$ \\
& $\varphi\varphi\to\varphi\varphi$ & $0$
& \multicolumn{2}{c}{$|2C_{\varphi\Box}+C_{\varphi D}|\leq16\pi$} \\
\midrule
\rowcolor{hdrcolor}
\boldmath$\psi^2\varphi^2D$
& \textbf{Channel} & \boldmath$J$
& \multicolumn{2}{c}{\textbf{Unitarity conditions}} \\
\hline
\multirow{3}{*}{$C_{\varphi\ell}^{(1)},\,C_{\varphi\ell}^{(3)}$}
& $\varphi\ell\to\varphi\ell$ & $1/2$
& $|C_{\varphi\ell}^{(1)}-3C_{\varphi\ell}^{(3)}|\leq8\pi$
& $|C_{\varphi\ell}^{(1)}+C_{\varphi\ell}^{(3)}|\leq8\pi$ \\
& $\varphi^\dagger\ell\to\varphi^\dagger\ell$ & $1/2$
& $|C_{\varphi\ell}^{(1)}-C_{\varphi\ell}^{(3)}|\leq8\pi$
& $|C_{\varphi\ell}^{(1)}+3C_{\varphi\ell}^{(3)}|\leq8\pi$ \\
& $\ell\overline\ell\to\varphi\varphi^\dagger$ & $1$
& $|C_{\varphi\ell}^{(1)}|\leq6\pi\sqrt{2}$
& $|C_{\varphi\ell}^{(3)}|\leq6\pi\sqrt{2}$ \\
\hline
\multirow{3}{*}{$C_{\varphi q}^{(1)},\,C_{\varphi q}^{(3)}$}
& $\varphi q\to\varphi q$ & $1/2$
& $|C_{\varphi q}^{(1)}-3C_{\varphi q}^{(3)}|\leq8\pi$
& $|C_{\varphi q}^{(1)}+C_{\varphi q}^{(3)}|\leq8\pi$ \\
& $\varphi^\dagger q\to\varphi^\dagger q$ & $1/2$
& $|C_{\varphi q}^{(1)}-C_{\varphi q}^{(3)}|\leq8\pi$
& $|C_{\varphi q}^{(1)}+3C_{\varphi q}^{(3)}|\leq8\pi$ \\
& $q\overline q\to\varphi\varphi^\dagger$ & $1$
& $|C_{\varphi q}^{(1)}|\leq2\pi\sqrt{6}$
& $|C_{\varphi q}^{(3)}|\leq2\pi\sqrt{6}$ \\
\bottomrule
\end{tabular}
\caption{
Partial-wave unitarity conditions on correlated pairs of Wilson
coefficients of the classes $\varphi^4D^2$ and $\psi^2\varphi^2D$ in the
single-flavor limit, for each scattering channel and total angular
momentum $J$.
The corresponding allowed regions are shown in
Figs.~\ref{fig:uniCHbox,D} (left) and~\ref{fig:uni CHq,l 13}.
The factors $s/\Lambda^2$ are left implicit.
}
\label{tab:phi2D_channels}
\end{table}

The $X^2\varphi^2$ operators provide another example in which the
coupled-channel structure is important.
Coefficients multiplying a field-strength tensor and its dual enter the
helicity amplitudes through the combinations
$C_{\varphi X}\pm iC_{\varphi\widetilde X}$ (and $C_{\varphi WB}\pm iC_{\varphi\widetilde WB}$), the sign depending on the
helicities of the gauge bosons.
The corresponding CP-even and CP-odd interactions must therefore be
treated simultaneously when deriving their marginalized bounds.
Figure~\ref{fig:uniCHbox,D} (right) shows the regions allowed by
Eq.~\eqref{eq:bosonic_unitarity_1} in the planes
$(C_{\varphi\Box},|C_{\varphi X}+iC_{\varphi\widetilde X}|)$: the singlet channel
correlates the $\varphi^4D^2$ and $X^2\varphi^2$ classes, and the bounds on $C_{\varphi X}$,
attained at $C_{\varphi\Box}=0$, scale as $1/\sqrt{n_X}$.

A separate remark concerns the $X^3$ operators.
In this case the relevant unitarity bounds are computed from amplitudes
containing the SM contribution and at most one insertion of a
dimension-six interaction, thereby truncating the amplitude consistently
at $\mathcal{O}(1/\Lambda^2)$.
Terms containing two insertions of dimension-six operators would enter at
$\mathcal{O}(1/\Lambda^4)$ and could, in principle, modify the numerical
constraint, especially when the corresponding gauge coupling is small.
Including such contributions while neglecting dimension-eight operators,
however, would not constitute a consistent EFT truncation.
Indeed, positivity arguments imply that the presence of a dimension-six
$X^3$ interaction is accompanied by dimension-eight $X^4$ operators,
which would contribute at the same order in the EFT expansion
\cite{Ghosh:2022qqq}.

The remaining purely scalar operator,
$O_\varphi=(\varphi^\dagger\varphi)^3$, deserves special treatment.
Its dimension-six $2\to2$ contribution does not exhibit the energy growth
relevant for the present analysis, so that the corresponding unitarity
constraint arises from higher-multiplicity scattering.
As discussed explicitly in Sec.~\ref{sec:example O_phi}, up to CP conjugation the relevant
$J=0$ channels are
\begin{equation}
    \varphi_i\varphi_j\varphi_k^{(\dagger)}
    \to
    \varphi_\ell\varphi_m\varphi_n^{(\dagger)},
    \qquad
    \varphi_i\varphi_j^{(\dagger)}
    \to
    \varphi_k\varphi_\ell
    \varphi_m^{(\dagger)}\varphi_n^\dagger ,
\end{equation}
and the strongest of these, $\varphi\varphi\varphi^\dagger\to\varphi\varphi\varphi^\dagger$, gives
\begin{equation}
    \frac{s}{\Lambda^2}|C_\varphi|
    \leq
    \frac{32\pi^3}{3}.
\end{equation}
This provides a concrete example of why restricting the analysis to
$2\to2$ scattering would fail to probe the complete dimension-six SMEFT
basis.

Finally, our bounds on the purely bosonic operators considered in
Ref.~\cite{Cao:2023qks} agree with those reported there, up to the factor-of-two difference associated
with the normalization of the partial-wave unitarity condition.
In Ref.~\cite{Cao:2023qks} the convention
$|a^J_{i\to f}|\leq2$ was adopted, whereas throughout this work we use
Eq.~\eqref{eq:unitarity_bounds}.

\subsubsection{Operators with a fermionic bilinear}
\label{sec:fermionic_bilinears}

We next consider dimension-six operators containing a single fermionic
bilinear. Depending on their field content, their leading unitarity
constraints arise either from higher-multiplicity scattering or from
$2\to2$ processes involving different allowed angular-momentum sectors.
We therefore discuss separately the classes $\psi^2\varphi^3$,
$\psi^2X\varphi$, and $\psi^2\varphi^2D$.

\paragraph{\texorpdfstring{$\psi^2\varphi^3$}{psi2phi3} operators.}

The operators of the class $\psi^2\varphi^3$ provide a second
application of the generalized $N\to M$ formalism, first discussed in Ref.~\cite{Bresciani:2025toe}. In the Warsaw basis they are
\begin{equation}
    O_{e\varphi}
    =
    (\varphi^\dagger\varphi)
    (\overline{\ell}_p e_r \varphi),
    \qquad
    O_{d\varphi}
    =
    (\varphi^\dagger\varphi)
    (\overline{q}_p d_r \varphi),
    \qquad
    O_{u\varphi}
    =
    (\varphi^\dagger\varphi)
    (\overline{q}_p u_r \widetilde{\varphi}),
\end{equation}
together with their Hermitian conjugates.
After electroweak symmetry breaking these interactions modify the SM Yukawa
couplings, but in the unbroken high-energy theory they also generate genuine
five-point contact amplitudes.
Their leading unitarity constraints are therefore obtained from
$2\to3$ and crossed $3\to2$ processes rather than from the conventional
$2\to2$ partial-wave analysis.

To make the construction explicit, it is convenient to decompose the Higgs
doublet into four real scalar components,
\begin{equation}
    \varphi
    =
    \frac{1}{\sqrt{2}}
    \begin{pmatrix}
        \phi_1+i\phi_3\\
        \phi_2+i\phi_4
    \end{pmatrix}.
\end{equation}
The $J=0$ coupled-channel analysis then contains all contact amplitudes of the
schematic form
\begin{equation}
    \phi_a\phi_b
    \to
    \phi_c\psi_i \psi_j,
    \qquad
    \psi_i\psi_j
    \to
    \phi_a\phi_b\phi_c,
\end{equation}
with $a,b,c=1,\ldots,4$ and with all allowed fermion species, gauge
configurations, and flavor indices included.

As a representative example, consider the contribution of $O_{e\varphi}$.
For a suitable choice of external Higgs components, one of the non-vanishing
amplitudes is
\begin{equation}
    \mathcal{A}
    \Big(
        1_{\phi_1},2_{\phi_1}
        \to
        3_{\phi_1},4_{\overline \nu_p}^{+},5_{e_r}^{+}
    \Big)
    =
    \frac{3}{\sqrt{2}}
    \frac{[C_{e\varphi}]_{pr}^{*}}{\Lambda^2}
    [54].
    \label{eq:psi2phi3_amp}
\end{equation}
The relevant normalized $J=0$ basis element for this helicity configuration is
\begin{equation}
    \mathcal{B}^{0}_{1_{\phi_1},2_{\phi_1}
    \to
    3_{\phi_1},4_{\overline \nu_p}^{+},5_{e_r}^{+}}
    =
    \frac{32\sqrt{6}\,\pi^2}{s}\,[54].
    \label{eq:psi2phi3_basis}
\end{equation}
Projecting Eq.~\eqref{eq:psi2phi3_amp} onto
Eq.~\eqref{eq:psi2phi3_basis}, and including the appropriate
$1/\sqrt{2}$ normalization factor for the two identical scalars in the
initial state, gives
\begin{equation}
    a^0_{
    \phi_1\phi_1
    \to
    \phi_1\overline \nu_p^+ e_r^+}
    =
    \frac{\sqrt{3}\,s}
         {64\sqrt{2}\,\pi^2\Lambda^2}
    [C_{e\varphi}]_{pr}^{*}.
    \label{eq:psi2phi3_pw}
\end{equation}
This illustrates explicitly how a five-point contact interaction is projected
onto a definite-angular-momentum partial wave without reducing the process to
an effective $2\to2$ description.

The strongest constraint follows from the complete $J=0$ coupled-channel
matrix containing all $\psi^2\varphi^3$ interactions.
It can be written compactly as
\begin{equation}
    \frac{s}{\Lambda^2}
    \sqrt{
        \operatorname{Tr}
        \!\Big[
            3\,C_{u\varphi}^\dagger C_{u\varphi}
            +
            3\,C_{d\varphi}^\dagger C_{d\varphi}
            +
            C_{e\varphi}^\dagger C_{e\varphi}
        \Big]
    }
    \leq
    \frac{32\pi^2}{\sqrt{3}}.
    \label{eq:psi2phi3_unitarity}
\end{equation}
The trace runs over flavor space, while the factors of three multiplying the
quark coefficients originate from the color multiplicity.
Equation~\eqref{eq:psi2phi3_unitarity} makes explicit that the different
Yukawa-like SMEFT interactions participate in the same coupled-channel
unitarity condition.

In the single-flavor limit, the marginalization over the remaining
$\psi^2\varphi^3$ coefficients gives
\begin{equation}
    \frac{s}{\Lambda^2}|C_{e\varphi}|
    \leq
    \frac{32\pi^2}{\sqrt{3}},
    \qquad
    \frac{s}{\Lambda^2}|C_{u\varphi}|
    \leq
    \frac{32\pi^2}{3},
    \qquad
    \frac{s}{\Lambda^2}|C_{d\varphi}|
    \leq
    \frac{32\pi^2}{3},
\end{equation}
in agreement with the results reported in
Table~\ref{tab:UniBounds}.

The class $\psi^2\varphi^3$ therefore provides, together with
$O_\varphi$, a direct illustration of why higher-multiplicity scattering is
needed for a complete unitarity analysis of the dimension-six SMEFT.
While $O_\varphi$ requires six-point amplitudes and probes
$3\to3$ and $2\to4$ channels, the Yukawa-like operators are instead
controlled by five-point amplitudes and $2\to3$ scattering.

\paragraph{\texorpdfstring{$\psi^2X\varphi$}{psi2Xphi} and \texorpdfstring{$\psi^2\varphi^2D$}
{psi2phi2D} operators.}

We now turn to the classes $\psi^2X\varphi$ and $\psi^2\varphi^2D$, 
whose leading high-energy constraints are instead captured by $2\to2$ 
amplitudes with two fermionic and two bosonic external states.
The allowed total angular momentum is determined by the helicity differences
of the initial and final states.
In particular, half-integer values of $J$ arise when one fermion belongs to
the initial state and the other to the final state, so that the lowest
allowed partial wave is $J=1/2$.
A representative set of such channels is
\begin{gather}
    \psi_R^{\alpha,p}\varphi_j
    \to
    \psi_L^{i,\beta,r}X^I,
    \qquad
    \psi_R^{\alpha,p}X^I
    \to
    \psi_L^{i,\beta,r}\varphi_j^\dagger,
    \\
    \psi_R^{\alpha,p}\varphi_i^{(\dagger)}
    \to
    \psi_R^{\beta,r}\varphi_j^{(\dagger)},
    \qquad
    \psi_L^{k,\alpha,p}\varphi_i^{(\dagger)}
    \to
    \psi_L^{\ell,\beta,r}\varphi_j^{(\dagger)} .
\end{gather}
Here $p,r$ denote flavor indices, $\alpha,\beta$ are
$\mathrm{SU}(3)_C$ fundamental or antifundamental indices whenever
applicable, $i,j,k,\ell$ denote $\mathrm{SU}(2)_L$ indices, and $L,R$
specify the fermion chirality.
The first line contains chirality-changing amplitudes generated by operators
of the class $\psi^2X\varphi$, while the second is representative of
chirality-preserving interactions of the class $\psi^2\varphi^2D$.
The remaining $J=1/2$ channels are related to these by crossing and by
discrete symmetries.

Integer angular momentum instead arises when the two fermions belong to the
same side of the scattering process.
The lowest such contribution is at $J=1$, with representative channels
\begin{equation}
    \psi_R^{\alpha,p}\overline\psi_L^{i,\beta,r}
    \to
    X^I\varphi_j^\dagger,
    \qquad
    \psi_R^{\alpha,p}\overline\psi_R^{\beta,r}
    \to
    \varphi_i\varphi_j^\dagger,
    \qquad
    \psi_L^{k,\alpha,p}\overline\psi_L^{\ell,\beta,r}
    \to
    \varphi_i\varphi_j^\dagger .
\end{equation}
For these configurations, the minimum value $J=1$ follows directly from
the corresponding helicity differences.

We compute the coupled-channel unitarity bounds in both angular-momentum
sectors.
The results are summarized in
Table~\ref{tab: UB for psi^2 Xphi and psi^2phi^2D}, while the operator
$O_{\varphi e}$ was discussed explicitly in
Sec.~\ref{sec:example OHe}.
For operators of the class $\psi^2X\varphi$, the $J=1/2$ constraints are
always stronger than those obtained from $J=1$.
For the class $\psi^2\varphi^2D$, instead, the situation is
operator-dependent.
Although one might naively expect the lowest allowed angular momentum to
yield the strongest constraint, the enhancement associated with the
multiplicity of internal gauge degrees of freedom can increase the largest
eigenvalue of the $J=1$ coupled-channel matrix and make the corresponding
bound more restrictive.
This occurs, in particular, for operators involving quark bilinears, where
the color multiplicity can play an important role.

\begin{table}[tbp]
\renewcommand{\arraystretch}{1.5}
\centering
\begin{tabular}{c|cc||c|cc}
\toprule
\rowcolor{hdrcolor}
\boldmath$\psi^2X\varphi$
& \boldmath$J=1/2$
& \boldmath$J=1$
& \boldmath$\psi^2\varphi^2D$
& \boldmath$J=1/2$
& \boldmath$J=1$
\\
\hline
$C_{eW}$ & $4\pi\sqrt{2/3}$ & $4\pi\sqrt{3}$
& $C_{\varphi e}$ & $8\pi$ & $12\pi$ \\

$C_{eB}$ & $4\pi$ & $12\pi$
& $C_{\varphi u}$ & $8\pi$ & $4\pi\sqrt{3}$ \\

$C_{uW}$ & $4\pi\sqrt{2/3}$ & $4\pi$
& $C_{\varphi d}$ & $8\pi$ & $4\pi\sqrt{3}$ \\

$C_{dW}$ & $4\pi\sqrt{2/3}$ & $4\pi$
& $C_{\varphi ud}$ & $8\pi$ & $4\pi\sqrt{6}$ \\

$C_{uB}$ & $4\pi$ & $4\pi\sqrt{3}$
& $C_{\varphi\ell}^{(1)}$ & $8\pi$ & $6\pi\sqrt{2}$ \\

$C_{dB}$ & $4\pi$ & $4\pi\sqrt{3}$
& $C_{\varphi\ell}^{(3)}$ & $8\pi/3$ & $6\pi\sqrt{2}$ \\

$C_{uG}$ & $2\pi\sqrt{3}$ & $12\pi\sqrt{2}$
& $C_{\varphi q}^{(1)}$ & $8\pi$ & $2\pi\sqrt{6}$ \\

$C_{dG}$ & $2\pi\sqrt{3}$ & $12\pi\sqrt{2}$
& $C_{\varphi q}^{(3)}$ & $8\pi/3$ & $2\pi\sqrt{6}$ \\
\bottomrule
\end{tabular}
\caption{
Comparison of the partial-wave unitarity bounds on the modulus of the
individual Wilson coefficients associated with the operator classes
$\psi^2X\varphi$ and $\psi^2\varphi^2D$, obtained from the
$J=1/2$ and $J=1$ sectors.
The factors $s/\Lambda^2$ are left implicit.
}
\label{tab: UB for psi^2 Xphi and psi^2phi^2D}
\end{table}

For the operators in these classes, the correlations among Wilson
coefficients do not modify the one-dimensional marginalized limits, which coincide with the strongest single-coefficient
bounds of Table~\ref{tab: UB for psi^2 Xphi and psi^2phi^2D}.
The correlations are nonetheless apparent in the two-dimensional allowed regions of the pairs
$(O_{\varphi q}^{(1)},O_{\varphi q}^{(3)})$ and
$(O_{\varphi\ell}^{(1)},O_{\varphi\ell}^{(3)})$, whose coefficients enter
common coupled-channel amplitudes.
The corresponding allowed regions in the single-flavor limit are shown in
Fig.~\ref{fig:uni CHq,l 13}.

\begin{figure}[tbp]
\centering
\includegraphics[width=0.4\textwidth]{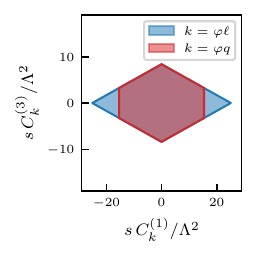}
\caption{
Parameter space allowed by partial-wave unitarity for the coefficients
$(C_{\varphi\ell}^{(1)},C_{\varphi\ell}^{(3)})$ (blue) and
$(C_{\varphi q}^{(1)},C_{\varphi q}^{(3)})$ (red) in the single-flavor limit.
Each region is the intersection of the $J=1/2$ and $J=1$ conditions
listed in Table~\ref{tab:phi2D_channels} and its extent along each axis
gives the marginalized bounds of Table~\ref{tab:UniBounds}.}
\label{fig:uni CHq,l 13}
\end{figure}

Compared with previous SMEFT unitarity analyses
\cite{Corbett:2014ora,Corbett:2017qgl,Cao:2023qks}, the fermionic sector
exhibits larger numerical differences than the purely bosonic one.
These arise from the full coupled-channel treatment, in particular of the $J=1$ sectors, in which the color and
weak-isospin multiplicities enhance the largest eigenvalue, while the marginalization plays no role.
After accounting for the different normalization of the unitarity condition discussed in Sec.~\ref{sec:bosonic_unitarity},
the differences for the operators of the classes $\psi^2X\varphi$ and
$\psi^2\varphi^2D$ are operator-dependent and can reach a
factor of two.
The resulting bounds are generally more stringent than those previously
reported, with the exception of $O_{\varphi ud}$, for which our limit is
weaker by a factor of $\sqrt{2}$.

\subsubsection{Four-fermion operators}
\label{sec:four_fermion_unitarity}

We finally turn to four-fermion operators.
In this case, the lowest angular momentum contributing to a given scattering
channel is determined by the chiralities, and hence by the helicities, of the
external fermions.
For the channels in which the helicity differences of both the initial and
final states vanish, the partial-wave expansion starts at $J=0$.
A representative set is
\begin{gather}
    \psi_R^{\alpha,p}\psi_R^{\beta,r}
    \to
    \psi_R^{\gamma,s}\psi_R^{\delta,t},
    \qquad
    \psi_L^{i,\alpha,p}\psi_L^{j,\beta,r}
    \to
    \psi_L^{k,\gamma,s}\psi_L^{\ell,\delta,t},
    \\
    \psi_R^{\alpha,p}\psi_R^{\beta,r}
    \to
    \psi_L^{i,\gamma,s}\psi_L^{j,\delta,t},
    \qquad
    \psi_R^{\alpha,p}\overline{\psi}_L^{i,\beta,r}
    \to
    \psi_R^{\gamma,s}\overline{\psi}_L^{j,\delta,t}.
\end{gather}
Here $p,r,s,t$ denote flavor indices, $\alpha,\beta,\gamma,\delta$
denote color indices whenever applicable, and
$i,j,k,\ell$ are $\mathrm{SU}(2)_L$ indices.

Other crossings of the same interactions instead start at $J=1$.
A representative set of such channels is
\begin{gather}
    \psi_R^{\alpha,p}\overline{\psi}_R^{\beta,r}
    \to
    \psi_R^{\gamma,s}\overline{\psi}_R^{\delta,t},
    \qquad
    \psi_L^{i,\alpha,p}\overline{\psi}_L^{j,\beta,r}
    \to
    \psi_L^{k,\gamma,s}\overline{\psi}_L^{\ell,\delta,t},
    \\
    \psi_R^{\alpha,p}\overline{\psi}_R^{\beta,r}
    \to
    \psi_L^{i,\gamma,s}\overline{\psi}_L^{j,\delta,t},
    \qquad
    \psi_R^{\alpha,p}\psi_L^{i,\beta,r}
    \to
    \psi_R^{\gamma,s}\psi_L^{j,\delta,t}.
\end{gather}
The remaining channels are related to these by crossing and discrete
symmetries and are included in the complete coupled-channel analysis.

To illustrate the relative importance of the two angular-momentum sectors,
Table~\ref{tab: UB for psi^4} compares the bounds obtained from $J=0$ and
$J=1$ for one Wilson coefficient at a time.
In most cases the $J=0$ partial wave gives the stronger constraint.
There are, however, several exceptions, most notably among operators carrying
singlet weak-isospin and color structures.
As in the case of operators with a single fermionic bilinear, the
multiplicity and coherent combination of internal gauge degrees of freedom
can enhance a higher-$J$ coupled-channel eigenvalue.
Consequently, determining the strongest unitarity constraint requires
examining all relevant angular-momentum sectors rather than stopping at the
smallest kinematically allowed value of $J$.

\begin{table}[tbp]
\renewcommand{\arraystretch}{1.5}
\centering
\begin{tabular}{c|cc||c|cc}
\toprule
\rowcolor{hdrcolor}
\boldmath$(\overline L L)(\overline L L)$
& \boldmath$J=0$ & \boldmath$J=1$
& {\boldmath$(\overline{L}R)(\overline{R}L)$}/{\boldmath$(\overline{L}R)(\overline{L}R)$}
& \boldmath$J=0$ & \boldmath$J=1$ \\
\hline
$C_{\ell\ell}$ & $2\pi$ & $2\pi$
& $C_{\ell e d q}$ & $8\pi/\sqrt{3}$ & $24\pi$ \\

$C_{qq}^{(1)}$ & $2\pi$ & $6\pi/7$
& $C_{quqd}^{(1)}$ & $16\pi/7$ & $24\pi$ \\

$C_{qq}^{(3)}$ & $2\pi$ & $6\pi/5$
& $C_{quqd}^{(8)}$ & $12\pi$ & $36\pi$ \\

$C_{\ell q}^{(1)}$ & $4\pi$ & $2\pi\sqrt{3}$
& $C_{\ell equ}^{(1)}$ & $8\pi/\sqrt{3}$ & $24\pi\sqrt{2}$ \\

$C_{\ell q}^{(3)}$ & $4\pi/3$ & $2\pi\sqrt{3}$
& $C_{\ell equ}^{(3)}$ & $2\pi\sqrt{2}/3$ & $2\pi\sqrt{3}$ \\
\midrule
\rowcolor{hdrcolor}
\boldmath$(\overline R R)(\overline R R)$
& \boldmath$J=0$ & \boldmath$J=1$
& \boldmath$(\overline L L)(\overline R R)$
& \boldmath$J=0$ & \boldmath$J=1$ \\
\hline
$C_{ee}$ & $2\pi$ & $3\pi$
& $C_{\ell e}$ & $4\pi$ & $6\pi\sqrt{2}$ \\

$C_{uu}$ & $2\pi$ & $3\pi/2$
& $C_{\ell u}$ & $4\pi$ & $2\pi\sqrt{6}$ \\

$C_{dd}$ & $2\pi$ & $3\pi/2$
& $C_{\ell d}$ & $4\pi$ & $2\pi\sqrt{6}$ \\

$C_{eu}$ & $4\pi$ & $4\pi\sqrt{3}$
& $C_{qe}$ & $4\pi$ & $2\pi\sqrt{6}$ \\

$C_{ed}$ & $4\pi$ & $4\pi\sqrt{3}$
& $C_{qu}^{(1)}$ & $4\pi$ & $2\pi\sqrt{2}$ \\

$C_{ud}^{(1)}$ & $4\pi$ & $4\pi$
& $C_{qu}^{(8)}$ & $3\pi$ & $12\pi\sqrt{2}$ \\

$C_{ud}^{(8)}$ & $6\pi$ & $9\pi$
& $C_{qd}^{(1)}$ & $4\pi$ & $2\pi\sqrt{2}$ \\

&&
& $C_{qd}^{(8)}$ & $3\pi$ & $12\pi\sqrt{2}$ \\
\bottomrule
\end{tabular}
\caption{
Comparison of the partial-wave unitarity bounds on the modulus of individual
four-fermion Wilson coefficients obtained from the $J=0$ and $J=1$ sectors.
For the purpose of this comparison, one coefficient at a time is kept
non-vanishing. The factors $s/\Lambda^2$ are left implicit.
}
\label{tab: UB for psi^4}
\end{table}

The four-fermion sector is characterized by a considerably richer pattern
of correlations among Wilson coefficients than the operator classes
discussed above.
Accordingly, the one-coefficient-at-a-time limits in
Table~\ref{tab: UB for psi^4} should only be used to identify which
angular-momentum sector is potentially dominant; they should not be
confused with the marginalized bounds reported in
Table~\ref{tab:UniBounds}.
The latter follow from the complete coupled-channel constraints, with all
correlated coefficients allowed to vary simultaneously.

\begin{figure}[tbp]
    \centering
    \includegraphics[width=0.48\linewidth]{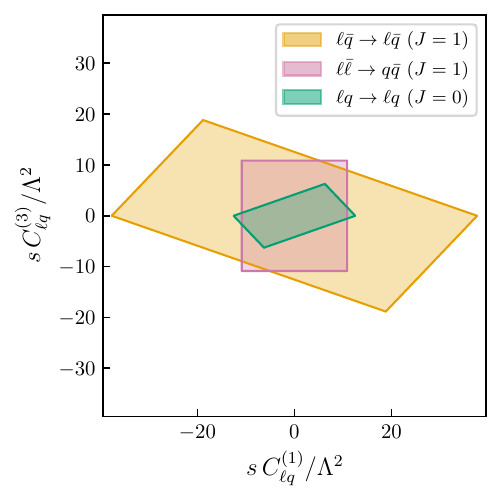}
    \hfill
    \includegraphics[width=0.48\linewidth]{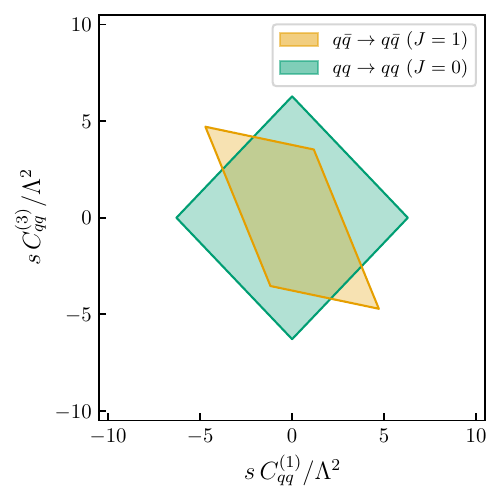}\\
    \includegraphics[width=0.48\linewidth]{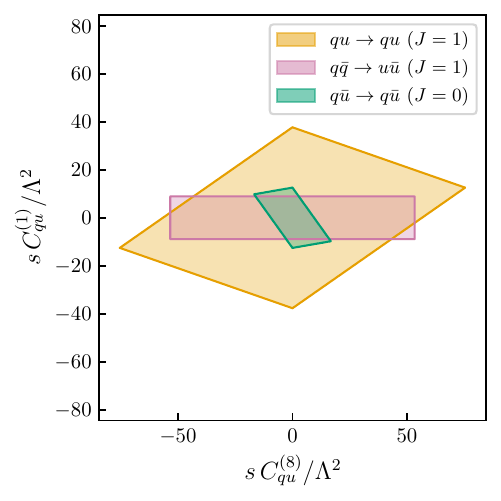}
    \hfill
    \includegraphics[width=0.48\linewidth]{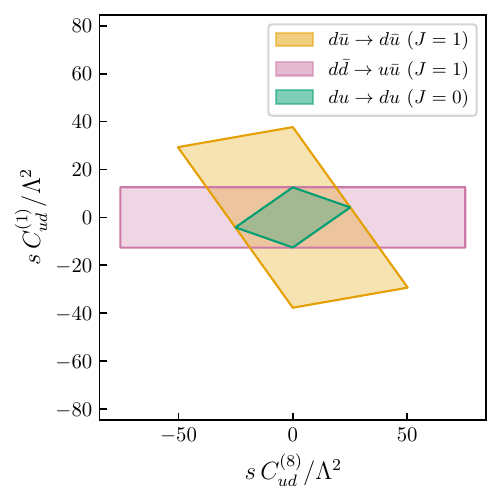}
\caption{Partial-wave unitarity constraints on correlated pairs of
four-fermion Wilson coefficients in the single-flavor limit.
Each shaded region is allowed by the scattering channel and total angular
momentum $J$ indicated in the legend, according to the conditions of
Table~\ref{tab:psi4_channels}.
The combined unitarity constraint is given by the intersection of all
regions. Its extent along each axis reproduces the marginalized bounds of
Table~\ref{tab:UniBounds}.
The allowed region for $(C_{qd}^{(1)},C_{qd}^{(8)})$ is identical to that
for $(C_{qu}^{(1)},C_{qu}^{(8)})$.}
    \label{fig:1-flavor unitary+sum rules}
\end{figure}

Representative examples of these correlations are shown in
Fig.~\ref{fig:1-flavor unitary+sum rules} and the corresponding unitarity conditions are listed in
Table~\ref{tab:psi4_channels}.
In particular, operators with different weak-isospin or color contractions,
such as $(O_{\ell q}^{(1)},O_{\ell q}^{(3)})$,
$(O_{qq}^{(1)},O_{qq}^{(3)})$,
$(O_{qu}^{(1)},O_{qu}^{(8)})$, and
$(O_{ud}^{(1)},O_{ud}^{(8)})$, contribute simultaneously to common
scattering channels and are therefore constrained jointly. 

\begin{table}[tbp]
\renewcommand{\arraystretch}{1.5}
\centering
\begin{tabular}{c|cc|cc}
\toprule
\rowcolor{hdrcolor}
\boldmath$(\overline L L)(\overline L L)$
& \textbf{Channel} & \boldmath$J$
& \multicolumn{2}{c}{\textbf{Unitarity conditions}} \\
\hline
\multirow{3}{*}{$C_{\ell q}^{(1)},\,C_{\ell q}^{(3)}$}
& $\ell q\to\ell q$ & $0$
& $|C_{\ell q}^{(1)}-3C_{\ell q}^{(3)}|\leq4\pi$
& $|C_{\ell q}^{(1)}+C_{\ell q}^{(3)}|\leq4\pi$ \\
& $\ell\overline q\to\ell\overline q$ & $1$
& $|C_{\ell q}^{(1)}-C_{\ell q}^{(3)}|\leq12\pi$
& $|C_{\ell q}^{(1)}+3C_{\ell q}^{(3)}|\leq12\pi$ \\
& $\ell\overline\ell\to q\overline q$ & $1$
& $|C_{\ell q}^{(1)}|\leq2\pi\sqrt{3}$
& $|C_{\ell q}^{(3)}|\leq2\pi\sqrt{3}$ \\
\hline
\multirow{2}{*}{$C_{qq}^{(1)},\,C_{qq}^{(3)}$}
& $qq\to qq$ & $0$
& $|C_{qq}^{(1)}-C_{qq}^{(3)}|\leq2\pi$
& $|C_{qq}^{(1)}+C_{qq}^{(3)}|\leq2\pi$ \\
& $q\overline q\to q\overline q$ & $1$
& $|7C_{qq}^{(1)}+3C_{qq}^{(3)}|\leq6\pi$
& $|C_{qq}^{(1)}+5C_{qq}^{(3)}|\leq6\pi$ \\
\midrule
\rowcolor{hdrcolor}
\boldmath$(\overline R R)(\overline R R)$
& \textbf{Channel} & \boldmath$J$
& \multicolumn{2}{c}{\textbf{Unitarity conditions}} \\
\hline
\multirow{3}{*}{$C_{ud}^{(1)},\,C_{ud}^{(8)}$}
& $du\to du$ & $0$
& $|3C_{ud}^{(1)}-2C_{ud}^{(8)}|\leq12\pi$
& $|3C_{ud}^{(1)}+C_{ud}^{(8)}|\leq12\pi$ \\
& $d\overline u\to d\overline u$ & $1$
& $|6C_{ud}^{(1)}-C_{ud}^{(8)}|\leq72\pi$
& $|3C_{ud}^{(1)}+4C_{ud}^{(8)}|\leq36\pi$ \\
& $d\overline d\to u\overline u$ & $1$
& $|C_{ud}^{(1)}|\leq4\pi$
& $|C_{ud}^{(8)}|\leq24\pi$ \\
\midrule
\rowcolor{hdrcolor}
\boldmath$(\overline L L)(\overline R R)$
& \textbf{Channel} & \boldmath$J$
& \multicolumn{2}{c}{\textbf{Unitarity conditions}} \\
\hline
\multirow{3}{*}{$C_{qu}^{(1)},\,C_{qu}^{(8)}$}
& $q\overline u\to q\overline u$ & $0$
& $|6C_{qu}^{(1)}-C_{qu}^{(8)}|\leq24\pi$
& $|3C_{qu}^{(1)}+4C_{qu}^{(8)}|\leq12\pi$ \\
& $qu\to qu$ & $1$
& $|3C_{qu}^{(1)}-2C_{qu}^{(8)}|\leq36\pi$
& $|3C_{qu}^{(1)}+C_{qu}^{(8)}|\leq36\pi$ \\
& $q\overline q\to u\overline u$ & $1$
& $|C_{qu}^{(1)}|\leq2\pi\sqrt{2}$
& $|C_{qu}^{(8)}|\leq12\pi\sqrt{2}$ \\
\midrule
\rowcolor{hdrcolor}
\boldmath$(\overline L R)(\overline L R)$
& \textbf{Channel} & \boldmath$J$
& \multicolumn{2}{c}{\textbf{Unitarity conditions}} \\
\hline
\multirow{2}{*}{$C_{quqd}^{(1)},\,C_{quqd}^{(8)}$}
& $ud\to qq$ & $0$
& \multicolumn{2}{c}{$|3C_{quqd}^{(1)}-2C_{quqd}^{(8)}|\leq24\pi\sqrt{2}$} \\
& $u\overline q\to q\overline d$ & $0$
& $|21C_{quqd}^{(1)}+4C_{quqd}^{(8)}|\leq48\pi$
& $|6C_{quqd}^{(1)}+5C_{quqd}^{(8)}|\leq96\pi$ \\
\hline
\multirow{6}{*}{$C_{\ell equ}^{(1)},\,C_{\ell equ}^{(3)}$}
& $eu\to\ell q$ & $0$
& \multicolumn{2}{c}{$|C_{\ell equ}^{(1)}-12C_{\ell equ}^{(3)}|\leq8\pi\sqrt{2}$} \\
& $e\overline\ell\to q\overline u$ & $0$
& \multicolumn{2}{c}{$|C_{\ell equ}^{(1)}|\leq8\pi/\sqrt{3}$} \\
& $e\overline q\to\ell\overline u$ & $0$
& \multicolumn{2}{c}{$|C_{\ell equ}^{(1)}+12C_{\ell equ}^{(3)}|\leq16\pi$} \\
& $eu\to\ell q$ & $1$
& \multicolumn{2}{c}{$|C_{\ell equ}^{(1)}+4C_{\ell equ}^{(3)}|\leq24\pi\sqrt{2}$} \\
& $e\overline\ell\to q\overline u$ & $1$
& \multicolumn{2}{c}{$|C_{\ell equ}^{(3)}|\leq2\pi\sqrt{3}$} \\
& $e\overline q\to\ell\overline u$ & $1$
& \multicolumn{2}{c}{$|C_{\ell equ}^{(1)}-4C_{\ell equ}^{(3)}|\leq48\pi$} \\
\bottomrule
\end{tabular}
\caption{
Partial-wave unitarity conditions on correlated pairs of four-fermion
Wilson coefficients in the single-flavor limit, for each scattering
channel and total angular momentum $J$.
The corresponding allowed regions are shown in
Figs.~\ref{fig:1-flavor unitary+sum rules}
and~\ref{fig:unicomplexphase Cquqd, Cellequ}.
Channels related by CP conjugation, and for $O_{quqd}^{(1,8)}$ by
$u\leftrightarrow d$, yield identical conditions and are not listed
separately.
The conditions for $(C_{qd}^{(1)},C_{qd}^{(8)})$ follow from those for
$(C_{qu}^{(1)},C_{qu}^{(8)})$ with $u\to d$.
The factors $s/\Lambda^2$ are left implicit.
}
\label{tab:psi4_channels}
\end{table}

For non-Hermitian four-fermion operators, the relative complex phases of
different Wilson coefficients provide an additional physical direction in
parameter space.
This is relevant, for example, for the pairs
$(C_{quqd}^{(1)},C_{quqd}^{(8)})$ and
$(C_{\ell equ}^{(1)},C_{\ell equ}^{(3)})$.
The corresponding unitarity regions depend on the relative phases
$\arg(C_{quqd}^{(1)}/C_{quqd}^{(8)})$ and
$\arg(C_{\ell equ}^{(1)}/C_{\ell equ}^{(3)})$, as illustrated in
Fig.~\ref{fig:unicomplexphase Cquqd, Cellequ} and reported in Table~\ref{tab:psi4_channels}.

\begin{figure}[tbp]
\centering
\includegraphics[width=0.8\textwidth]
{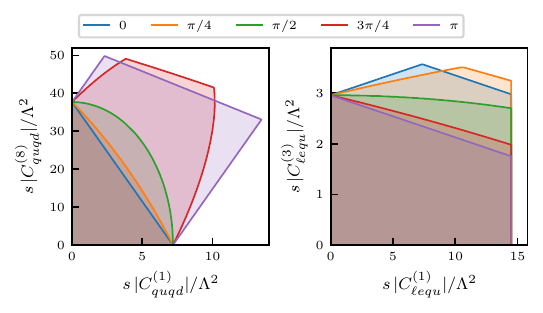}
\caption{
Regions allowed by partial-wave unitarity in the planes of the moduli $(|C_{quqd}^{(1)}|,|C_{quqd}^{(8)}|)$ (left) and
$(|C_{\ell equ}^{(1)}|,|C_{\ell equ}^{(3)}|)$ (right), in the single-flavor limit, for the values of the relative complex
phases
$\arg(C_{quqd}^{(1)}/C_{quqd}^{(8)})$ and
$\arg(C_{\ell equ}^{(1)}/C_{\ell equ}^{(3)})$ indicated in the legend.
The corresponding conditions are listed in Table~\ref{tab:psi4_channels}.
}
\label{fig:unicomplexphase Cquqd, Cellequ}
\end{figure}

\section{Spinning sum rules}
\label{sec:sum_rules}

Perturbative unitarity constrains the magnitude of effective interactions,
but by itself provides limited information on the nature of the heavy states
responsible for them.
Additional information can be obtained by combining unitarity with
analyticity and crossing symmetry of the $S$ matrix.
Dispersion relations can then relate low-energy Wilson coefficients to
weighted sums over UV partial waves, thereby retaining information on the
angular momentum of the states exchanged in the UV
\cite{Gu:2020thj,Remmen:2020uze,Remmen:2022orj,Azatov:2021ygj}.
Under suitable assumptions, these so-called spinning sum rules correlate the
sign of specific combinations of SMEFT Wilson coefficients with the spin of
the heavy states that generate them.

In this section we formulate these constraints directly in the Warsaw basis,
with particular emphasis on four-fermion interactions.
We first review the relevant dispersion relations and identify the
Wilson-coefficient combinations associated with scalar- and vector-dominated
UV dynamics.
We then show how the resulting sign conditions can be imposed exactly in
flavor space and, finally, discuss the assumptions under which this
interpretation is valid and provide explicit UV examples illustrating how
the corresponding inequalities can be evaded.

\subsection{Spinning sum rules in the SMEFT}
\label{sec:sum_rules_SMEFT}

Following Refs.~\cite{Gu:2020thj,Remmen:2020uze,Remmen:2022orj,
Azatov:2021ygj}, we consider the elastic amplitude
$\mathcal{A}_{\alpha\beta}(s,t)$ for two fermionic states of opposite
helicity.
The labels $\alpha$ and $\beta$ collectively denote their internal quantum
numbers, including flavor and gauge charges, while $\bar\alpha$ denotes the
crossed state, with conjugated quantum numbers and reversed helicity.
The two crossed channels have different helicity structures: the
$\alpha\beta$ channel carries one unit of net helicity and therefore starts
at $J=1$, whereas the $\bar\alpha\beta$ channel has vanishing net helicity
and admits a $J=0$ partial wave.

Expanding the UV amplitudes in partial waves of definite total angular
momentum, the forward dispersion relation can be written as\footnote{Our normalization of the partial-wave coefficients differs from that adopted
in Ref.~\cite{Remmen:2020uze}.}
\begin{multline}
        \lim_{s\to0}\partial_s
    \mathcal{A}_{\alpha\beta}(s,0)
    +
    C_{\alpha\beta}^{(s,\infty)}
    =
    -\lim_{s\to0}\partial_s
    \mathcal{A}_{\bar\alpha\beta}(s,0)
    -
    C_{\bar\alpha\beta}^{(s,\infty)}
    \\
    =
    8\int_{s_0}^{\infty}\frac{ds}{s^2}
    \left[
        -\,\mathrm{Im}\,a_{\bar\alpha\beta}^{0}(s)
        +
        \sum_{J=1}^{\infty}(2J+1)
        \left(
            \mathrm{Im}\,a_{\alpha\beta}^{J}(s)
            -
            \mathrm{Im}\,a_{\bar\alpha\beta}^{J}(s)
        \right)
    \right],
    \label{eq:spinning_forward}
\end{multline}
where $s_0$ denotes the characteristic mass-squared scale at which the UV
degrees of freedom enter.
The boundary contribution is
\begin{equation}
    C_{\alpha\beta}^{(s,\infty)}
    =
    \underset{s=\infty}{\mathrm{Res}}\,
    \frac{\mathcal{A}_{\alpha\beta}(s,0)}{s^2}
    =
    -\lim_{|s|\to\infty}
    \frac{\mathcal{A}_{\alpha\beta}(s,0)}{s}.
    \label{eq:boundary_s}
\end{equation}

A second dispersion relation follows by applying the same contour argument
to the coefficient of $t$ in
$\mathcal{A}_{\alpha\beta}(0,t)$.
This gives a beyond-forward relation with an independent boundary
contribution, which we denote by
$C_{\alpha\beta}^{(t,\infty)}$.
For a dimension-six interaction involving two opposite-helicity fermions,
the local amplitude is proportional to the spinor structure
$[13]\langle24\rangle$.
As a consequence, an appropriate combination of the forward and
beyond-forward dispersion relations has a vanishing low-energy side, up to
boundary contributions, while all UV partial waves enter with
non-negative weights.

If the relevant boundary terms vanish and the partial-wave expansion
converges, positivity then requires the higher-spin contributions to vanish
individually.
In particular, the absorptive parts of the relevant channels with
$J\geq2$ vanish, together with the crossed $J=1$ contribution.
The remaining dispersion relation therefore involves only a scalar
$J=0$ channel and a vector $J=1$ channel:
\begin{equation}
    \lim_{s\to0}\partial_s
    \mathcal{A}_{\bar\alpha\beta}(s,0)
    =
    -\lim_{s\to0}\partial_s
    \mathcal{A}_{\alpha\beta}(s,0)
    =
    8\int_{s_0}^{\infty}\frac{ds}{s^2}
    \left[
        \mathrm{Im}\,a_{\bar\alpha\beta}^{0}(s)
        -
        3\,\mathrm{Im}\,a_{\alpha\beta}^{1}(s)
    \right].
    \label{eq:spinning_scalar_vector}
\end{equation}
By partial-wave unitarity, the absorptive parts appearing in
Eq.~\eqref{eq:spinning_scalar_vector} are non-negative.
Consequently, if the UV dynamics is dominated by spin-0 states, the
left-hand side is non-negative for any choice of the external-state
quantum numbers. Conversely, dominance by spin-1 states implies a
non-positive left-hand side.
This is the central observation underlying the spinning sum rules.

For four-fermion SMEFT operators, the derivative of the low-energy amplitude
is linear in the corresponding Wilson coefficients.
Equation~\eqref{eq:spinning_scalar_vector} can therefore be translated into
sign conditions on specific combinations of their flavor-dependent entries.
Since the basis adopted in Ref.~\cite{Remmen:2020uze} differs from the
Warsaw basis, we collect in Table~\ref{tab:spinning_sum_rules} the
corresponding Warsaw-basis combinations.
Denoting any such combination by $b^\psi_{prst}$, its contraction
\begin{equation}
    b^\psi_{prst}\,
    \alpha_p\alpha_r^*
    \beta_s^*\beta_t
    \label{eq:bpsi_contraction}
\end{equation}
is positive for scalar-dominated UV dynamics and negative for
vector-dominated UV dynamics, for arbitrary complex flavor vectors
$\alpha$ and $\beta$.

\begin{table}[tbp]
\renewcommand{\arraystretch}{1.5}
    \centering
    \begin{tabular}{cc}
    \toprule
    \rowcolor{hdrcolor}
    \textbf{Operator class} & \textbf{Spinning sum rules} \\
    \midrule
    \multirow{6}{*}{\boldmath$(\overline{R}R)(\overline{R}R)$} &
    $[C_{ee}]_{prst}$ \\
    & $[C_{uu}]_{prst}$\,, \qquad $[C_{uu}]_{prst} + [C_{uu}]_{ptsr}$ \\
    & $[C_{dd}]_{prst}$\,, \qquad $[C_{dd}]_{prst} + [C_{dd}]_{ptsr}$ \\
    & $[C_{eu}]_{prst}$\,, \qquad $[C_{ed}]_{prst}$ \\
    & $[C_{ud}^{(1)}]_{prst}+\tfrac{1}{3}[C_{ud}^{(8)}]_{prst}$ \\
    & $[C_{ud}^{(1)}]_{prst}-\tfrac{1}{6}[C_{ud}^{(8)}]_{prst}$ \\
    \midrule
    \multirow{5}{*}{\boldmath$(\overline{L}L)(\overline{L}L)$} &
    $[C_{\ell\ell}]_{prst}$\,, \qquad $[C_{\ell\ell}]_{prst} + [C_{\ell\ell}]_{ptsr}$ \\
    & $[C_{qq}^{(1)}]_{prst} + [C_{qq}^{(1)}]_{ptsr} + [C_{qq}^{(3)}]_{prst} + [C_{qq}^{(3)}]_{ptsr}$ \\
    & $[C_{qq}^{(1)}]_{prst} - [C_{qq}^{(3)}]_{prst} + 2[C_{qq}^{(3)}]_{ptsr}$ \\
    & $[C_{\ell q}^{(1)}]_{prst} + [C_{\ell q}^{(3)}]_{prst}$ \\
    & $[C_{\ell q}^{(1)}]_{prst} - [C_{\ell q}^{(3)}]_{prst}$ \\
    \midrule
    \multirow{6}{*}{\boldmath$(\overline{L}L)(\overline{R}R)$} &
    $-[C_{\ell e}]_{prst}$\,, \qquad $-[C_{\ell u}]_{prst}$ \\
    & $-[C_{\ell d}]_{prst}$\,, \qquad $-[C_{qe}]_{prst}$ \\
    & $-[C_{qu}^{(1)}]_{prst} - \tfrac{1}{3}[C_{qu}^{(8)}]_{prst}$ \\
    & $-[C_{qu}^{(1)}]_{prst} + \tfrac{1}{6}[C_{qu}^{(8)}]_{prst}$ \\
    & $-[C_{qd}^{(1)}]_{prst} - \tfrac{1}{3}[C_{qd}^{(8)}]_{prst}$ \\
    & $-[C_{qd}^{(1)}]_{prst} + \tfrac{1}{6}[C_{qd}^{(8)}]_{prst}$ \\
    \bottomrule
    \end{tabular}
    \caption{Combinations of four-fermion Wilson coefficients that, when contracted with $\alpha_p\alpha_r^*\beta_s^*\beta_t$, are positive for scalar-dominated UV
    dynamics and negative for vector-dominated UV dynamics, under the assumptions
    leading to Eq.~\eqref{eq:spinning_scalar_vector}.}
    \label{tab:spinning_sum_rules}
\end{table}

% Spinning sum rules are not restricted in principle to four-fermion
% operators~\cite{Gu:2020thj,Remmen:2022orj}, but it is in this sector that they lead most directly to
% sign-definite constraints on Wilson coefficients.
% For purely bosonic operators of the class $\varphi^4D^2$, analogous
% relations can be derived only for restricted combinations of coefficients,
% with the correspondence between sign and UV spin reversed relative to the
% fermionic case \cite{Remmen:2022orj}.
% For operators involving gauge-boson field strengths, boundary contributions
% are generically present and depend on the gauge configuration, preventing a
% model-independent sign assignment.
% If these boundary terms vanish, however, the corresponding dispersion
% relations imply the stronger statement that the UV amplitudes mediating
% elastic vector scattering must vanish.

Spinning sum rules are not restricted in principle to four-fermion
operators, but it is in this sector that they lead most directly to
sign-definite constraints on Wilson coefficients.
For purely bosonic operators of the class $\varphi^4D^2$, analogous
relations can be derived only for restricted combinations of
coefficients~\cite{Remmen:2022orj}.
For operators of the classes $X^2\varphi^2$ and $X^3$, the relevant elastic
amplitudes vanish identically at dimension six, so that the corresponding
Wilson coefficients are constrained by neither conventional nor spinning
sum rules~\cite{Gu:2020thj,Remmen:2022orj}.
If the boundary terms vanish, however, the dispersion relations imply that
the UV amplitudes mediating the elastic scattering of these states must
vanish~\cite{Remmen:2022orj}.

% The situation is also different for operators of the class
% $\psi^2\varphi^2D$.
% After fixing the Higgs configuration, no scalar-versus-vector spinning sum
% rule remains.
% The corresponding dispersion relations can still constrain the UV
% structure --- in particular, vanishing boundary terms imply dominance by a
% fermionic mode --- but they do not provide the scalar/vector discrimination
% available for four-fermion interactions.
% For this reason, in the remainder of this work we focus on the
% four-fermion sector, for which Eq.~\eqref{eq:bpsi_contraction} can be turned
% into explicit inequalities in flavor space.

The situation is also different for operators of the class
$\psi^2\varphi^2D$.
Since the corresponding fermion--Higgs amplitudes vanish at $s=0$, the
beyond-forward relation does not involve their Wilson coefficients, and no
spinning sum rule can be constructed.
The corresponding dispersion relations can still constrain the UV
structure --- in particular, vanishing boundary terms imply dominance by
spin-$1/2$ states --- but they do not provide the scalar/vector discrimination
available for four-fermion interactions~\cite{Remmen:2022orj}.
For this reason, in the remainder of this work we focus on the
four-fermion sector, for which Eq.~\eqref{eq:bpsi_contraction} can be turned
into explicit inequalities in flavor space.

\subsection{Application to flavor space}
\label{sec:sum_rules_flavor}

We now explain how the spinning sum rules derived above can be translated
into explicit inequalities among flavor-dependent Wilson coefficients.
For any of the four-fermion combinations $b^\psi_{prst}$ collected in
Table~\ref{tab:spinning_sum_rules}, scalar dominance requires
\begin{equation}
    b^\psi_{prst}\,
    \alpha_p\alpha_r^*
    \beta_s^*\beta_t
    \geq 0
    \qquad
    \forall\,\alpha,\beta ,
    \label{eq:scalar_flavor_condition}
\end{equation}
where $\alpha$ and $\beta$ are arbitrary complex vectors in flavor space.
For vector-dominated UV dynamics the inequality is reversed.
Since Eq.~\eqref{eq:scalar_flavor_condition} is homogeneous in both flavor
vectors, their overall normalization is irrelevant and they can be taken to
have unit norm.

The condition in Eq.~\eqref{eq:scalar_flavor_condition} is a biquadratic
form in flavor space.
For fixed $\beta$, we define the Hermitian matrix
\begin{equation}
    \left[M^\psi(\beta)\right]_{rp}
    \equiv
    b^\psi_{prst}\,\beta_s^*\beta_t .
    \label{eq:Mpsi_definition}
\end{equation}
Equation~\eqref{eq:scalar_flavor_condition} can then be written as
\begin{equation}
    \alpha^\dagger M^\psi(\beta)\alpha\geq0
    \qquad
    \forall\,\alpha,\beta .
\end{equation}
Scalar dominance is therefore equivalent to requiring
$M^\psi(\beta)$ to be positive semidefinite for every direction $\beta$ in
flavor space, or, equivalently,
\begin{equation}
    \lambda_i\!\left[M^\psi(\beta)\right]\geq0
    \qquad
    \forall\,i,\beta ,
    \label{eq:Mpsi_eigenvalues}
\end{equation}
where $\lambda_i$ denote the eigenvalues of $M^\psi(\beta)$. 
For vector dominance, the same construction applies with
$M^\psi(\beta)$ required to be negative semidefinite.

For three generations, a convenient parametrization of a normalized flavor
direction is
\begin{equation}
    \beta=
    \begin{pmatrix}
        e^{i\delta_1}\beta_1\\[1mm]
        e^{i\delta_2}\beta_2\\[1mm]
        \sqrt{1-\beta_1^2-\beta_2^2}
    \end{pmatrix},
    \qquad
    \beta_1^2+\beta_2^2\leq1 ,
    \label{eq:beta_parameterization}
\end{equation}
where $\beta_{1,2}\geq0$ and $\delta_{1,2}\in[0,2\pi)$, and the third component has been taken real and non-negative
by removing an overall phase.
For all flavor structures encountered in our analysis, the non-trivial
eigenvalues of $M^\psi(\beta)$ can be reduced to functions of the form
\begin{equation}
    f_\pm(x)
    =
    a+bx\pm\sqrt{c^2+dx+ex^2},
    \qquad
    x\equiv\beta_1^2+\beta_2^2\in[0,1],
    \label{eq:fpm_sumrules}
\end{equation}
with $a,b,c,d,e\in\mathbb{R}$ functions of the Wilson coefficients.
The remaining eigenvalues, when present, are linear functions and are
obtained as special cases of the same analysis.
Since $f_+(x)\geq f_-(x)$, positive semidefiniteness is controlled by the
lowest eigenvalue, and scalar dominance reduces to the condition
\begin{equation}
    f_-(x)\geq0
    \qquad
    \forall\,x\in[0,1].
    \label{eq:fminus_positive}
\end{equation}

A numerical scan over flavor directions would test
Eq.~\eqref{eq:fminus_positive} only on a finite set of points and would
therefore provide, in general, only an approximate characterization of the
allowed region.
Instead, the requirement of positivity over the complete interval can be
translated exactly into a finite set of algebraic conditions.
For
$f_-(x)$,
assumed real for $x\in[0,1]$, one has
$f_-(x)\geq0$ throughout the interval if and only if
the system of inequalities
\begin{align}
    \Delta(a,b,c,d,e) &= 
    a \ge 0 
    \ \land \
    a+b \ge 0
    \ \land \ 
    a^2 - c^2 \ge 0
    \ \land \ 
    (a+b)^2 - c^2 -d-e \ge 0 \nonumber
    \\& \quad
    \land \
    [e - b^2 \ge 0 \ \lor\  (2ab-d)(2ab-d+2b^2-2e) \ge 0 \nonumber 
    \\& \quad
    \lor \ 4(b^2-e)(a^2-c^2)-(2ab-d)^2 \ge 0]
    \label{eq:Delta_sumrules}
\end{align}
is true.
A proof of Eq.~\eqref{eq:Delta_sumrules} is given in
App.~\ref{sec:appendix, logical expression}.
This converts the requirement that the sum rule hold for an infinite set of
flavor directions into a finite system of inequalities among Wilson
coefficients.

The vector-dominance conditions follow without repeating the analysis:
one simply applies the scalar-dominance criterion to
$-M^\psi(\beta)$, or equivalently to $-b^\psi_{prst}$.
In this way, the two UV hypotheses define two different regions in the
space of flavor-dependent Wilson coefficients.

In the single-flavor limit, the flavor vectors become trivial and the
conditions above reduce directly to sign inequalities on the combinations
listed in Table~\ref{tab:spinning_sum_rules}.
With several generations, instead, requiring the sum rule to hold for every
possible flavor direction can generate non-trivial correlations among the
independent coefficients and can be considerably more restrictive than
testing only individual flavor entries.
This observation will be particularly important in
Sec.~\ref{sec:flavorful}, where we apply the procedure to the MFV and
$\mathrm{U}(2)^5$ flavor structures.

\subsection{Limitations and UV examples}
\label{sec:sum_rules_limitations}

Unlike the perturbative unitarity bounds derived in Sec.~\ref{sec:unitarity},
the spinning sum-rule inequalities do not follow from EFT consistency alone.
Their interpretation in terms of the spin of the UV completion
relies on a set of additional assumptions and can therefore fail whenever
one of these assumptions is violated.
In particular, the scalar- and vector-dominance conditions discussed above
rest on three main ingredients
\cite{Gu:2020thj,Remmen:2020uze,Remmen:2022orj,Azatov:2021ygj}:
\begin{itemize}
    \item the effective interaction is generated at tree level;
    \item the UV dynamics is dominated by states of a definite spin, so that
    the right-hand side of
    Eq.~\eqref{eq:spinning_scalar_vector} has a definite sign;
    \item the boundary contributions entering the relevant dispersion
    relations vanish.
\end{itemize}
Violating any of these conditions can move an otherwise consistent UV
completion outside the scalar- or vector-dominated regions derived in
Sec.~\ref{sec:sum_rules_flavor}.
The three possibilities are physically distinct and we discuss them in turn.

Loop-generated interactions provide the first source of violation.
The simple sign classification derived above assumes a tree-level UV
origin of the dimension-six four-fermion interactions and need not hold
when the matching is generated at loop level.
Explicit examples have been discussed in Refs.~\cite{Remmen:2020uze,Azatov:2021ygj}, and we do not reconsider them here.

A second possibility is a non-trivial interplay between UV states of
different spin.
If both spin-0 and spin-1 exchanges contribute with comparable strength,
the two terms on the right-hand side of
Eq.~\eqref{eq:spinning_scalar_vector}
compete, and the resulting low-energy coefficient need not be sign
definite for arbitrary flavor directions.
As a simple illustration, consider the schematic interactions
\begin{equation}
    \mathcal{L}
    \supset
    \left(
        c\,A_\mu\,\overline{\mu}_R\gamma^\mu e_R
        +\text{h.c.}
    \right)
    +
    \left(
        y\,\phi\,\overline{\tau}_R\,\mu_R^{\,c}
        +\text{h.c.}
    \right),
    \label{eq:mixed_spin_example}
\end{equation}
where $A_\mu$ is a heavy complex vector and $\phi$ a heavy complex scalar.
The purpose of this example is only to illustrate the loss of
sign-definiteness when both scalar and vector mediators contribute.
Both exchanges contribute to the same four-lepton operator $O_{ee}$. 

Indeed, integrating out the heavy mediators gives the low-energy
Lagrangian
\begin{equation}
    \mathcal{L}_{\rm EFT}\supset
    \frac{|y|^2}{2M_\phi^2}
    (\overline \mu_R\gamma^\mu \mu_R)
    (\overline \tau_R\gamma_\mu \tau_R)
    -
    \frac{|c|^2}{M_A^2}
    (\overline \mu_R\gamma^\mu e_R)
    (\overline e_R\gamma_\mu \mu_R).
\end{equation}
Using the combination entering the spinning sum rule in
Table~\ref{tab:spinning_sum_rules}, we obtain
\begin{align}
    b^e_{\alpha\beta}
    &\equiv
    \frac{1}{\Lambda^2}[C_{ee}]_{prst}\,
    \alpha_p\alpha_r^*\beta_s^*\beta_t
    =
    \frac{|y|^2}{8M_\phi^2}
    \left|
        \alpha_2^*\beta_3
        +
        \alpha_3^*\beta_2
    \right|^2
    -
    \frac{|c|^2}{4M_A^2}
    \left|
        \alpha_1^*\beta_2
        +
        \alpha_2^*\beta_1
    \right|^2,
\end{align}
where the Fierz-symmetric convention $[C_{ee}]_{prst}=[C_{ee}]_{stpr}=[C_{ee}]_{ptsr}=[C_{ee}]_{srpt}$ is used. For non-vanishing $y$ and $c$, different choices of the flavor vectors $\alpha$ and $\beta$ can make $b^e_{\alpha\beta}$ either positive or negative. Hence, the resulting flavor tensor $[C_{ee}]_{prst}$ is in general neither positive nor negative semidefinite. Explicitly, $\alpha=(0,1,0)$ with $\beta=(0,0,1)$ or $\beta=(1,0,0)$ isolates the positive or the negative contribution, respectively.

The corresponding EFT point therefore lies outside both the purely
scalar- and purely vector-dominated regions, despite arising from a
perfectly consistent UV completion.
If one mediator dominates parametrically, the expected sign-definite
behavior is recovered approximately.

The third possibility is the presence of non-vanishing boundary terms.
This is particularly relevant for tree-level vector exchange in the
$t$ channel, for which the boundary contribution can carry the entire
low-energy amplitude.
Consider, for example, a heavy neutral vector
$B_\mu$ with interactions
\begin{equation}
    \mathcal{L}
    \supset
    -[y^e]_{pr}\,
    \overline e_R^{\,p}\gamma^\mu e_R^{\,r} B_\mu
    -
    [y^u]_{pr}\,
    \overline u_R^{\,p}\gamma^\mu u_R^{\,r} B_\mu .
    \label{eq:tchannel_vector_UV}
\end{equation}
Integrating out $B_\mu$ generates
\begin{equation}
    \mathcal{L}_{\rm EFT}
    \supset
    \frac{1}{M_B^2}
    [C_{eu}]_{prst}
    \left(
        \overline e_R^{\,p}\gamma^\mu e_R^{\,r}
    \right)
    \left(
        \overline u_R^{\,s}\gamma_\mu u_R^{\,t}
    \right),
\end{equation}
with
\begin{equation}
    [C_{eu}]_{prst}
    =
    -[y^e]_{pr}[y^u]_{st}.
    \label{eq:Ceu_vector_matching}
\end{equation}
For the corresponding opposite-helicity elastic amplitude one finds
\begin{equation}
    \lim_{s\to0}
    \partial_s\mathcal{A}_{\alpha\beta}(s,0)
    =
    -\frac{2}{M_B^2}
    [C_{eu}]_{prst}\,
    \alpha_p\alpha_r^*
    \beta_s^*\beta_t .
    \label{eq:tchannel_lowenergy}
\end{equation}
In this case the same contribution is entirely reproduced by the boundary
term,
\begin{equation}
    C_{\alpha\beta}^{(s,\infty)}
    =
    -\lim_{|s|\to\infty}
    \frac{\mathcal{A}_{\alpha\beta}(s,0)}{s}
    =
    \frac{2}{M_B^2}
    [C_{eu}]_{prst}\,
    \alpha_p\alpha_r^*
    \beta_s^*\beta_t ,
    \label{eq:tchannel_boundary}
\end{equation}
while the absorptive $s$-channel partial wave vanishes.
The sign of the low-energy Wilson coefficient is therefore no longer tied
to the spin-1 contribution appearing on the right-hand side of
Eq.~\eqref{eq:spinning_scalar_vector}, because the assumption of vanishing
boundary terms is not satisfied.

This behavior is characteristic of tree-level vector exchange in the
$t$ channel and occurs generically in the corresponding four-fermion UV
completions.
It should therefore not be interpreted as a violation of analyticity or
unitarity, but rather as a failure of the assumptions required for the
simple spinning sum-rule classification.
Conversely, scalar tree-level completions with vanishing boundary
contributions satisfy the corresponding scalar-dominance inequalities,
while vector mediators exchanged in the $s$ channel are classified
according to their physical spin.

Non-vanishing boundary terms do not necessarily imply that the resulting
Wilson coefficients lie outside the nominal scalar- or vector-dominated
regions.
For example, the same neutral vector can generate
$O_{ee}$ through both crossed and direct channels.
In that case, part of the low-energy amplitude is carried by the boundary
term while another part is associated with a physical vector partial wave.
Depending on the flavor structure of the couplings, the resulting
coefficient may still satisfy the vector-dominance inequalities.
Thus, satisfying the corresponding spinning sum-rule inequalities is
consistent with a simple spin-dominated completion under the assumptions above, but does not uniquely identify the UV dynamics.

These examples make clear that the spinning sum rules should be interpreted as UV discriminants rather than as model-independent consistency conditions. Their violation does not signal an inconsistent EFT, but rather that the assumptions underlying the simple scalar/vector classification are not realized. In this sense, the failure of the sign constraints can itself provide information on the structure of the UV completion.

\section{Flavorful four-fermion operators}
\label{sec:flavorful}

The results presented so far were obtained either in the single-flavor
limit or, in the case of the spinning sum rules, in terms of generic flavor
tensors.
We now turn to the full three-generation structure of the four-fermion
sector.
Once all flavor indices are retained, the number of independent Wilson
coefficients becomes very large, making a completely unrestricted
coupled-channel analysis impractical.
At the same time, flavor non-universality is a central feature of many
well-motivated new-physics scenarios, in particular those with enhanced
couplings to third-generation fermions.
It is therefore useful to impose theoretically motivated flavor symmetries
that reduce the number of independent structures while preserving
non-trivial flavor information.

In this section we consider two standard hypotheses: MFV and an exact $\mathrm{U}(2)^5$ flavor symmetry.
We apply both the perturbative unitarity bounds of Sec.~\ref{sec:unitarity}
and the spinning sum rules of Sec.~\ref{sec:sum_rules} directly in the
resulting flavor space.
A representative example will be discussed in the main text, while the
complete set of bounds for all current--current four-fermion operators is collected in
App.~\ref{app:flavor_results}.

\subsection{Flavor symmetries}
\label{sec:flavor_symmetries}

We first specify the flavor structures retained in the analysis.

\subsubsection{Minimal Flavor Violation}
\label{sec:MFV}

MFV assumes that the flavor structure of new
physics is governed by the same sources of flavor-symmetry breaking that
appear in the SM.
In the limit of vanishing Yukawa couplings, the SM fermion sector is
invariant under the global flavor symmetry
\begin{equation}
    G_{\rm MFV}
    \equiv \mathrm{U}(3)^5
    =
    \mathrm{U}(3)_q\times \mathrm{U}(3)_u\times \mathrm{U}(3)_d
    \times \mathrm{U}(3)_\ell\times \mathrm{U}(3)_e ,
\end{equation}
under which the fermion multiplets transform as
\begin{equation}
    q\to U_q\, q,
    \qquad
    u\to U_u\, u,
    \qquad
    d\to U_d\, d,
    \qquad
    \ell\to U_\ell\,\ell,
    \qquad
    e\to U_e\, e .
    \label{eq:MFV_fermion_transformations}
\end{equation}
The Yukawa matrices are promoted to spurions transforming as
\begin{equation}
    Y_u\to U_qY_uU_u^\dagger,
    \qquad
    Y_d\to U_qY_dU_d^\dagger,
    \qquad
    Y_e\to U_\ell Y_eU_e^\dagger ,
    \label{eq:MFV_spurions}
\end{equation}
so that the SM Yukawa interactions are formally invariant under
$G_{\rm MFV}$.

The MFV hypothesis requires the SMEFT Wilson coefficients to transform in
such a way that all flavor-symmetry breaking can be expressed through
appropriate insertions of these spurions.
Consequently, the four flavor indices carried by a generic four-fermion
coefficient are not independent, but can be organized into a finite set of
$\mathrm{U}(3)^5$-invariant tensors constructed from Kronecker deltas and Yukawa
combinations.

In the present analysis, we neglect structures involving insertions of
$Y_d$ and $Y_e$, and retain the exact $\mathrm{U}(3)^5$ structures together with
the leading flavor-breaking terms up to $\mathcal O(Y_u^2)$. This captures
the dominant source of flavor non-universality while keeping the number
of independent structures tractable.
In the fermion mass basis, these flavor-breaking terms are controlled predominantly by the
top Yukawa coupling through
\begin{equation}
    Y_uY_u^\dagger
    \simeq
    y_t^2\begin{pmatrix}
        0 & 0 & 0 \\ 0 & 0 & 0 \\ 0 & 0 & 1
    \end{pmatrix},
    \qquad
    Y_u^\dagger Y_u
    \simeq
    y_t^2\begin{pmatrix}
        0 & 0 & 0 \\ 0 & 0 & 0 \\ 0 & 0 & 1
    \end{pmatrix},
    \label{eq:MFV_top_limit}
\end{equation}
up to the appropriate flavor-basis rotations.
Since $y_t$ is of order one, the truncation at $\mathcal O(Y_u^2)$ should be regarded as a definite choice of flavor expansion rather than as a parametric expansion in a small quantity. Higher powers of $Y_u$ would progressively resolve the third generation from the first two and would effectively approach a $\mathrm{U}(2)_q\times \mathrm{U}(2)_u$ flavor structure~\cite{Faroughy:2020ina,Greljo:2022cah}.

\subsubsection{\texorpdfstring{$\mathrm{U}(2)^5$}{U(2)\textasciicircum 5} flavor symmetry}
\label{sec:U2}

As a second benchmark, we consider the flavor symmetry
\begin{equation}
    G_F
    \equiv
    \mathrm{U}(2)^5
    =
    \mathrm{U}(2)_q\times \mathrm{U}(2)_u\times \mathrm{U}(2)_d
    \times \mathrm{U}(2)_\ell\times \mathrm{U}(2)_e ,
    \label{eq:GU2}
\end{equation}
under which the first two generations of each fermion species transform as
a doublet, while the third generation is a singlet.
Accordingly, we decompose
\begin{equation}
    \psi = (\psi_a,\psi_3),
    \qquad
    a=1,2,
    \qquad
    \psi=q,u,d,\ell,e .
    \label{eq:U2_decomposition}
\end{equation}

This symmetry pattern is particularly well suited to new-physics scenarios in which the first two generations are approximately universal while the third generation is allowed to couple differently, see for instance Refs.~\cite{Bordone:2017bld,Greljo:2018tuh,Allwicher:2020esa,Fuentes-Martin:2020bnh,Fuentes-Martin:2022xnb,Barbieri:2023qpf,Barbieri:2024zkh,Davighi:2023evx,Davighi:2023iks,Fuentes-Martin:2024fpx,Covone:2024elw,Lizana:2024jby,Beneito:2025fzf,FernandezNavarro:2023rhv,FernandezNavarro:2024hnv,FernandezNavarro:2025zmb,Isidori:2025rci,Greljo:2025mwj,Masiero:2026qon}.
The independent flavor structures of dimension-six SMEFT operators under $\mathrm{U}(2)^5$ have been systematically classified in Refs.~\cite{Faroughy:2020ina,Greljo:2022cah}.
There is no unique or universally preferred choice of spurions describing the breaking of $\mathrm{U}(2)^5$. We therefore restrict our analysis to the exact-symmetry limit, without introducing symmetry-breaking spurion insertions. Our results are consequently independent of the specific realization of $\mathrm{U}(2)^5$ breaking.

The resulting Wilson coefficients are therefore organized according to whether the fermion bilinears involve fields belonging to the light $\mathrm{U}(2)$ doublets or to the third-generation singlets.
For example, a coefficient involving two different fermion species can contain independent structures corresponding schematically to
\begin{equation}
    (\overline\psi_a\psi_a)(\overline\chi_b\chi_b),
    \qquad
    (\overline\psi_3\psi_3)(\overline\chi_a\chi_a),
    \qquad
    (\overline\psi_a\psi_a)(\overline\chi_3\chi_3),
    \qquad
    (\overline\psi_3\psi_3)(\overline\chi_3\chi_3),
    \label{eq:U2_generic_structures}
\end{equation}
with additional exchange structures whenever allowed by the fermion
content of the operator.
The precise decomposition is operator dependent and will be given where
needed below, while the complete set of flavor structures is collected in App.~\ref{app:flavor_results}.

\subsection{Theoretical bounds in flavor space}
\label{sec:flavor_bounds}

We now apply the perturbative unitarity bounds derived in
Sec.~\ref{sec:unitarity} and the spinning sum rules of
Sec.~\ref{sec:sum_rules} to the multiflavor structures introduced above.
For each four-fermion operator, the MFV or $\mathrm{U}(2)^5$ decomposition expresses
the full flavor tensor in terms of a reduced set of independent
coefficients.
These coefficients are then inserted into the complete coupled-channel
scattering matrices, retaining all flavor configurations allowed by the
chosen symmetry hypothesis.

The unitarity constraints are imposed on the eigenvalues of the resulting
partial-wave matrices exactly as in the single-flavor analysis.
Since several independent flavor structures can contribute to the same
scattering channels, the corresponding coefficients are treated
simultaneously and the one-dimensional limits are obtained by marginalizing
over the remaining coefficients.
The resulting bounds should therefore be interpreted as projections of the
full multiflavor allowed region, rather than as one-coefficient-at-a-time
constraints.

The spinning sum rules are implemented in the same flavor space.
For each relevant combination $b^\psi_{prst}$ listed in
Table~\ref{tab:spinning_sum_rules}, the MFV or $\mathrm{U}(2)^5$ decomposition is
substituted into the biquadratic form
\begin{equation}
    b^\psi_{prst}\,
    \alpha_p\alpha_r^*\beta_s^*\beta_t ,
\end{equation}
and the scalar- or vector-dominance conditions are imposed for arbitrary
flavor directions $\alpha$ and $\beta$, following the procedure described
in Sec.~\ref{sec:sum_rules_flavor}.
The resulting inequalities therefore constrain correlated combinations of
the independent flavor coefficients, rather than individual tensor entries.

To illustrate the resulting constraints, we focus in the following 
on the semileptonic operators $O_{\ell q}^{(1)}$ and $O_{\ell q}^{(3)}$,
while the complete results are collected in App.~\ref{app:flavor_results}.

\subsubsection{Representative example:
\texorpdfstring{$O_{\ell q}^{(1)}$ and $O_{\ell q}^{(3)}$}{Olq1 and Olq3}}
\label{sec:flavor_example_lq}

As a representative example of the multiflavor analysis, we consider the
semileptonic operators
\begin{equation}
    O_{\ell q}^{(1)}
    =
    (\overline\ell_p\gamma_\mu\ell_r)
    (\overline q_s\gamma^\mu q_t),
    \qquad
    O_{\ell q}^{(3)}
    =
    (\overline\ell_p\gamma_\mu\sigma^I\ell_r)
    (\overline q_s\gamma^\mu\sigma^I q_t).
    \label{eq:Olq_definition_flavor}
\end{equation}
These operators provide a useful benchmark because their singlet and
triplet Wilson coefficients contribute simultaneously to common scattering
channels and are therefore correlated by both perturbative unitarity and
the spinning sum rules.

Under MFV, retaining the exact flavor structures and terms up to
$\mathcal O(Y_u^2)$, each coefficient contains two independent structures:
\begin{equation}
    [C_{\ell q}^{(i)}]_{prst}
    =
    C_{\ell q}^{(i,1)}\,\delta_{pr}\delta_{st}
    +
    C_{\ell q}^{(i,2)}\,
    \delta_{pr}(Y_uY_u^\dagger)_{st},
    \qquad
    i=1,3.
    \label{eq:Clq_MFV}
\end{equation}
The first term is flavor universal, whereas the second introduces the
leading non-universality associated with the up-type Yukawa spurion.

In the exact $\mathrm{U}(2)^5$ limit, four independent structures arise for each
of the two operators:
\begin{align}
    [C_{\ell q}^{(i)}]_{prst}
    ={}&
    C_{\ell q}^{(i,1)}
    \delta_{pa}\delta_{ar}\delta_{sb}\delta_{bt}
    +
    C_{\ell q}^{(i,2)}
    \delta_{p3}\delta_{3r}\delta_{sa}\delta_{at}
    \nonumber\\
    &+
    C_{\ell q}^{(i,3)}
    \delta_{pa}\delta_{ar}\delta_{s3}\delta_{3t}
    +
    C_{\ell q}^{(i,4)}
    \delta_{p3}\delta_{3r}\delta_{s3}\delta_{3t},
    \qquad
    i=1,3,
    \label{eq:Clq_U2}
\end{align}
where $a,b=1,2$ are summed $\mathrm{U}(2)$ indices.
The four terms correspond, respectively, to light--light,
third--light, light--third, and third--third fermion bilinears.

Substituting Eqs.~\eqref{eq:Clq_MFV} and~\eqref{eq:Clq_U2} into the
complete coupled-channel matrices gives the marginalized unitarity bounds
reported in Table~\ref{tab:lq_flavor_bounds}. In the numerical MFV analysis, here and in App.~\ref{app:flavor_results}, we set $y_t=1$ in
Eq.~\eqref{eq:MFV_top_limit}, so that the bounds on the coefficients of the $Y_u$-dependent structures, such as
$C_{\ell q}^{(1,2)}$ and $C_{\ell q}^{(3,2)}$, apply to $y_t^2$ times these coefficients.
For comparison, we also show the bounds obtained after imposing the
spinning sum rules under scalar or vector dominance.
As throughout this work, these latter regions should be interpreted under
the assumptions discussed in Sec.~\ref{sec:sum_rules_limitations}.

\begin{table}[tbp]
\centering
\renewcommand{\arraystretch}{1.2}
\begin{tabular}{c c c c c}
\toprule
\rowcolor{hdrcolor}
\textbf{Symmetry}
&
\textbf{Coefficient}
&
\textbf{Unitarity}
&
\textbf{Scalar dominance}
&
\textbf{Vector dominance}
\\
\midrule

\multirow{4}{*}{MFV}
&
$C_{\ell q}^{(1,1)}$
&
$[-4.44,\,4.44]$
&
$[0,\,4.44]$
&
$[-4.44,\,0]$
\\

&
$C_{\ell q}^{(1,2)}$
&
$[-7.70,\,7.70]$
&
$[-4.44,\,6.28]$
&
$[-6.28,\,4.44]$
\\

&
$C_{\ell q}^{(3,1)}$
&
$[-4.44,\,4.44]$
&
$[-3.14,\,4.44]$
&
$[-4.44,\,3.14]$
\\

&
$C_{\ell q}^{(3,2)}$
&
$[-7.70,\,7.70]$
&
$[-6.99,\,7.70]$
&
$[-7.70,\,6.99]$
\\
\midrule

\multirow{8}{*}{$\mathrm{U}(2)^5$}
&
$C_{\ell q}^{(1,1)}$
&
$[-5.44,\,5.44]$
&
$[0,\,5.44]$
&
$[-5.44,\,0]$
\\

&
$C_{\ell q}^{(1,2)}$
&
$[-7.70,\,7.70]$
&
$[0,\,7.70]$
&
$[-7.70,\,0]$
\\

&
$C_{\ell q}^{(1,3)}$
&
$[-7.70,\,7.70]$
&
$[0,\,7.70]$
&
$[-7.70,\,0]$
\\

&
$C_{\ell q}^{(1,4)}$
&
$[-10.9,\,10.9]$
&
$[0,\,10.9]$
&
$[-10.9,\,0]$
\\

&
$C_{\ell q}^{(3,1)}$
&
$[-5.44,\,5.44]$
&
$[-3.14,\,5.44]$
&
$[-5.44,\,3.14]$
\\

&
$C_{\ell q}^{(3,2)}$
&
$[-6.28,\,6.28]$
&
$[-3.14,\,6.28]$
&
$[-6.28,\,3.14]$
\\

&
$C_{\ell q}^{(3,3)}$
&
$[-6.28,\,6.28]$
&
$[-3.14,\,6.28]$
&
$[-6.28,\,3.14]$
\\

&
$C_{\ell q}^{(3,4)}$
&
$[-6.28,\,6.28]$
&
$[-3.14,\,6.28]$
&
$[-6.28,\,3.14]$
\\

\bottomrule
\end{tabular}
\caption{
Marginalized bounds on the independent coefficients of
$O_{\ell q}^{(1)}$ and $O_{\ell q}^{(3)}$ under MFV and $\mathrm{U}(2)^5$.
The intervals $[a,b]$ correspond to
$(s/\Lambda^2)C_{\ell q}^{(i,j)}\in[a,b]$.
The first column of bounds follows from perturbative unitarity alone,
whereas the last two additionally impose the spinning sum rules under
scalar or vector dominance, respectively.
}
\label{tab:lq_flavor_bounds}
\end{table}

\begin{figure}[tbp]
    \centering
    \includegraphics[scale = 0.75]{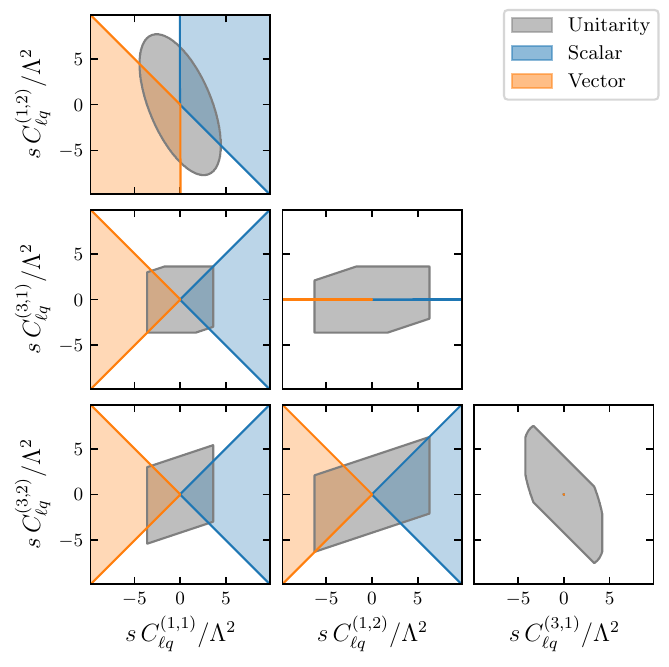}
\caption{
Partial-wave unitarity bounds and spinning sum-rule constraints for
$O_{\ell q}^{(1,3)}$ under MFV, where
$[C_{\ell q}^{(i)}]_{prst}
=
C_{\ell q}^{(i,1)}\delta_{pr}\delta_{st}
+
C_{\ell q}^{(i,2)}\delta_{pr}(Y_uY_u^\dagger)_{st}$,
$i=1,3$.
Gray denotes the region allowed by perturbative unitarity, while blue and orange correspond to the additional spinning sum-rule constraints under scalar- and vector-dominated UV dynamics, respectively. The panels are two-dimensional slices of the full allowed region, obtained by setting all other independent coefficients to zero.
}
    \label{fig:Olq_mfv}
\end{figure}

\begin{figure}[tbp]
    \centering
    \includegraphics[width=1\linewidth]{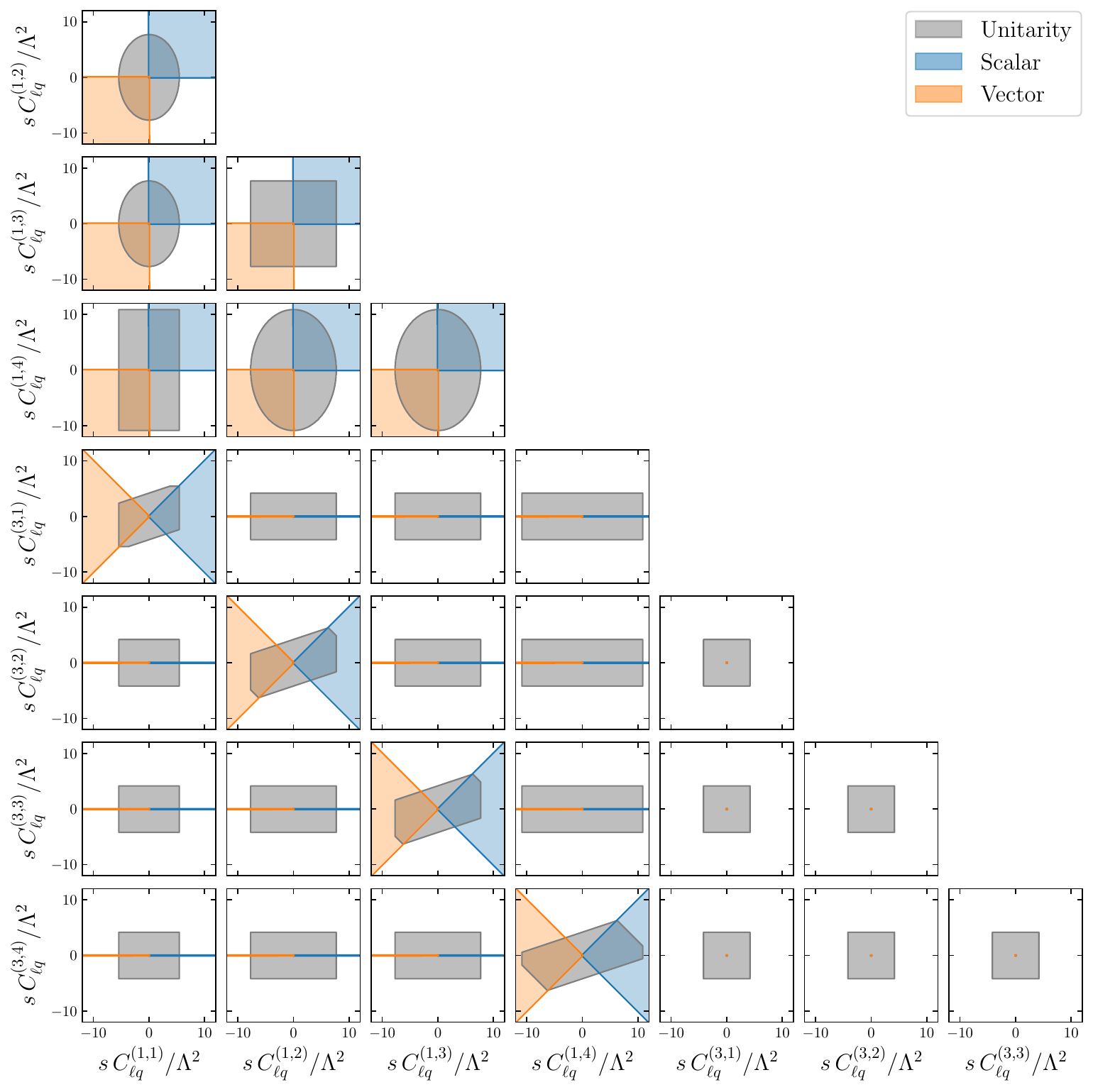}
\caption{
Partial-wave unitarity bounds and spinning sum-rule constraints for
$O_{\ell q}^{(1,3)}$ under $\mathrm{U}(2)^5$ flavor symmetry, where
$[C_{\ell q}^{(i)}]_{prst}
=
C_{\ell q}^{(i,1)}\Pij{p}{r}{a}\Pij{s}{t}{b}
+
C_{\ell q}^{(i,2)}\delta_{p3}\delta_{3r}\Pij{s}{t}{a}
+
C_{\ell q}^{(i,3)}\Pij{p}{r}{a}\delta_{s3}\delta_{3t}
+
C_{\ell q}^{(i,4)}\delta_{p3}\delta_{3r}\delta_{s3}\delta_{3t}$,
$i=1,3$.
Gray denotes the region allowed by perturbative unitarity, while blue and orange correspond to the additional spinning sum-rule constraints under scalar- and vector-dominated UV dynamics, respectively. The panels are two-dimensional slices of the full allowed region, obtained by setting all other independent coefficients to zero.
}
\label{fig:Olq_u2}
\end{figure}

The correlations among the independent coefficients are more transparent in the two-dimensional sections of the allowed regions shown in Figs.~\ref{fig:Olq_mfv} and~\ref{fig:Olq_u2}.
Each panel is a two-dimensional slice of the full multidimensional region, obtained by setting all independent coefficients other than the two displayed to zero.
This should be distinguished from the one-dimensional intervals in Table~\ref{tab:lq_flavor_bounds}, which are obtained by marginalizing over the remaining coefficients.

The comparison illustrates the complementary information carried by
perturbative unitarity and the spinning sum rules.
Unitarity bounds the overall size of the interactions and generally
produces regions symmetric under an overall sign reversal, whereas the
sum rules select particular directions in coefficient space according to
the assumed spin of the UV dynamics.
In some two-dimensional slices, the flavor-space sum-rule conditions are
sufficiently restrictive that setting the remaining coefficients to zero
forces one or both displayed coefficients onto lower-dimensional loci,
so that the allowed region collapses to a line or even to a point.
In the multiflavor case, this behavior follows from the requirement that the positivity or negativity condition
hold for every flavor direction, which correlates coefficients that would appear independent in a single-flavor
analysis.

\subsubsection{General features}
\label{sec:flavor_general_features}

The example above illustrates several features that are common to the
multiflavor analysis of four-fermion operators. In particular:
\begin{itemize}
    \item Imposing a flavor symmetry does not reduce the problem to a set 
of independent single-flavor constraints. Different flavor structures generally contribute to the same coupled-channel amplitudes and are therefore constrained simultaneously. As a result, both the perturbative unitarity bounds and the spinning sum-rule conditions generate non-trivial correlations among the independent flavor coefficients.
\item The dimensionality of the scattering problem can remain sizable
even after the flavor reduction.
For instance, in the $\mathrm{U}(2)^5$ analysis of the $qq\to qq$ sector, the
coupled-channel matrix has dimension $324\times324$, reflecting the combined
multiplicity of color, weak isospin, and flavor degrees of freedom.
This provides a representative example of the computational complexity of
the full three-generation analysis.
\item Spinning sum rules become increasingly restrictive once their
definiteness conditions are imposed over the complete flavor space.
As seen in the $O_{\ell q}^{(1,3)}$ example, correlations among the
independent flavor coefficients can force some two-dimensional slices of
the allowed region to collapse onto lower-dimensional loci.
In the multiflavor case, this behavior arises from requiring the sum-rule inequalities to hold for arbitrary flavor directions.
\item Finally, the spinning sum rules act as discriminants of simple UV
structures rather than as model-independent consistency conditions.
Regions outside both the scalar- and vector-dominated domains would
indicate that the assumptions underlying the simple spin-dominance
classification are not realized, pointing to a more structured UV origin
involving, for instance, non-vanishing boundary terms, mixed-spin dynamics,
or loop-generated matching.
\end{itemize}

The same procedure has been applied systematically to all remaining
current--current four-fermion operators under both MFV and $\mathrm{U}(2)^5$.
For each operator, we determine the independent flavor structures, construct
the corresponding coupled-channel matrices, derive the marginalized
perturbative unitarity bounds, and impose the scalar- and vector-dominance
sum-rule conditions whenever applicable.

Since the complete set of flavor decompositions, bounds, and allowed regions
is extensive, we do not reproduce it in the main text.
The full results are collected in
App.~\ref{app:flavor_results}, where the bounds on each independent
coefficient and, where more than one coefficient is present, the corresponding two-dimensional slices are reported for all current--current four-fermion operators.  
An accompanying Mathematica notebook~\cite{Bresciani:2026acy} provides a machine-readable implementation of the flavor decompositions and theoretical constraints, together with the analytic and numerical results used to obtain the bounds reported in App.~\ref{app:flavor_results}.

\subsection{Matching to UV completions}
\label{sec:UV_matching}

To illustrate more explicitly both the diagnostic power and the limitations
of the theoretical constraints derived above, we now consider representative
single-field UV completions that generate the semileptonic operators
$O_{\ell q}^{(1)}$ and $O_{\ell q}^{(3)}$ at tree level.
The complete tree-level dictionary of renormalizable heavy mediators
generating dimension-six SMEFT operators was classified in
Ref.~\cite{deBlas:2017xtg}.
Restricting to mediators that generate the left-handed semileptonic
structures relevant here leads to the scalar leptoquarks
$\omega_1$ and $\zeta$, the vector leptoquarks $\mathcal U_2$ and $\mathcal X$,
and the color-singlet vectors $\mathcal B$ and $\mathcal W$.
Their gauge quantum numbers, relevant interactions, and tree-level matching onto $[O_{\ell q}^{(1)}]_{3333}$ and 
$[O_{\ell q}^{(3)}]_{3333}$ are summarized in Table~\ref{tab:UV-completions-summary}.

For definiteness, we consider the exact $\mathrm{U}(2)^5$ limit and assume that the new mediators couple predominantly to third-generation fermions.
The matching then populates only the components $[C_{\ell q}^{(1)}]_{3333}$ and $[C_{\ell q}^{(3)}]_{3333}$, corresponding to the coefficients
$C_{\ell q}^{(1,4)}$ and $C_{\ell q}^{(3,4)}$ in the decomposition of
Eq.~\eqref{eq:Clq_U2}. 
This choice allows the UV completions to be compared directly with the
two-dimensional slice of Fig.~\ref{fig:Olq_u2} spanned by
$C_{\ell q}^{(1,4)}$ and $C_{\ell q}^{(3,4)}$.

From the viewpoint of the dispersion relations, the six mediators fall
into two classes. 
The scalar leptoquarks $\omega_1$ and $\zeta$ are exchanged in the $s$ channel of $\ell q\to\ell q$, and the vector leptoquarks $\mathcal U_2$ and $\mathcal X$ in the $s$ channel of $\ell\overline q\to\ell\overline q$, so that, in the absence of boundary terms, the sum-rule sign tracks their spin. The color-singlet vectors $\mathcal B$ and
$\mathcal W$ instead contribute through $t$-channel exchange, for which
non-vanishing boundary terms invalidate the simple scalar/vector
classification~\cite{Gu:2020thj,Remmen:2020uze,Remmen:2022orj,Azatov:2021ygj}.

This can be seen directly from the two combinations entering the spinning
sum rules,
\begin{equation}
    [C_{\ell q}^{(1)}]_{3333}
    +[C_{\ell q}^{(3)}]_{3333},
    \qquad
    [C_{\ell q}^{(1)}]_{3333}
    -[C_{\ell q}^{(3)}]_{3333}.
\end{equation}
For the scalar leptoquarks $\omega_1$ and $\zeta$, both combinations are
non-negative, whereas for the vector leptoquarks $\mathcal U_2$ and $\mathcal X$
they are both non-positive.
The scalar and vector leptoquarks are therefore placed precisely in the
scalar- and vector-dominated regions predicted by the spinning sum rules.

The two $t$-channel vectors exhibit a qualitatively different behavior.
For the singlet vector $\mathcal B$,
$[C_{\ell q}^{(3)}]_{3333}=0$, so that the two sum-rule combinations
coincide. Depending on the sign of
$[g_{\mathcal B}^{\ell}]_{33}[g_{\mathcal B}^{q}]_{33}$,
the matched point can therefore lie either in the nominal
vector-dominated region or in the scalar-dominated one.
In the latter case, a spin-1 mediator would be incorrectly classified
as scalar if the boundary contribution were neglected.

For the triplet vector $\mathcal W$, instead, $[C_{\ell q}^{(1)}]_{3333}=0$.
The two combinations $C_{\ell q}^{(1)}\pm C_{\ell q}^{(3)}$ consequently
have opposite signs for any non-vanishing coupling product.
The corresponding UV trajectory therefore lies outside both the
scalar- and vector-dominated regions, except at the origin.
This does not signal an inconsistency of the UV theory: it is precisely
the behavior expected when the non-vanishing boundary term associated
with vector $t$-channel exchange invalidates the simple spin
classification.

The perturbative unitarity constraints have a different status.
Substituting the matching relations of Table~\ref{tab:UV-completions-summary} into the analytic bounds derived above gives the conditions listed in Table~\ref{tab:models}.
They constrain combinations of the form
\begin{equation}
    |g_{\rm UV}|^2\,s \lesssim {\cal O}(4\pi)\,M_{\rm UV}^2,
\end{equation}
with representation-dependent numerical coefficients.
Therefore, throughout the regime in which the mediator can consistently
be integrated out, $s\ll M_{\rm UV}^2$, perturbative couplings naturally
satisfy the EFT unitarity bounds.
Conversely, extrapolating the effective interaction toward
$s\sim M_{\rm UV}^2$ translates the unitarity condition into an
upper bound on the UV coupling of the expected perturbative size.
At such energies, however, the EFT expansion itself approaches its
breakdown scale and the explicit mediator should be restored.

\begin{figure}[tbp]
    \centering
    \includegraphics[width=0.32\textwidth]{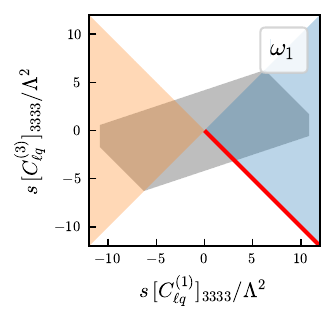}
    \hfill
    \includegraphics[width=0.32\textwidth]{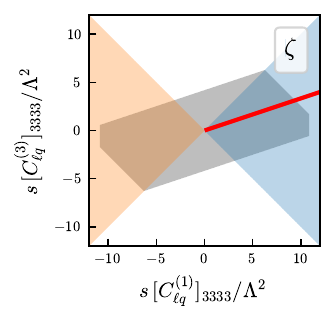}
    \hfill
    \includegraphics[width=0.32\textwidth]{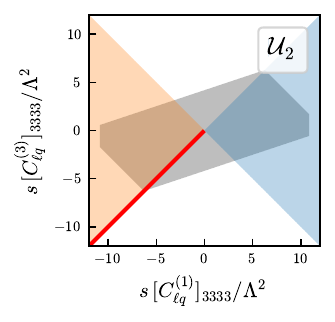}\\%
    \includegraphics[width=0.32\textwidth]{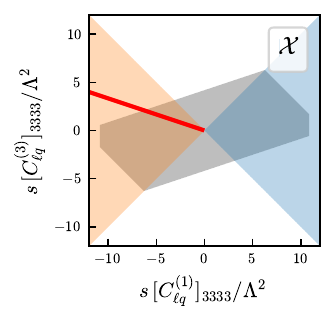}
        \hfill
    \includegraphics[width=0.32\textwidth]{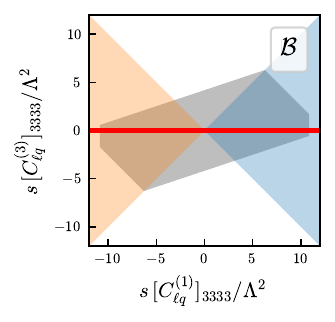}%
    \hfill
    \includegraphics[width=0.32\textwidth]{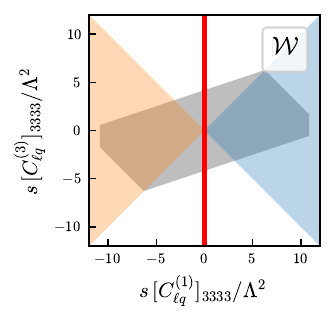}
    \caption{Single-field UV completions in the $\big(s[C_{\ell q}^{(1)}]_{3333}/\Lambda^2, s[C_{\ell q}^{(3)}]_{3333}/\Lambda^2\big)$ plane. For each mediator, the red line denotes the one-dimensional trajectory selected by the tree-level matching relation, the gray region is allowed by perturbative unitarity, and the regions shaded in blue and orange show the scalar- and vector-dominated domains implied by the spinning sum rules, respectively. The scalar leptoquarks $\omega_1$ and $\zeta$ and the vector leptoquarks $\mathcal U_2$ and $\mathcal X$, exchanged in the $s$ channel, are classified according to their physical spin. By contrast, the $t$-channel vectors $\mathcal B$ and 
    $\mathcal W$ can evade this classification because of non-vanishing boundary contributions.}
    \label{fig:UV-plots}
\end{figure}

Figure~\ref{fig:UV-plots} summarizes these results geometrically.
For a fixed single-field completion, the matching relation fixes a
direction in the $(C_{\ell q}^{(1,4)},C_{\ell q}^{(3,4)})$ plane, while the coupling-to-mass ratio sets the distance from the origin. The figure makes manifest the correct spin classification of the $s$-channel
leptoquarks and its possible evasion by the $t$-channel vectors.

This exercise makes explicit the interpretation discussed in Sec.~\ref{sec:sum_rules_limitations}.
A point outside both spin-dominated regions does not imply an inconsistent
UV theory, but rather the failure of at least one assumption underlying
the sign-definite sum rules. The $t$-channel vectors provide a
concrete tree-level realization through non-vanishing boundary terms.

\begin{table}[tbp]
    \centering
    \begin{adjustbox}{max width=\textwidth}
    \renewcommand{\arraystretch}{1.5}
    \begin{tabular}{cccccc}
    \toprule
    \rowcolor{hdrcolor}
    \textbf{Field} & \textbf{Irrep} & \textbf{Spin} & \textbf{UV Lagrangian} &
    \boldmath$[C_{\ell q}^{(1)}]_{3333}/\Lambda^2$ &
    \boldmath$[C_{\ell q}^{(3)}]_{3333}/\Lambda^2$ \\
    \midrule
    $\omega_1$ & $(3,1)_{-\frac{1}{3}}$ & $0$ &
    $-[y_{\omega_1}^{q\ell}]_{33}\omega_1^\dag
    \overline{q}_L^{3c}\,i\sigma^2\ell_L^3 + \text{h.c.}$ &
    $\displaystyle\frac{1}{4M_{\omega_1}^2}|[y^{q\ell}_{\omega_1}]_{33}|^2$ &
    $\displaystyle-\frac{1}{4M_{\omega_1}^2}|[y^{q\ell}_{\omega_1}]_{33}|^2$ \\[0.5cm]

    $\zeta$ & $(3,3)_{-\frac{1}{3}}$ & $0$ &
    $-[y_{\zeta}^{q\ell}]_{33}\zeta^{I\dag}
    \overline{q}_L^{3c}i\sigma^2\sigma^I\ell_L^3 + \text{h.c.}$ &
    $\displaystyle\frac{3}{4M_{\zeta}^2}|[y^{q\ell}_{\zeta}]_{33}|^2$ &
    $\displaystyle\frac{1}{4M_{\zeta}^2}|[y^{q\ell}_{\zeta}]_{33}|^2$ \\[0.5cm]

    $\mathcal{U}_2$ & $(3,1)_{\frac{2}{3}}$ & $1$ &
    $-[g_{\mathcal{U}_2}^{\ell q}]_{33}\mathcal{U}_2^{\mu\dag}
    \overline\ell_L^3\gamma_\mu q_L^3 + \text{h.c.}$ &
    $\displaystyle-\frac{1}{2M_{\mathcal U_2}^2}|[g_{\mathcal U_2}^{\ell q}]_{33}|^2$ &
    $\displaystyle-\frac{1}{2M_{\mathcal U_2}^2}|[g_{\mathcal U_2}^{\ell q}]_{33}|^2$ \\[0.5cm]

    $\mathcal{X}$ & $(3,3)_{\frac{2}{3}}$ & $1$ &
    $-\frac{1}{2}[g_{\mathcal{X}}]_{33}\mathcal{X}^{I\mu\dag}
    \overline\ell_L^3\gamma_\mu\sigma^I q_L^3 + \text{h.c.}$ &
    $\displaystyle -\frac{3}{8M_{\mathcal X}^2}|[g_{\mathcal X}]_{33}|^2$ &
    $\displaystyle \frac{1}{8M_{\mathcal X}^2}|[g_{\mathcal X}]_{33}|^2$ \\[0.5cm]
    $\mathcal{B}$ & $(1,1)_0$ & $1$ &
$\makecell[c]{
-[g_{\mathcal{B}}^\ell]_{33}\mathcal{B}^\mu
\overline\ell_L^3\gamma_\mu\ell_L^3\\
\qquad-[g_{\mathcal{B}}^q]_{33}\mathcal{B}^\mu
\overline q_L^3\gamma_\mu q_L^3
}$ 
% $-[g_{\mathcal{B}}^\ell]_{33}\mathcal{B}^\mu
% \overline\ell_L^3\gamma_\mu\ell_L^3-[g_{\mathcal{B}}^q]_{33}\mathcal{B}^\mu
% \overline q_L^3\gamma_\mu q_L^3$ 
&
$\displaystyle-\frac{1}{M_{\mathcal B}^2}
[g_{\mathcal B}^\ell]_{33}[g_{\mathcal B}^q]_{33}$ &
$\displaystyle 0$ \\[0.5cm]

$\mathcal{W}$ & $(1,3)_0$ & $1$ &
$\makecell[c]{
-\frac{1}{2}[g_{\mathcal{W}}^\ell]_{33}\mathcal{W}^{I\mu}
\overline\ell_L^3\gamma_\mu\sigma^I\ell_L^3\\
\qquad-\frac{1}{2}[g_{\mathcal{W}}^q]_{33}\mathcal{W}^{I\mu}
\overline q_L^3\gamma_\mu\sigma^I q_L^3
}$ 
% $
% -\frac{1}{2}[g_{\mathcal{W}}^\ell]_{33}\mathcal{W}^{I\mu}
% \overline\ell_L^3\gamma_\mu\sigma^I\ell_L^3-\frac{1}{2}[g_{\mathcal{W}}^q]_{33}\mathcal{W}^{I\mu}
% \overline q_L^3\gamma_\mu\sigma^I q_L^3$ 
&$\displaystyle 0$ &
$\displaystyle -\frac{1}{4M_{\mathcal W}^2}
[g_{\mathcal W}^\ell]_{33}[g_{\mathcal W}^q]_{33}$ \\
    \bottomrule
    \end{tabular}
    \end{adjustbox}
    \caption{Single-field tree-level UV completions generating
$[O_{\ell q}^{(1)}]_{3333}$ and $[O_{\ell q}^{(3)}]_{3333}$.
We report the SM gauge representation, spin, relevant interaction, 
and tree-level matching onto the Warsaw-basis Wilson coefficients.
The mediator classification follows the tree-level SMEFT dictionary 
of Ref.~\cite{deBlas:2017xtg}.}
    \label{tab:UV-completions-summary}
\end{table}

\begin{table}[tbp]
    \centering
    \renewcommand{\arraystretch}{1.5}
    \begin{tabular}{cccc}
    \toprule
    \rowcolor{hdrcolor}
    \textbf{Field} & \textbf{Channel} & \textbf{Unitarity bound} & \textbf{Sum rules} \\
    \midrule
    $\omega_1$ & $s$ &
    $|[y^{q\ell}_{\omega_1}]_{33}|^2\,s
    \leq 4\pi M^2_{\omega_1}$ &
    $|[y^{q\ell}_{\omega_1}]_{33}|^2\geq 0$ \\[0.3cm]

    $\zeta$ & $s$ &
    $|[y^{q\ell}_{\zeta}]_{33}|^2\,s
    \leq 4\pi M^2_{\zeta}$ &
    $|[y^{q\ell}_{\zeta}]_{33}|^2\geq 0$ \\[0.3cm]

    $\mathcal{U}_2$ & $s$ &
    $|[g_{\mathcal U_2}^{\ell q}]_{33}|^2\,s
    \leq 4\pi M^2_{\mathcal U_2}$ &
    $-|[g_{\mathcal U_2}^{\ell q}]_{33}|^2\leq 0$ \\[0.3cm]

    $\mathcal{X}$ & $s$ &
    $|[g_{\mathcal X}]_{33}|^2\,s
    \leq \frac{16}{3}\pi M^2_{\mathcal X}$ &
    $-|[g_{\mathcal X}]_{33}|^2\leq 0$ \\[0.3cm]

    $\mathcal{B}$ & $t$ &
    $|[g_{\mathcal B}^\ell]_{33}[g_{\mathcal B}^q]_{33}|\,s
    \leq 2\sqrt{3}\,\pi M^2_{\mathcal B}$ &
    $-[g_{\mathcal B}^\ell]_{33}[g_{\mathcal B}^q]_{33}\leq 0$ \\[0.3cm]

    $\mathcal{W}$ & $t$ &
    $|[g_{\mathcal W}^\ell]_{33}[g_{\mathcal W}^q]_{33}|\,s
    \leq \frac{16}{3}\pi M^2_{\mathcal W}$ &
    $[g_{\mathcal W}^\ell]_{33}[g_{\mathcal W}^q]_{33}=0$ \\
    \bottomrule
    \end{tabular}
\caption{Perturbative unitarity bounds and spinning sum-rule conditions for the single-field UV completions of Table~\ref{tab:UV-completions-summary}, after substituting the corresponding tree-level matching relations. 
For the $s$-channel leptoquarks, the sum-rule conditions reduce to
manifestly satisfied inequalities with the sign dictated by the mediator
spin. For the $t$-channel vectors, the entries show the conditions that
would follow from the naive spin classification if boundary contributions
were neglected; the latter are non-vanishing and invalidate this
interpretation.}
    \label{tab:models}
\end{table}

\section{Conclusions}
\label{sec:conclusion}

We have presented a systematic analysis of perturbative unitarity bounds and spinning sum rules in the baryon-number-conserving dimension-six SMEFT, 
with particular emphasis on the correlations among Wilson coefficients and their flavor structure. 
Our results provide complementary information on the range of validity of the effective description and, under suitable assumptions, on the possible ultraviolet origin of four-fermion interactions.

Using the on-shell partial-wave formalism for generic $N\to M$ scattering, we have performed a 
fully coupled-channel analysis of the complete dimension-six operator basis. 
By accounting for all relevant helicity and gauge configurations, and retaining correlations among operators contributing to the same scattering amplitudes, 
we have obtained perturbative unitarity bounds on individual Wilson coefficients through marginalization over the remaining interactions. 
This procedure yields conservative one-dimensional limits that can differ from conventional single-operator bounds. The generalized formalism also allows higher-multiplicity processes to be treated consistently, 
which is essential for the operator $(\varphi^\dagger\varphi)^3$ and the Yukawa-like $\psi^2\varphi^3$ interactions. 
Our analysis includes the complete four-fermion basis and operators involving gluon fields, extending the scope of previous systematic studies.

We have complemented these results by expressing spinning sum rules for dimension-six four-fermion operators in the Warsaw basis. 
We have formulated their implications as definiteness conditions in flavor space and translated them into inequalities among Wilson coefficients. 
When the relevant ultraviolet assumptions are satisfied, these relations connect the signs of particular coefficient combinations to the spin of the exchanged heavy states, providing a means of distinguishing scalar- from vector-dominated dynamics beyond the information accessible through perturbative unitarity alone.

A central result of this work is the extension of both classes of constraints to the three-generation four-fermion sector under Minimal Flavor Violation and $\mathrm{U}(2)^5$ flavor symmetry. 
These hypotheses reduce the number of independent flavor structures while preserving non-trivial correlations among different generations. 
Within these frameworks, we have derived the corresponding unitarity and sum-rule constraints, identifying restrictions on combinations of flavor coefficients that cannot be inferred from a single-flavor analysis. 
The resulting allowed regions exhibit non-trivial correlations and, in certain sections of coefficient space, collapse onto lower-dimensional loci.
Such features are particularly relevant for scenarios with non-universal new-physics couplings, including interactions preferentially involving third-generation fermions.

The interpretation of spinning sum rules nevertheless requires particular care. 
Unlike perturbative unitarity bounds, their sign constraints are not universal consistency conditions on the SMEFT. 
They rely on assumptions concerning the ultraviolet dynamics, including the relevant spin content and the absence of boundary contributions to the dispersion relations. 
Through explicit matching to representative ultraviolet completions, we have illustrated both the discriminating power and the limitations of the spinning sum rules. 
In particular, vector exchange in the $t$ channel can generate non-vanishing boundary terms and invalidate a naive identification of the sign of a Wilson coefficient with the spin of the mediator. 
Departures from both the scalar- and vector-dominated sum-rule regions therefore need not signal an inconsistent ultraviolet completion, but can instead point to more general underlying dynamics.

Taken together, our results provide a systematic set of theoretical constraints that can complement experimental SMEFT analyses. 
The unitarity bounds offer quantitative criteria for assessing perturbative validity at the hard-scattering scale, while the spinning sum rules, where applicable, 
supply additional information about the structure of possible ultraviolet completions. 
Incorporating these constraints alongside theoretically motivated flavor hypotheses can help characterize the viable parameter space of new-physics scenarios and sharpen the interpretation of present and future high-energy measurements.

\begin{acknowledgments}
We thank Giulio Crisanti for valuable help with the diagonalization of the largest multiflavor partial-wave matrices of the four-fermion operators. This work received funding from the INFN Iniziativa Specifica APINE and from the European Union's Horizon 2020 research and innovation programme under the Marie Sk\l{}odowska-Curie grant agreements n.\ 860881 --- HIDDeN, n.\ 101086085 --- ASYMMETRY. This work was also partially supported by the Italian MUR Departments of Excellence grant 2023-2027 ``Quantum Frontiers'' and by the European Union --- Next Generation EU and by the Italian Ministry of University and Research (MUR) via the PRIN 2022 project n.\ 2022K4B58X --- AxionOrigins. 
\end{acknowledgments}

\appendix

\section{Mathematical tools\label{app:mathtools}}

\subsection{Logical expression\label{sec:appendix, logical expression}}

The function $f(x) = a + bx - \sqrt{c^2 + dx + ex^2}$ --- assumed real on $[0,1]$ and with $a,b,c,d,e\in\mathbb{R}$ --- is such that $f(x) \ge 0$ for every $x \in [0,1]$ if and only if the system of inequalities
\begin{align}
    \Delta(a,b,c,d,e) &= 
    a \ge 0 
    \ \land \
    a+b \ge 0
    \ \land \ 
    a^2 - c^2 \ge 0
    \ \land \ 
    (a+b)^2 - c^2 -d-e \ge 0 \nonumber
    \\& \quad
    \land \
    [e - b^2 \ge 0 \ \lor\  (2ab-d)(2ab-d+2b^2-2e) \ge 0 \nonumber 
    \\& \quad
    \lor \ 4(b^2-e)(a^2-c^2)-(2ab-d)^2 \ge 0]
\end{align}
is true.

To prove this, note that,
since $g(x)=c^2 + dx + ex^2 \ge 0$ on $[0,1]$ by assumption, $f(x) \ge 0$ on $[0,1]$ is equivalent to $\ell(x) = a + bx \ge 0$ and $h(x) = \ell(x)^2 -g(x) \ge 0$ on $[0,1]$. The latter reads $h(x) = \alpha x^2 + \beta x + \gamma$, with $\alpha = b^2-e$, $\beta = 2ab-d$, and $\gamma=a^2-c^2$.
Thus, it is sufficient to prove that $p(x) = \alpha x^2+\beta x + \gamma \ge 0$ on $[0,1]$ if and only if
    \begin{equation}\label{eq:lemma}
        \gamma\ge0 \ \land \ \alpha+\beta+\gamma\ge0 \ \land \
    [\alpha\le0 \ \lor \ \beta(\beta+2\alpha)\ge0 \ \lor \ 4\alpha\gamma-\beta^2\ge0 ] .
    \end{equation}
Indeed, $\Delta$ is the conjunction of Eq.~\eqref{eq:lemma} applied to $\ell$ and $h$.
The first two conditions in Eq.~\eqref{eq:lemma} are $p(0)\ge 0$ and $p(1)\ge 0$, thus they are necessary; assume them in what follows. If $\alpha \le 0$, then $p$ is concave and has its minimum on $[0,1]$ at an endpoint, so $p\ge 0$; instead, if $\alpha >0$, the vertex is at $x_\ast = -\beta/(2\alpha)$, and
\begin{equation}
    x_\ast\le0 \iff \beta\ge0 , \qquad x_\ast\ge1 \iff \beta+2\alpha\le0 ,
\end{equation}
so $x_\ast \notin (0,1)$ if and only if $\beta(\beta+2\alpha)\ge 0$, in which case the minimum is again at an endpoint and $p\ge 0$. Otherwise, $x_\ast \in (0,1)$ and $\min_{[0,1]} p = p(x_\ast) = (4\alpha\gamma-\beta^2)/(4\alpha)$, which has the same sign as $4\alpha\gamma-\beta^2$. The last condition is then both necessary and sufficient. 

\subsection{Eigenvalues of Jordan--Wielandt matrices\label{app:JW}}
Given the Hermitian Jordan--Wielandt matrix
\begin{equation}
    M=\begin{pmatrix}
        \mathbb{O}_{m\times m} & A \\ A^\dag & \mathbb{O}_{n\times n}
    \end{pmatrix}
\end{equation}
of size $m+n$,
with $A\in \mathbb{C}^{m \times n}$,
we have that its characteristic polynomial --- with the convention $p_X(x) = \det[x\mathbb{1}-X]$ --- is
\begin{equation}
    p_M(x)=x^{n-m}\,p_{AA^\dag}(x^2) = x^{m-n}\,p_{A^\dag A}(x^2) ,
\end{equation}
which means that the eigenvalues of $M$ can be computed as the positive and negative square roots of the eigenvalues of $AA^\dag$ if $m\le n$ (of $A^\dag A$ if $m > n$), together with additional $\abs{n-m}$ vanishing eigenvalues. 

To prove this, consider a generic block matrix that takes the form
\begin{equation}
    \begin{pmatrix}
        P & Q \\ R & S
    \end{pmatrix}.
\end{equation}
If $S$ is invertible, then
\begin{equation}
    \begin{pmatrix}
        P & Q \\ R & S
    \end{pmatrix}
    \begin{pmatrix} \mathbb{1} & \mathbb{O} \\ -S^{-1}R & \mathbb{1} 
    \end{pmatrix}
    =
    \begin{pmatrix} P-QS^{-1}R & Q \\ \mathbb{O} & S 
    \end{pmatrix},
\end{equation}
which implies that 
\begin{equation}
    \det[\begin{pmatrix} P & Q \\ R & S \end{pmatrix}]=\det[P-QS^{-1}R] \det[S].
\end{equation}
By applying this to the matrix $x\mathbb{1}_{(m+n)\times(m+n)}-M$, we find the result above.

\section{Complete results for flavorful four-fermion operators\label{app:flavor_results}}

Here we collect the complete perturbative-unitarity and spinning sum-rule constraints on the current--current four-fermion SMEFT operators under the MFV and $\mathrm{U}(2)^5$ flavor hypotheses. Table~\ref{tab:summary} indexes the tables of marginalized bounds and the corresponding figures.
\begin{table}[tbp]
\renewcommand{\arraystretch}{1.2}
    \centering
    \begin{tabular}{ccccc}
    \toprule
    \rowcolor{hdrcolor} \textbf{Operator class} & \textbf{Operator(s)} & \textbf{Table} & \textbf{Figure (MFV)} & \textbf{Figure} \boldmath$(\mathrm{U}(2)^5)$ \\
    \midrule
    \multirow{3}{*}{\boldmath$(\overline LL)(\overline LL)$} & $O_{\ell \ell}$ & \ref{tab:Oll_UB} & \ref{fig:Oll_mfv} & \ref{fig:Oll_u2} \\
    & $O_{qq}^{(1,3)}$ & \ref{tab:Oqq_UB} & \ref{fig:Oqq_mfv} & \ref{fig:Oqq_u2} \\
    & $O_{\ell q}^{(1,3)}$ & \ref{tab:lq_flavor_bounds} & \ref{fig:Olq_mfv} & \ref{fig:Olq_u2} \\
    \midrule
    \multirow{6}{*}{\boldmath$(\overline RR)(\overline RR)$} & $O_{ee}$ & \ref{tab:Oee_UB} & --- & \ref{fig:Oee_u2} \\
    & $O_{uu}$ & \ref{tab:Ouu_UB} & \ref{fig:Ouu_mfv} & \ref{fig:Ouu_u2} \\
    & $O_{dd}$ & \ref{tab:Odd_UB} & \ref{fig:Odd_mfv} & \ref{fig:Ouu_u2} \\
    & $O_{eu}$ & \ref{tab:Oeu_UB} & \ref{fig:Oeu_mfv} & \ref{fig:Oeu_u2} \\
    & $O_{ed}$ & \ref{tab:Oed_UB} & --- & \ref{fig:Oeu_u2} \\
    & $O_{ud}^{(1,8)}$ & \ref{tab:Oud_UB} & \ref{fig:Oud_mfv} & \ref{fig:Oud_u2} \\
    \midrule
    \multirow{6}{*}{\boldmath$(\overline LL)(\overline RR)$} & $O_{\ell e}$ & \ref{tab:Ole_UB} & --- & \ref{fig:Ole_u2} \\
    & $O_{\ell u}$ & \ref{tab:Olu_UB} & \ref{fig:Olu_mfv} & \ref{fig:Olu_u2} \\
    & $O_{\ell d}$ & \ref{tab:Old_UB} & --- & \ref{fig:Olu_u2} \\
    & $O_{qe}$ & \ref{tab:Oqe_UB} & \ref{fig:Oqe_mfv} & \ref{fig:Oqe_u2} \\
    & $O_{qu}^{(1,8)}$ & \ref{tab:Oqu_UB} & \ref{fig:Oqu_mfv} & \ref{fig:Oqu_u2} \\
    & $O_{qd}^{(1,8)}$ & \ref{tab:Oqd_UB} & \ref{fig:Oqd_mfv} & \ref{fig:Oqu_u2} \\
    \bottomrule
    \end{tabular}
    \caption{References to the tables of marginalized bounds (from partial-wave unitarity alone and combined with the spinning sum rules) and to the
    figures showing the corresponding allowed regions, for each four-fermion operator under the MFV
    and $\mathrm{U}(2)^5$ flavor hypotheses. A dash indicates that the operator admits a single independent
    coefficient under MFV, so that the bound is one-dimensional and no region is displayed. Under $\mathrm{U}(2)^5$, $O_{dd}$, $O_{ed}$, $O_{\ell d}$, and $O_{qd}^{(1,8)}$ have the same decomposition and the same allowed regions as $O_{uu}$, $O_{eu}$, $O_{\ell u}$, and $O_{qu}^{(1,8)}$, respectively, and share the corresponding figure.}
    \label{tab:summary}
\end{table}

% ============================================================
%  TABLE 1 — Q_{ell ell}
% ============================================================
\begin{table}[p]
\centering
{\small
% ---- info block ----
\begin{tabular}{ll}
    \textbf{Operator:} & $\boxed{O_{\ell\ell} = (\overline\ell_p \gamma^\mu \ell_r)(\overline\ell_s \gamma_\mu \ell_t)}$ \\[0.2cm]
    \textbf{MFV decomposition:} &
        $[C_{\ell\ell}]_{prst} = C_{\ell\ell}^{(1)}\delta_{pr}\delta_{st} + C_{\ell\ell}^{(2)}\delta_{pt}\delta_{rs}$ \\[0.15cm]
    \textbf{$\mathrm{U}(2)^5$ decomposition:} &
        $[C_{\ell\ell}]_{prst} =
            C_{\ell\ell}^{(1)}\Pij{p}{r}{a}\Pij{s}{t}{b}
            + C_{\ell\ell}^{(2)}(\Pij{s}{t}{a}\delta_{p3}\delta_{3r} + \Pij{p}{r}{a}\delta_{s3}\delta_{3t})$ \\
        & $\phantom{[C_{\ell\ell}]_{prst} ={}}
            + C_{\ell\ell}^{(3)}\delta_{p3}\delta_{3r}\delta_{s3}\delta_{3t}
            + C_{\ell\ell}^{(4)}\Pij{s}{r}{a}\Pij{p}{t}{b}
            $ \\
        & $\phantom{[C_{\ell\ell}]_{prst} ={}}+ C_{\ell\ell}^{(5)}(\Pij{p}{t}{a}\delta_{s3}\delta_{3r} + \Pij{s}{r}{a}\delta_{p3}\delta_{3t})$ \\
\end{tabular}
\\[0.3cm]
% ---- bounds table ----
\renewcommand{\arraystretch}{1.3}
\begin{tabular}{ccccc}
    \toprule
    \rowcolor{hdrcolor}
    \textbf{Symmetry} & \textbf{Coefficient}
        & \textbf{Unitarity bound}
        & \textbf{Scalar dominance}
        & \textbf{Vector dominance} \\
    \midrule
    % --- MFV section ---
    %\rowcolor{seccolor}
    %\multicolumn{5}{l}{\textit{Minimal Flavor Violation}} 
    %\\
    \multirow{2}{*}{MFV}
        & $C_{\ell\ell}^{(1)}$ & $[-6.28,\, 6.28]$ & $[0,\, 6.28]$      & $[-6.28,\, 0]$    \\
        & $C_{\ell\ell}^{(2)}$ & $[-8.38,\, 8.38]$ & $[-6.28,\, 3.77]$  & $[-3.77,\, 6.28]$ \\
    \midrule
    % --- \mathrm{U}(2)^5 section ---
    %\rowcolor{seccolor}
    %\multicolumn{5}{l}{\textit{$\mathrm{U}(2)^5$ Flavor Symmetry}} \\
    \multirow{5}{*}{$\mathrm{U}(2)^5$}
        & $C_{\ell\ell}^{(1)}$ & $[-6.28,\, 6.28]$ & $[0,\, 6.28]$     & $[-6.28,\, 0]$    \\
        & $C_{\ell\ell}^{(2)}$ & $[-6.28,\, 6.28]$ & $[0,\, 6.28]$     & $[-6.28,\, 0]$    \\
        & $C_{\ell\ell}^{(3)}$ & $[-6.28,\, 6.28]$ & $[0,\, 6.28]$     & $[-6.28,\, 0]$    \\
        & $C_{\ell\ell}^{(4)}$ & $[-12.6,\, 12.6]$ & $[-6.28,\, 4.71]$ & $[-4.71,\, 6.28]$ \\
        & $C_{\ell\ell}^{(5)}$ & $[-12.6,\, 12.6]$ & $[-6.28,\, 4.14]$ & $[-4.14,\, 6.28]$ \\
    \bottomrule
\end{tabular}}
\caption{Marginalized partial-wave unitarity bounds for $O_{\ell \ell}$ under MFV and $\mathrm{U}(2)^5$
flavor symmetry, with and without spinning sum rules under scalar or vector dominance in the UV.
Bounds are intervals $[a,b]$ with $(s/\Lambda^2)C \in [a,b]$, rounded to 3 significant digits.}
\label{tab:Oll_UB}
\end{table}

% ============================================================
%  TABLE 2 — Q_{qq}
% ============================================================
\begin{table}[p]
\centering
{\small
% ---- info block ----
\begin{tabular}{ll}
    \textbf{Operators:} &
        $\boxed{O_{qq}^{(1)} = (\overline q_p \gamma^\mu q_r)(\overline q_s \gamma_\mu q_t),\quad O_{qq}^{(3)} = (\overline q_p \gamma^\mu \sigma^I q_r)(\overline q_s \gamma_\mu \sigma^I q_t)}$ \\[0.2cm]
    \textbf{MFV decomposition:} &
        $[C_{qq}^{(i)}]_{prst} =
            C_{qq}^{(i,1)}\delta_{pr}\delta_{st}
            + C_{qq}^{(i,2)}\delta_{pt}\delta_{sr}
            $ \\
        & $\phantom{[C_{qq}^{(i)}]_{prst} ={}}+ C_{qq}^{(i,3)}((Y_u Y_u^\dagger)_{st}\delta_{pr} + (Y_u Y_u^\dagger)_{pr}\delta_{st})$ \\
        & $\phantom{[C_{qq}^{(i)}]_{prst} ={}}
            + C_{qq}^{(i,4)}((Y_u Y_u^\dagger)_{sr}\delta_{pt} + (Y_u Y_u^\dagger)_{pt}\delta_{sr})$,\quad $i=1,3$ \\[0.15cm]
    \textbf{$\mathrm{U}(2)^5$ decomposition:} &
        $[C_{qq}^{(i)}]_{prst} =
            C_{qq}^{(i,1)}\Pij{p}{r}{a}\Pij{s}{t}{b}
            + C_{qq}^{(i,2)}(\Pij{s}{t}{a}\delta_{p3}\delta_{3r} + \Pij{p}{r}{a}\delta_{s3}\delta_{3t})
            $ \\
        & $\phantom{[C_{qq}^{(i)}]_{prst} ={}}+ C_{qq}^{(i,3)}\delta_{p3}\delta_{3r}\delta_{s3}\delta_{3t}
            + C_{qq}^{(i,4)}\Pij{p}{t}{a}\Pij{s}{r}{b}
            $ \\
        & $\phantom{[C_{qq}^{(i)}]_{prst} ={}}+ C_{qq}^{(i,5)}(\Pij{s}{r}{a}\delta_{p3}\delta_{3t} + \Pij{p}{t}{a}\delta_{s3}\delta_{3r})$,\quad $i=1,3$ \\
\end{tabular}
\\[0.3cm]
% ---- bounds table ----
\renewcommand{\arraystretch}{1.3}
\begin{tabular}{ccccc}
    \toprule
    \rowcolor{hdrcolor}
    \textbf{Symmetry} & \textbf{Coefficient}
        & \textbf{Unitarity bound}
        & \textbf{Scalar dominance}
        & \textbf{Vector dominance} \\
    \midrule
    % --- MFV section ---
    \multirow{8}{*}{MFV}
        & $C_{qq}^{(1,1)}$ & $[-5.39,\, 5.39]$ & $[-5.39,\, 4.71]$ & $[-4.71,\, 5.39]$ \\
        & $C_{qq}^{(3,1)}$ & $[-3.46,\, 3.46]$ & $[-3.46,\, 2.51]$ & $[-2.51,\, 3.46]$ \\
        & $C_{qq}^{(1,2)}$ & $[-5.34,\, 5.34]$ & $[-5.03,\, 5.34]$ & $[-5.34,\, 5.03]$ \\
        & $C_{qq}^{(3,2)}$ & $[-3.67,\, 3.67]$ & $[-1.57,\, 3.67]$ & $[-3.67,\, 1.57]$ \\
        & $C_{qq}^{(1,3)}$ & $[-7.68,\, 7.68]$ & $[-7.10,\, 7.26]$ & $[-7.26,\, 7.10]$ \\
        & $C_{qq}^{(3,3)}$ & $[-5.01,\, 5.01]$ & $[-4.03,\, 5.01]$ & $[-5.01,\, 4.03]$ \\
        & $C_{qq}^{(1,4)}$ & $[-8.20,\, 8.20]$ & $[-8.20,\, 7.67]$ & $[-7.67,\, 8.20]$ \\
        & $C_{qq}^{(3,4)}$ & $[-5.31,\, 5.31]$ & $[-4.56,\, 3.79]$ & $[-3.79,\,4.56]$ \\
    \midrule
    % --- \mathrm{U}(2)^5 section ---
    \multirow{10}{*}{$\mathrm{U}(2)^5$}
        & $C_{qq}^{(1,1)}$ & $[-5.39,\, 5.39]$ & $[-5.39,\, 4.71]$ & $[-4.71,\, 5.39]$ \\
        & $C_{qq}^{(3,1)}$ & $[-3.46,\, 3.46]$ & $[-3.46,\, 2.51]$ & $[-2.51,\, 3.46]$ \\
        & $C_{qq}^{(1,2)}$ & $[-4.44,\, 4.44]$ & $[-4.44,\, 3.33]$ & $[-3.33,\, 4.44]$ \\
        & $C_{qq}^{(3,2)}$ & $[-3.68,\, 3.68]$ & $[-3.48,\, 3.68]$ & $[-3.68,\, 3.48]$ \\
        & $C_{qq}^{(1,3)}$ & $[-3.77,\, 3.77]$ & $[-3.14,\, 3.77]$ & $[-3.77,\, 3.14]$ \\
        & $C_{qq}^{(3,3)}$ & $[-4.19,\, 4.19]$ & $[-3.14,\, 4.19]$ & $[-4.19,\, 3.14]$ \\
        & $C_{qq}^{(1,4)}$ & $[-5.34,\, 5.34]$ & $[-5.03,\, 5.34]$ & $[-5.34,\, 5.03]$ \\
        & $C_{qq}^{(3,4)}$ & $[-3.67,\, 3.67]$ & $[-1.57,\, 3.67]$ & $[-3.67,\, 1.57]$ \\
        & $C_{qq}^{(1,5)}$ & $[-5.20,\, 5.20]$ & $[-5.20,\, 5.15]$ & $[-5.15,\, 5.20]$ \\
        & $C_{qq}^{(3,5)}$ & $[-4.19,\, 4.19]$ & $[-3.14,\, 4.19]$ & $[-4.19,\, 3.14]$ \\
    \bottomrule
\end{tabular}}
\caption{Same as Table~\ref{tab:Oll_UB} but for $O_{qq}^{(1,3)}$.}
\label{tab:Oqq_UB}
\end{table}

\begin{table}[p]
\centering
{\small
% ---- info block ----
\begin{tabular}{ll}
    \textbf{Operator:} &
        $\boxed{O_{ee} = (\overline e_p \gamma^\mu e_r)(\overline e_s \gamma_\mu e_t)}$ \\[0.2cm]
    \textbf{MFV decomposition:} &
        $[C_{ee}]_{prst} = C_{ee}^{(1)}\delta_{pr}\delta_{st}$ \\[0.15cm]
    \textbf{$\mathrm{U}(2)^5$ decomposition:} &
        $[C_{ee}]_{prst} =
            C_{ee}^{(1)}\Pij{p}{r}{a}\Pij{s}{t}{b}
            + C_{ee}^{(2)}(\Pij{s}{t}{a}\delta_{p3}\delta_{3r} + \Pij{p}{r}{a}\delta_{s3}\delta_{3t})$ \\
        & $\phantom{[C_{ee}]_{prst} ={}}+ C_{ee}^{(3)}\delta_{p3}\delta_{3r}\delta_{s3}\delta_{3t}$ \\
\end{tabular}
\\[0.3cm]
% ---- bounds table ----
\renewcommand{\arraystretch}{1.3}
\begin{tabular}{ccccc}
    \toprule
    \rowcolor{hdrcolor}
    \textbf{Symmetry} & \textbf{Coefficient}
        & \textbf{Unitarity bound}
        & \textbf{Scalar dominance}
        & \textbf{Vector dominance} \\
    \midrule
    % --- MFV section ---
    \multirow{1}{*}{MFV}
        & $C_{ee}^{(1)}$ & $[-4.71,\, 4.71]$ & $[0,\, 4.71]$ & $[-4.71,\, 0]$ \\
    \midrule
    % --- \mathrm{U}(2)^5 section ---
    \multirow{3}{*}{$\mathrm{U}(2)^5$}
        & $C_{ee}^{(1)}$ & $[-6.28,\, 6.28]$ & $[0,\, 6.28]$ & $[-6.28,\, 0]$ \\
        & $C_{ee}^{(2)}$ & $[-6.28,\, 6.28]$ & $[0,\, 6.28]$ & $[-6.28,\, 0]$ \\
        & $C_{ee}^{(3)}$ & $[-6.28,\, 6.28]$ & $[0,\, 6.28]$ & $[-6.28,\, 0]$ \\
    \bottomrule
\end{tabular}}
\caption{Same as Table~\ref{tab:Oll_UB} but for $O_{ee}$.}
\label{tab:Oee_UB}
\end{table}

% ============================================================
%  TABLE 5 — Q_{uu}
% ============================================================
\begin{table}[p]
\centering
{\small
% ---- info block ----
\begin{tabular}{ll}
    \textbf{Operator:} &
        $\boxed{O_{uu} = (\overline u_p \gamma^\mu u_r)(\overline u_s \gamma_\mu u_t)}$ \\[0.2cm]
    \textbf{MFV decomposition:} &
        $[C_{uu}]_{prst} =
            C_{uu}^{(1)}\delta_{pr}\delta_{st}
            + C_{uu}^{(2)}\delta_{pt}\delta_{sr}
            $ \\
        & $\phantom{[C_{uu}]_{prst} ={}}+ C_{uu}^{(3)}((Y_u^\dagger Y_u)_{st}\delta_{pr} + (Y_u^\dagger Y_u)_{pr}\delta_{st})$ \\
        & $\phantom{[C_{uu}]_{prst} ={}}
            + C_{uu}^{(4)}((Y_u^\dagger Y_u)_{sr}\delta_{pt} + (Y_u^\dagger Y_u)_{pt}\delta_{sr})$ \\[0.15cm]
    \textbf{$\mathrm{U}(2)^5$ decomposition:} &
        $[C_{uu}]_{prst} =
            C_{uu}^{(1)}\Pij{p}{r}{a}\Pij{s}{t}{b}
            + C_{uu}^{(2)}(\Pij{s}{t}{a}\delta_{p3}\delta_{3r} + \Pij{p}{r}{a}\delta_{s3}\delta_{3t})
            $ \\
        & $\phantom{[C_{uu}]_{prst} ={}}+ C_{uu}^{(3)}\delta_{p3}\delta_{3r}\delta_{s3}\delta_{3t}
            + C_{uu}^{(4)}\Pij{p}{t}{a}\Pij{s}{r}{b}
            $ \\
        & $\phantom{[C_{uu}]_{prst} ={}}+ C_{uu}^{(5)}(\Pij{s}{r}{a}\delta_{p3}\delta_{3t} + \Pij{p}{t}{a}\delta_{s3}\delta_{3r})$ \\
\end{tabular}
\\[0.3cm]
% ---- bounds table ----
\renewcommand{\arraystretch}{1.3}
\begin{tabular}{ccccc}
    \toprule
    \rowcolor{hdrcolor}
    \textbf{Symmetry} & \textbf{Coefficient}
        & \textbf{Unitarity bound}
        & \textbf{Scalar dominance}
        & \textbf{Vector dominance} \\
    \midrule
    % --- MFV section ---
    \multirow{4}{*}{MFV}
        & $C_{uu}^{(1)}$ & $[-6.28,\, 6.28]$  & $[0,\, 6.28]$      & $[-6.28,\, 0]$    \\
        & $C_{uu}^{(2)}$ & $[-8.38,\, 8.38]$  & $[-6.28,\, 3.59]$  & $[-3.59,\, 6.28]$ \\
        & $C_{uu}^{(3)}$ & $[-12.3,\, 12.3]$  & $[-6.28,\, 6.28]$  & $[-6.28,\, 6.28]$ \\
        & $C_{uu}^{(4)}$ & $[-13.2,\, 13.2]$  & $[-6.45,\, 7.53]$  & $[-7.53,\, 6.45]$ \\
    \midrule
    % --- \mathrm{U}(2)^5 section ---
    \multirow{5}{*}{$\mathrm{U}(2)^5$}
        & $C_{uu}^{(1)}$ & $[-6.28,\, 6.28]$ & $[0,\, 6.28]$      & $[-6.28,\, 0]$    \\
        & $C_{uu}^{(2)}$ & $[-6.28,\, 6.28]$ & $[0,\, 6.28]$      & $[-6.28,\, 0]$    \\
        & $C_{uu}^{(3)}$ & $[-4.71,\, 4.71]$ & $[0,\, 4.71]$      & $[-4.71,\, 0]$    \\
        & $C_{uu}^{(4)}$ & $[-8.38,\, 8.38]$ & $[-6.28,\, 3.77]$  & $[-3.77,\, 6.28]$ \\
        & $C_{uu}^{(5)}$ & $[-8.38,\, 8.38]$ & $[-6.28,\, 3.32]$  & $[-3.32,\, 6.28]$ \\
    \bottomrule
\end{tabular}}
\caption{Same as Table~\ref{tab:Oll_UB} but for $O_{uu}$.}
\label{tab:Ouu_UB}
\end{table}

% ============================================================
%  TABLE 6 — Q_{dd}
% ============================================================
\begin{table}[p]
\centering
{\small
% ---- info block ----
\begin{tabular}{ll}
    \textbf{Operator:} &
        $\boxed{O_{dd} = (\overline d_p \gamma^\mu d_r)(\overline d_s \gamma_\mu d_t)}$ \\[0.2cm]
    \textbf{MFV decomposition:} &
        $[C_{dd}]_{prst} =
            C_{dd}^{(1)}\delta_{pr}\delta_{st}
            + C_{dd}^{(2)}\delta_{pt}\delta_{sr}
            $ \\[0.15cm]
    \textbf{$\mathrm{U}(2)^5$ decomposition:} &
        $[C_{dd}]_{prst} =
            C_{dd}^{(1)}\Pij{p}{r}{a}\Pij{s}{t}{b}
            + C_{dd}^{(2)}(\Pij{s}{t}{a}\delta_{p3}\delta_{3r} + \Pij{p}{r}{a}\delta_{s3}\delta_{3t})
            $ \\
        & $\phantom{[C_{dd}]_{prst} ={}}+ C_{dd}^{(3)}\delta_{p3}\delta_{3r}\delta_{s3}\delta_{3t}
            + C_{dd}^{(4)}\Pij{p}{t}{a}\Pij{s}{r}{b}
            $ \\
        & $\phantom{[C_{dd}]_{prst} ={}}+ C_{dd}^{(5)}(\Pij{s}{r}{a}\delta_{p3}\delta_{3t} + \Pij{p}{t}{a}\delta_{s3}\delta_{3r})$ \\
\end{tabular}
\\[0.3cm]
% ---- bounds table ----
\renewcommand{\arraystretch}{1.3}
\begin{tabular}{ccccc}
    \toprule
    \rowcolor{hdrcolor}
    \textbf{Symmetry} & \textbf{Coefficient}
        & \textbf{Unitarity bound}
        & \textbf{Scalar dominance}
        & \textbf{Vector dominance} \\
    \midrule
    % --- MFV section ---
    \multirow{2}{*}{MFV}
        & $C_{dd}^{(1)}$ & $[-6.28,\, 6.28]$  & $[0,\, 4.71]$      & $[-4.71,\, 0]$    \\
        & $C_{dd}^{(2)}$ & $[-8.38,\, 8.38]$  & $[-4.71,\, 3.14]$  & $[-3.14,\, 4.71]$ \\
    \midrule
    % --- \mathrm{U}(2)^5 section ---
    \multirow{5}{*}{$\mathrm{U}(2)^5$}
        & $C_{dd}^{(1)}$ & $[-6.28,\, 6.28]$ & $[0,\, 6.28]$      & $[-6.28,\, 0]$    \\
        & $C_{dd}^{(2)}$ & $[-6.28,\, 6.28]$ & $[0,\, 6.28]$      & $[-6.28,\, 0]$    \\
        & $C_{dd}^{(3)}$ & $[-4.71,\, 4.71]$ & $[0,\, 4.71]$      & $[-4.71,\, 0]$    \\
        & $C_{dd}^{(4)}$ & $[-8.38,\, 8.38]$ & $[-6.28,\, 3.77]$  & $[-3.77,\, 6.28]$ \\
        & $C_{dd}^{(5)}$ & $[-8.38,\, 8.38]$ & $[-6.28,\, 3.32]$  & $[-3.32,\, 6.28]$ \\
    \bottomrule
\end{tabular}}
\caption{Same as Table~\ref{tab:Oll_UB} but for $O_{dd}$.}
\label{tab:Odd_UB}
\end{table}

% ============================================================
%  TABLE 7 — Q_{eu}
% ============================================================
\begin{table}[p]
\centering
{\small
% ---- info block ----
\begin{tabular}{ll}
    \textbf{Operator:} &
        $\boxed{O_{eu} = (\overline e_p \gamma^\mu e_r)(\overline u_s \gamma_\mu u_t)}$ \\[0.2cm]
    \textbf{MFV decomposition:} &
        $[C_{eu}]_{prst} =
            C_{eu}^{(1)}\delta_{pr}\delta_{st}
            + C_{eu}^{(2)}\delta_{pr}(Y_u^\dagger Y_u)_{st}
            $ \\[0.15cm]
    \textbf{$\mathrm{U}(2)^5$ decomposition:} &
        $[C_{eu}]_{prst} =
            C_{eu}^{(1)}\Pij{p}{r}{a}\Pij{s}{t}{b}
            + C_{eu}^{(2)}\delta_{p3}\delta_{3r}\Pij{s}{t}{a} + C_{eu}^{(3)}\delta_{s3}\delta_{3t}\Pij{p}{r}{a}
            $ \\
        & $\phantom{[C_{eu}]_{prst} ={}}+ C_{eu}^{(4)}\delta_{p3}\delta_{3r}\delta_{s3}\delta_{3t}$ \\
\end{tabular}
\\[0.3cm]
% ---- bounds table ----
\renewcommand{\arraystretch}{1.3}
\begin{tabular}{ccccc}
    \toprule
    \rowcolor{hdrcolor}
    \textbf{Symmetry} & \textbf{Coefficient}
        & \textbf{Unitarity bound}
        & \textbf{Scalar dominance}
        & \textbf{Vector dominance} \\
    \midrule
    % --- MFV section ---
    \multirow{2}{*}{MFV}
        & $C_{eu}^{(1)}$ & $[-8.89,\, 8.89]$  & $[0,\, 8.89]$      & $[-8.89,\, 0]$    \\
        & $C_{eu}^{(2)}$ & $[-15.4,\, 15.4]$  & $[-8.89,\, 12.6]$  & $[-12.6,\, 8.89]$ \\
    \midrule
    % --- \mathrm{U}(2)^5 section ---
    \multirow{4}{*}{$\mathrm{U}(2)^5$}
        & $C_{eu}^{(1)}$ & $[-10.9,\, 10.9]$  & $[0,\, 10.9]$      & $[-10.9,\, 0]$    \\
        & $C_{eu}^{(2)}$ & $[-12.6,\, 12.6]$  & $[0,\, 12.6]$  & $[-12.6,\, 0]$ \\
        & $C_{eu}^{(3)}$ & $[-12.6,\, 12.6]$  & $[0,\, 12.6]$  & $[-12.6,\, 0]$ \\
        & $C_{eu}^{(4)}$ & $[-12.6,\, 12.6]$  & $[0,\, 12.6]$  & $[-12.6,\, 0]$ \\
    \bottomrule
\end{tabular}}
\caption{Same as Table~\ref{tab:Oll_UB} but for $O_{eu}$.}
\label{tab:Oeu_UB}
\end{table}

% ============================================================
%  TABLE 8 — Q_{ed}
% ============================================================
\begin{table}[p]
\centering
{\small
% ---- info block ----
\begin{tabular}{ll}
    \textbf{Operator:} &
        $\boxed{O_{ed} = (\overline e_p \gamma^\mu e_r)(\overline d_s \gamma_\mu d_t)}$ \\[0.2cm]
    \textbf{MFV decomposition:} &
        $[C_{ed}]_{prst} =
            C_{ed}^{(1)}\delta_{pr}\delta_{st}$ \\[0.15cm]
    \textbf{$\mathrm{U}(2)^5$ decomposition:} &
        $[C_{ed}]_{prst} =
            C_{ed}^{(1)}\Pij{p}{r}{a}\Pij{s}{t}{b}
            + C_{ed}^{(2)}\delta_{p3}\delta_{3r}\Pij{s}{t}{a} + C_{ed}^{(3)}\delta_{s3}\delta_{3t}\Pij{p}{r}{a}
            $ \\
        & $\phantom{[C_{ed}]_{prst} ={}}+ C_{ed}^{(4)}\delta_{p3}\delta_{3r}\delta_{s3}\delta_{3t}$ \\
\end{tabular}
\\[0.3cm]
% ---- bounds table ----
\renewcommand{\arraystretch}{1.3}
\begin{tabular}{ccccc}
    \toprule
    \rowcolor{hdrcolor}
    \textbf{Symmetry} & \textbf{Coefficient}
        & \textbf{Unitarity bound}
        & \textbf{Scalar dominance}
        & \textbf{Vector dominance} \\
    \midrule
    % --- MFV section ---
    \multirow{1}{*}{MFV}
        & $C_{ed}^{(1)}$ & $[-7.26,\, 7.26]$  & $[0,\, 7.26]$      & $[-7.26,\, 0]$    \\
    \midrule
    % --- \mathrm{U}(2)^5 section ---
    \multirow{4}{*}{$\mathrm{U}(2)^5$}
        & $C_{ed}^{(1)}$ & $[-10.9,\, 10.9]$  & $[0,\, 10.9]$      & $[-10.9,\, 0]$    \\
        & $C_{ed}^{(2)}$ & $[-12.6,\, 12.6]$  & $[0,\, 12.6]$  & $[-12.6,\, 0]$ \\
        & $C_{ed}^{(3)}$ & $[-12.6,\, 12.6]$  & $[0,\, 12.6]$  & $[-12.6,\, 0]$ \\
        & $C_{ed}^{(4)}$ & $[-12.6,\, 12.6]$  & $[0,\, 12.6]$  & $[-12.6,\, 0]$ \\
    \bottomrule
\end{tabular}}
\caption{Same as Table~\ref{tab:Oll_UB} but for $O_{ed}$.}
\label{tab:Oed_UB}
\end{table}

% ============================================================
%  TABLE 9 — Q_{ud}
% ============================================================
\begin{table}[p]
\centering
{\small
% ---- info block ----
\begin{tabular}{ll}
    \textbf{Operators:} &
        $\boxed{O_{ud}^{(1)} = (\overline u_p \gamma^\mu u_r)(\overline d_s \gamma_\mu d_t),\quad O_{ud}^{(8)} = (\overline u_p \gamma^\mu T^A u_r)(\overline d_s \gamma_\mu T^A d_t)}$ \\[0.2cm]
    \textbf{MFV decomposition:} &
        $[C_{ud}^{(i)}]_{prst} =
            C_{ud}^{(i,1)}\delta_{pr}\delta_{st}
            + C_{ud}^{(i,2)}(Y_u^\dagger Y_u)_{pr}\delta_{st}$,\quad $i=1,8$ \\[0.15cm]
    \textbf{$\mathrm{U}(2)^5$ decomposition:} &
        $[C_{ud}^{(i)}]_{prst} =
            C_{ud}^{(i,1)}\Pij{p}{r}{a}\Pij{s}{t}{b}
            + C_{ud}^{(i,2)}\Pij{s}{t}{a}\delta_{p3}\delta_{3r}$ \\
        & $\phantom{[C_{ud}^{(i)}]_{prst} ={}}+ C_{ud}^{(i,3)}\Pij{p}{r}{a}\delta_{s3}\delta_{3t}
            + C_{ud}^{(i,4)}\delta_{p3}\delta_{3r}\delta_{s3}\delta_{3t}$,\quad $i=1,8$ \\
\end{tabular}
\\[0.3cm]
% ---- bounds table ----
\renewcommand{\arraystretch}{1.3}
\begin{tabular}{ccccc}
    \toprule
    \rowcolor{hdrcolor}
    \textbf{Symmetry} & \textbf{Coefficient}
        & \textbf{Unitarity bound}
        & \textbf{Scalar dominance}
        & \textbf{Vector dominance} \\
    \midrule
    % --- MFV section ---
    \multirow{4}{*}{MFV}
        & $C_{ud}^{(1,1)}$ & $[-5.13,\, 5.13]$  & $[0,\, 5.13]$      & $[-5.13,\, 0]$    \\
        & $C_{ud}^{(1,2)}$ & $[-8.89,\, 8.89]$  & $[-5.13,\, 7.26]$  & $[-7.26,\, 5.13]$ \\
        & $C_{ud}^{(8,1)}$ & $[-25.1,\, 25.1]$  & $[-12.6,\, 25.1]$  & $[-25.1,\, 12.6]$ \\
        & $C_{ud}^{(8,2)}$ & $[-50.3,\, 50.3]$  & $[-37.7,\, 37.7]$  & $[-37.7,\, 37.7]$ \\
    \midrule
    % --- \mathrm{U}(2)^5 section ---
    \multirow{8}{*}{$\mathrm{U}(2)^5$}
        & $C_{ud}^{(1,1)}$ & $[-6.28,\, 6.28]$ & $[0,\, 6.28]$     & $[-6.28,\, 0]$    \\
        & $C_{ud}^{(1,2)}$ & $[-8.89,\, 8.89]$ & $[0,\, 8.89]$     & $[-8.89,\, 0]$    \\
        & $C_{ud}^{(1,3)}$ & $[-8.89,\, 8.89]$ & $[0,\, 8.89]$     & $[-8.89,\, 0]$    \\
        & $C_{ud}^{(1,4)}$ & $[-12.6,\, 12.6]$ & $[0,\, 12.6]$     & $[-12.6,\, 0]$    \\
        & $C_{ud}^{(8,1)}$ & $[-25.1,\, 25.1]$ & $[-12.6,\, 25.1]$ & $[-25.1,\, 12.6]$ \\
        & $C_{ud}^{(8,2)}$ & $[-25.1,\, 25.1]$ & $[-12.6,\, 25.1]$ & $[-25.1,\, 12.6]$ \\
        & $C_{ud}^{(8,3)}$ & $[-25.1,\, 25.1]$ & $[-12.6,\, 25.1]$ & $[-25.1,\, 12.6]$ \\
        & $C_{ud}^{(8,4)}$ & $[-25.1,\, 25.1]$ & $[-12.6,\, 25.1]$ & $[-25.1,\, 12.6]$ \\
    \bottomrule
\end{tabular}}
\caption{Same as Table~\ref{tab:Oll_UB} but for $O_{ud}^{(1,8)}$.}
\label{tab:Oud_UB}
\end{table}

% ============================================================
%  TABLE 10 — Q_{le}
% ============================================================
\begin{table}[p]
\centering
{\small
% ---- info block ----
\begin{tabular}{ll}
    \textbf{Operator:} &
        $\boxed{O_{\ell e} = (\overline\ell_p \gamma^\mu \ell_r)(\overline e_s \gamma_\mu e_t)}$ \\[0.2cm]
    \textbf{MFV decomposition:} &
        $[C_{\ell e}]_{prst} = C_{\ell e}^{(1)}\delta_{pr}\delta_{st}$ \\[0.15cm]
    \textbf{$\mathrm{U}(2)^5$ decomposition:} &
        $[C_{\ell e}]_{prst} =
            C_{\ell e}^{(1)}\Pij{p}{r}{a}\Pij{s}{t}{b}
            + C_{\ell e}^{(2)}\Pij{s}{t}{a}\delta_{p3}\delta_{3r}$ \\
        & $\phantom{[C_{\ell e}]_{prst} ={}}+ C_{\ell e}^{(3)}\Pij{p}{r}{a}\delta_{s3}\delta_{3t}
            + C_{\ell e}^{(4)}\delta_{p3}\delta_{3r}\delta_{s3}\delta_{3t}$ \\
\end{tabular}
\\[0.3cm]
% ---- bounds table ----
\renewcommand{\arraystretch}{1.3}
\begin{tabular}{ccccc}
    \toprule
    \rowcolor{hdrcolor}
    \textbf{Symmetry} & \textbf{Coefficient}
        & \textbf{Unitarity bound}
        & \textbf{Scalar dominance}
        & \textbf{Vector dominance} \\
    \midrule
    % --- MFV section ---
    \multirow{1}{*}{MFV}
        & $C_{\ell e}^{(1)}$ & $[-8.89,\, 8.89]$ & $[-8.89,\, 0]$ & $[0,\, 8.89]$ \\
    \midrule
    % --- \mathrm{U}(2)^5 section ---
    \multirow{4}{*}{$\mathrm{U}(2)^5$}
        & $C_{\ell e}^{(1)}$ & $[-12.6,\, 12.6]$ & $[-12.6,\, 0]$ & $[0,\, 12.6]$ \\
        & $C_{\ell e}^{(2)}$ & $[-12.6,\, 12.6]$ & $[-12.6,\, 0]$ & $[0,\, 12.6]$ \\
        & $C_{\ell e}^{(3)}$ & $[-12.6,\, 12.6]$ & $[-12.6,\, 0]$ & $[0,\, 12.6]$ \\
        & $C_{\ell e}^{(4)}$ & $[-12.6,\, 12.6]$ & $[-12.6,\, 0]$ & $[0,\, 12.6]$ \\
    \bottomrule
\end{tabular}}
\caption{Same as Table~\ref{tab:Oll_UB} but for $O_{\ell e}$.}
\label{tab:Ole_UB}
\end{table}

% ============================================================
%  TABLE 11 — Q_{lu}
% ============================================================
\begin{table}[p]
\centering
{\small
% ---- info block ----
\begin{tabular}{ll}
    \textbf{Operator:} &
        $\boxed{O_{\ell u} = (\overline\ell_p \gamma^\mu \ell_r)(\overline u_s \gamma_\mu u_t)}$ \\[0.2cm]
    \textbf{MFV decomposition:} &
        $[C_{\ell u}]_{prst} = C_{\ell u}^{(1)}\delta_{pr}\delta_{st} + C_{\ell u}^{(2)}\delta_{pr}(Y_u^\dagger Y_u)_{st}$ \\[0.15cm]
    \textbf{$\mathrm{U}(2)^5$ decomposition:} &
        $[C_{\ell u}]_{prst} =
            C_{\ell u}^{(1)}\Pij{p}{r}{a}\Pij{s}{t}{b}
            + C_{\ell u}^{(2)}\Pij{s}{t}{a}\delta_{p3}\delta_{3r}$ \\
        & $\phantom{[C_{\ell u}]_{prst} ={}}+ C_{\ell u}^{(3)}\Pij{p}{r}{a}\delta_{s3}\delta_{3t}
            + C_{\ell u}^{(4)}\delta_{p3}\delta_{3r}\delta_{s3}\delta_{3t}$ \\
\end{tabular}
\\[0.3cm]
% ---- bounds table ----
\renewcommand{\arraystretch}{1.3}
\begin{tabular}{ccccc}
    \toprule
    \rowcolor{hdrcolor}
    \textbf{Symmetry} & \textbf{Coefficient}
        & \textbf{Unitarity bound}
        & \textbf{Scalar dominance}
        & \textbf{Vector dominance} \\
    \midrule
    % --- MFV section ---
    \multirow{2}{*}{MFV}
        & $C_{\ell u}^{(1)}$ & $[-6.28,\, 6.28]$  & $[-6.28,\, 0]$    & $[0,\, 6.28]$    \\
        & $C_{\ell u}^{(2)}$ & $[-10.9,\, 10.9]$  & $[-8.89,\, 6.28]$ & $[-6.28,\, 8.89]$ \\
    \midrule
    % --- \mathrm{U}(2)^5 section ---
    \multirow{4}{*}{$\mathrm{U}(2)^5$}
        & $C_{\ell u}^{(1)}$ & $[-7.70,\, 7.70]$ & $[-7.70,\, 0]$ & $[0,\, 7.70]$ \\
        & $C_{\ell u}^{(2)}$ & $[-10.9,\, 10.9]$ & $[-10.9,\, 0]$ & $[0,\, 10.9]$ \\
        & $C_{\ell u}^{(3)}$ & $[-10.9,\, 10.9]$ & $[-10.9,\, 0]$ & $[0,\, 10.9]$ \\
        & $C_{\ell u}^{(4)}$ & $[-12.6,\, 12.6]$ & $[-12.6,\, 0]$ & $[0,\, 12.6]$ \\
    \bottomrule
\end{tabular}}
\caption{Same as Table~\ref{tab:Oll_UB} but for $O_{\ell u}$.}
\label{tab:Olu_UB}
\end{table}

% ============================================================
%  TABLE 12 — Q_{ld}
% ============================================================
\begin{table}[p]
\centering
{\small
% ---- info block ----
\begin{tabular}{ll}
    \textbf{Operator:} &
        $\boxed{O_{\ell d} = (\overline\ell_p \gamma^\mu \ell_r)(\overline d_s \gamma_\mu d_t)}$ \\[0.2cm]
    \textbf{MFV decomposition:} &
        $[C_{\ell d}]_{prst} = C_{\ell d}^{(1)}\delta_{pr}\delta_{st}$ \\[0.15cm]
    \textbf{$\mathrm{U}(2)^5$ decomposition:} &
        $[C_{\ell d}]_{prst} =
            C_{\ell d}^{(1)}\Pij{p}{r}{a}\Pij{s}{t}{b}
            + C_{\ell d}^{(2)}\Pij{s}{t}{a}\delta_{p3}\delta_{3r}$ \\
        & $\phantom{[C_{\ell d}]_{prst} ={}}+ C_{\ell d}^{(3)}\Pij{p}{r}{a}\delta_{s3}\delta_{3t}
            + C_{\ell d}^{(4)}\delta_{p3}\delta_{3r}\delta_{s3}\delta_{3t}$ \\
\end{tabular}
\\[0.3cm]
% ---- bounds table ----
\renewcommand{\arraystretch}{1.3}
\begin{tabular}{ccccc}
    \toprule
    \rowcolor{hdrcolor}
    \textbf{Symmetry} & \textbf{Coefficient}
        & \textbf{Unitarity bound}
        & \textbf{Scalar dominance}
        & \textbf{Vector dominance} \\
    \midrule
    % --- MFV section ---
    \multirow{1}{*}{MFV}
        & $C_{\ell d}^{(1)}$ & $[-5.13,\, 5.13]$ & $[-5.13,\, 0]$ & $[0,\, 5.13]$ \\
    \midrule
    % --- \mathrm{U}(2)^5 section ---
    \multirow{4}{*}{$\mathrm{U}(2)^5$}
        & $C_{\ell d}^{(1)}$ & $[-7.70,\, 7.70]$ & $[-7.70,\, 0]$ & $[0,\, 7.70]$ \\
        & $C_{\ell d}^{(2)}$ & $[-10.9,\, 10.9]$ & $[-10.9,\, 0]$ & $[0,\, 10.9]$ \\
        & $C_{\ell d}^{(3)}$ & $[-10.9,\, 10.9]$ & $[-10.9,\, 0]$ & $[0,\, 10.9]$ \\
        & $C_{\ell d}^{(4)}$ & $[-12.6,\, 12.6]$ & $[-12.6,\, 0]$ & $[0,\, 12.6]$ \\
    \bottomrule
\end{tabular}}
\caption{Same as Table~\ref{tab:Oll_UB} but for $O_{\ell d}$.}
\label{tab:Old_UB}
\end{table}

% ============================================================
%  TABLE 13 — Q_{qe}
% ============================================================
\begin{table}[p]
\centering
{\small
% ---- info block ----
\begin{tabular}{ll}
    \textbf{Operator:} &
        $\boxed{O_{qe} = (\overline q_p \gamma^\mu q_r)(\overline e_s \gamma_\mu e_t)}$ \\[0.2cm]
    \textbf{MFV decomposition:} &
        $[C_{qe}]_{prst} = C_{qe}^{(1)}\delta_{pr}\delta_{st} + C_{qe}^{(2)}(Y_u Y_u^\dagger)_{pr}\delta_{st}$ \\[0.15cm]
    \textbf{$\mathrm{U}(2)^5$ decomposition:} &
        $[C_{qe}]_{prst} =
            C_{qe}^{(1)}\Pij{p}{r}{a}\Pij{s}{t}{b}
            + C_{qe}^{(2)}\Pij{s}{t}{a}\delta_{p3}\delta_{3r}$ \\
        & $\phantom{[C_{qe}]_{prst} ={}}+ C_{qe}^{(3)}\Pij{p}{r}{a}\delta_{s3}\delta_{3t}
            + C_{qe}^{(4)}\delta_{p3}\delta_{3r}\delta_{s3}\delta_{3t}$ \\
\end{tabular}
\\[0.3cm]
% ---- bounds table ----
\renewcommand{\arraystretch}{1.3}
\begin{tabular}{ccccc}
    \toprule
    \rowcolor{hdrcolor}
    \textbf{Symmetry} & \textbf{Coefficient}
        & \textbf{Unitarity bound}
        & \textbf{Scalar dominance}
        & \textbf{Vector dominance} \\
    \midrule
    % --- MFV section ---
    \multirow{2}{*}{MFV}
        & $C_{qe}^{(1)}$ & $[-6.28,\, 6.28]$  & $[-6.28,\, 0]$    & $[0,\, 6.28]$    \\
        & $C_{qe}^{(2)}$ & $[-10.9,\, 10.9]$  & $[-8.89,\, 6.28]$ & $[-6.28,\, 8.89]$ \\
    \midrule
    % --- \mathrm{U}(2)^5 section ---
    \multirow{4}{*}{$\mathrm{U}(2)^5$}
        & $C_{qe}^{(1)}$ & $[-7.70,\, 7.70]$ & $[-7.70,\, 0]$ & $[0,\, 7.70]$ \\
        & $C_{qe}^{(2)}$ & $[-10.9,\, 10.9]$ & $[-10.9,\, 0]$ & $[0,\, 10.9]$ \\
        & $C_{qe}^{(3)}$ & $[-10.9,\, 10.9]$ & $[-10.9,\, 0]$ & $[0,\, 10.9]$ \\
        & $C_{qe}^{(4)}$ & $[-12.6,\, 12.6]$ & $[-12.6,\, 0]$ & $[0,\, 12.6]$ \\
    \bottomrule
\end{tabular}}
\caption{Same as Table~\ref{tab:Oll_UB} but for $O_{qe}$.}
\label{tab:Oqe_UB}
\end{table}

% ============================================================
%  TABLE 14 — Q_{qu}
% ============================================================
\begin{table}[p]
\centering
{\small
% ---- info block ----
\begin{tabular}{ll}
    \textbf{Operators:} &
        $\boxed{O_{qu}^{(1)} = (\overline q_p \gamma^\mu q_r)(\overline u_s \gamma_\mu u_t),\quad O_{qu}^{(8)} = (\overline q_p \gamma^\mu T^A q_r)(\overline u_s \gamma_\mu T^A u_t)}$ \\[0.2cm]
    \textbf{MFV decomposition:} &
        $[C_{qu}^{(i)}]_{prst} =
            C_{qu}^{(i,1)}\delta_{pr}\delta_{st}
            + C_{qu}^{(i,2)}(Y_u Y_u^\dagger)_{pr}\delta_{st}$ \\
        & $\phantom{[C_{qu}^{(i)}]_{prst} ={}}+ C_{qu}^{(i,3)}\delta_{pr}(Y_u^\dagger Y_u)_{st}
            + C_{qu}^{(i,4)}(Y_u)_{pt}(Y_u^\dagger)_{sr}$,\quad $i=1,8$ \\[0.15cm]
    \textbf{$\mathrm{U}(2)^5$ decomposition:} &
        $[C_{qu}^{(i)}]_{prst} =
            C_{qu}^{(i,1)}\Pij{p}{r}{a}\Pij{s}{t}{b}
            + C_{qu}^{(i,2)}\Pij{s}{t}{a}\delta_{p3}\delta_{3r}$ \\
        & $\phantom{[C_{qu}^{(i)}]_{prst} ={}}+ C_{qu}^{(i,3)}\Pij{p}{r}{a}\delta_{s3}\delta_{3t}
            + C_{qu}^{(i,4)}\delta_{p3}\delta_{3r}\delta_{s3}\delta_{3t}$,\quad $i=1,8$ \\
\end{tabular}
\\[0.3cm]
% ---- bounds table ----
\renewcommand{\arraystretch}{1.3}
\begin{tabular}{ccccc}
    \toprule
    \rowcolor{hdrcolor}
    \textbf{Symmetry} & \textbf{Coefficient}
        & \textbf{Unitarity bound}
        & \textbf{Scalar dominance}
        & \textbf{Vector dominance} \\
    \midrule
    % --- MFV section ---
    \multirow{8}{*}{MFV}
        & $C_{qu}^{(1,1)}$ & $[-4.44,\, 4.44]$ & $[-4.44,\, 0]$    & $[0,\, 4.44]$    \\
        & $C_{qu}^{(1,2)}$ & $[-7.70,\, 7.70]$ & $[-6.28,\, 4.44]$ & $[-4.44,\, 6.28]$ \\
        & $C_{qu}^{(1,3)}$ & $[-7.70,\, 7.70]$ & $[-6.28,\, 4.44]$ & $[-4.44,\, 6.28]$ \\
        & $C_{qu}^{(1,4)}$ & $[-13.3,\, 13.3]$ & $[-13.3,\, 12.6]$ & $[-12.6,\, 13.3]$ \\
        & $C_{qu}^{(8,1)}$ & $[-12.8,\, 12.8]$ & $[-8.38,\, 12.8]$ & $[-12.8,\, 8.38]$ \\
        & $C_{qu}^{(8,2)}$ & $[-24.6,\, 24.6]$ & $[-21.1,\, 22.3]$ & $[-22.3,\, 21.1]$ \\
        & $C_{qu}^{(8,3)}$ & $[-24.6,\, 24.6]$ & $[-21.1,\, 22.3]$ & $[-22.3,\, 21.1]$ \\
        & $C_{qu}^{(8,4)}$ & $[-47.7,\, 47.7]$ & $[-42.8,\, 42.0]$ & $[-42.0,\, 42.8]$ \\
    \midrule
    % --- \mathrm{U}(2)^5 section ---
    \multirow{8}{*}{$\mathrm{U}(2)^5$}
        & $C_{qu}^{(1,1)}$ & $[-4.44,\, 4.44]$ & $[-4.44,\, 0]$    & $[0,\, 4.44]$    \\
        & $C_{qu}^{(1,2)}$ & $[-6.28,\, 6.28]$ & $[-6.28,\, 0]$    & $[0,\, 6.28]$    \\
        & $C_{qu}^{(1,3)}$ & $[-6.28,\, 6.28]$ & $[-6.28,\, 0]$    & $[0,\, 6.28]$    \\
        & $C_{qu}^{(1,4)}$ & $[-8.89,\, 8.89]$ & $[-8.89,\, 0]$    & $[0,\, 8.89]$    \\
        & $C_{qu}^{(8,1)}$ & $[-12.8,\, 12.8]$ & $[-8.38,\, 12.8]$ & $[-12.8,\, 8.38]$ \\
        & $C_{qu}^{(8,2)}$ & $[-14.1,\, 14.1]$ & $[-8.38,\, 14.1]$ & $[-14.1,\, 8.38]$ \\
        & $C_{qu}^{(8,3)}$ & $[-14.1,\, 14.1]$ & $[-8.38,\, 14.1]$ & $[-14.1,\, 8.38]$ \\
        & $C_{qu}^{(8,4)}$ & $[-16.1,\, 16.1]$ & $[-8.38,\, 16.1]$ & $[-16.1,\, 8.38]$ \\
    \bottomrule
\end{tabular}}
\caption{Same as Table~\ref{tab:Oll_UB} but for $O_{qu}^{(1,8)}$.}
\label{tab:Oqu_UB}
\end{table}

% ============================================================
%  TABLE 15 — Q_{qd}
% ============================================================
\begin{table}[p]
\centering
{\small
% ---- info block ----
\begin{tabular}{ll}
    \textbf{Operators:} &
        $\boxed{O_{qd}^{(1)} = (\overline q_p \gamma^\mu q_r)(\overline d_s \gamma_\mu d_t),\quad O_{qd}^{(8)} = (\overline q_p \gamma^\mu T^A q_r)(\overline d_s \gamma_\mu T^A d_t)}$ \\[0.2cm]
    \textbf{MFV decomposition:} &
        $[C_{qd}^{(i)}]_{prst} =
            C_{qd}^{(i,1)}\delta_{pr}\delta_{st}
            + C_{qd}^{(i,2)}(Y_u Y_u^\dagger)_{pr}\delta_{st}$,\quad $i=1,8$ \\[0.15cm]
    \textbf{$\mathrm{U}(2)^5$ decomposition:} &
        $[C_{qd}^{(i)}]_{prst} =
            C_{qd}^{(i,1)}\Pij{p}{r}{a}\Pij{s}{t}{b}
            + C_{qd}^{(i,2)}\Pij{s}{t}{a}\delta_{p3}\delta_{3r}$ \\
        & $\phantom{[C_{qd}^{(i)}]_{prst} ={}}+ C_{qd}^{(i,3)}\Pij{p}{r}{a}\delta_{s3}\delta_{3t}
            + C_{qd}^{(i,4)}\delta_{p3}\delta_{3r}\delta_{s3}\delta_{3t}$,\quad $i=1,8$ \\
\end{tabular}
\\[0.3cm]
% ---- bounds table ----
\renewcommand{\arraystretch}{1.3}
\begin{tabular}{ccccc}
    \toprule
    \rowcolor{hdrcolor}
    \textbf{Symmetry} & \textbf{Coefficient}
        & \textbf{Unitarity bound}
        & \textbf{Scalar dominance}
        & \textbf{Vector dominance} \\
    \midrule
    % --- MFV section ---
    \multirow{4}{*}{MFV}
        & $C_{qd}^{(1,1)}$ & $[-3.63,\, 3.63]$ & $[-3.63,\, 0]$     & $[0,\, 3.63]$     \\
        & $C_{qd}^{(1,2)}$ & $[-6.28,\, 6.28]$ & $[-5.13,\, 3.63]$  & $[-3.63,\, 5.13]$ \\
        & $C_{qd}^{(8,1)}$ & $[-12.1,\, 12.1]$ & $[-8.38,\, 10.9]$  & $[-10.9,\, 8.38]$ \\
        & $C_{qd}^{(8,2)}$ & $[-23.6,\, 23.6]$ & $[-18.8,\, 21.4]$  & $[-21.4,\, 18.8]$ \\
    \midrule
    % --- \mathrm{U}(2)^5 section ---
    \multirow{8}{*}{$\mathrm{U}(2)^5$}
        & $C_{qd}^{(1,1)}$ & $[-4.44,\, 4.44]$ & $[-4.44,\, 0]$    & $[0,\, 4.44]$    \\
        & $C_{qd}^{(1,2)}$ & $[-6.28,\, 6.28]$ & $[-6.28,\, 0]$    & $[0,\, 6.28]$    \\
        & $C_{qd}^{(1,3)}$ & $[-6.28,\, 6.28]$ & $[-6.28,\, 0]$    & $[0,\, 6.28]$    \\
        & $C_{qd}^{(1,4)}$ & $[-8.89,\, 8.89]$ & $[-8.89,\, 0]$    & $[0,\, 8.89]$    \\
        & $C_{qd}^{(8,1)}$ & $[-12.8,\, 12.8]$ & $[-8.38,\, 12.8]$ & $[-12.8,\, 8.38]$ \\
        & $C_{qd}^{(8,2)}$ & $[-14.1,\, 14.1]$ & $[-8.38,\, 14.1]$ & $[-14.1,\, 8.38]$ \\
        & $C_{qd}^{(8,3)}$ & $[-14.1,\, 14.1]$ & $[-8.38,\, 14.1]$ & $[-14.1,\, 8.38]$ \\
        & $C_{qd}^{(8,4)}$ & $[-16.1,\, 16.1]$ & $[-8.38,\, 16.1]$ & $[-16.1,\, 8.38]$ \\
    \bottomrule
\end{tabular}}
\caption{Same as Table~\ref{tab:Oll_UB} but for $O_{qd}^{(1,8)}$.}
\label{tab:Oqd_UB}
\end{table}
\clearpage
\begin{figure}[p]
    \centering
    \includegraphics[scale = 0.75]{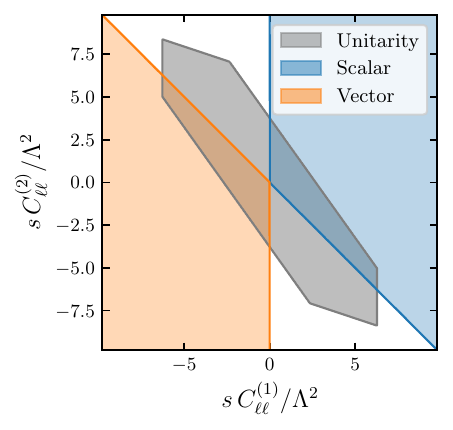}
    \caption{Partial-wave unitarity bounds and sum-rule constraints (under scalar and vector dominance) for the operator $O_{\ell\ell}$ under MFV, where $[C_{\ell\ell}]_{prst} = C_{\ell\ell}^{(1)}\delta_{pr}\delta_{st} + C_{\ell\ell}^{(2)}\delta_{pt}\delta_{sr}$.}
    \label{fig:Oll_mfv}
\end{figure}

\begin{figure}[p]
    \centering
    \includegraphics[scale = 0.75]{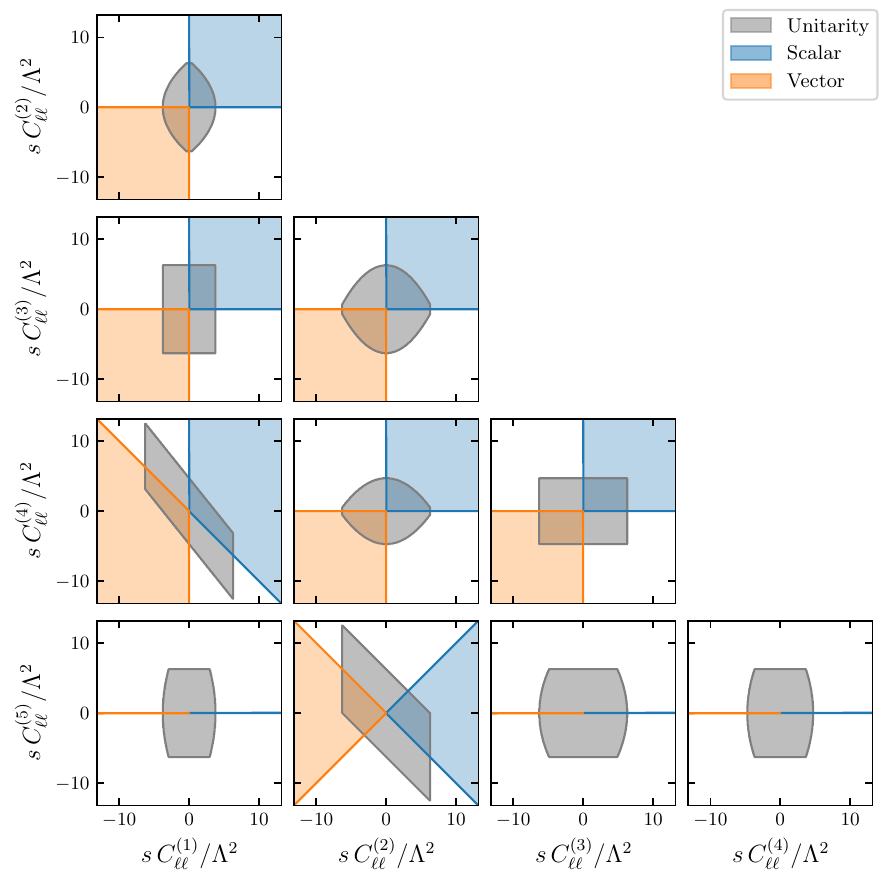}
    \caption{Partial-wave unitarity bounds and sum-rule constraints (under scalar and vector dominance) for the operator $O_{\ell\ell}$ under $\mathrm{U}(2)^5$ flavor symmetry, where $[C_{\ell\ell}]_{prst} =
            C_{\ell\ell}^{(1)}\Pij{p}{r}{a}\Pij{s}{t}{b}
            + C_{\ell\ell}^{(2)}(\Pij{s}{t}{a}\delta_{p3}\delta_{3r} + \Pij{p}{r}{a}\delta_{s3}\delta_{3t})
            + C_{\ell\ell}^{(3)}\delta_{p3}\delta_{3r}\delta_{s3}\delta_{3t}
            + C_{\ell\ell}^{(4)}\Pij{s}{r}{a}\Pij{p}{t}{b}
            + C_{\ell\ell}^{(5)}(\Pij{p}{t}{a}\delta_{s3}\delta_{3r} + \Pij{s}{r}{a}\delta_{p3}\delta_{3t})$. The regions shown are two-dimensional slices of the full allowed region, obtained by setting to zero all the independent coefficients other than the two displayed.}
    \label{fig:Oll_u2}
\end{figure}

\begin{figure}[p]
    \centering
    \includegraphics[width=1\linewidth]{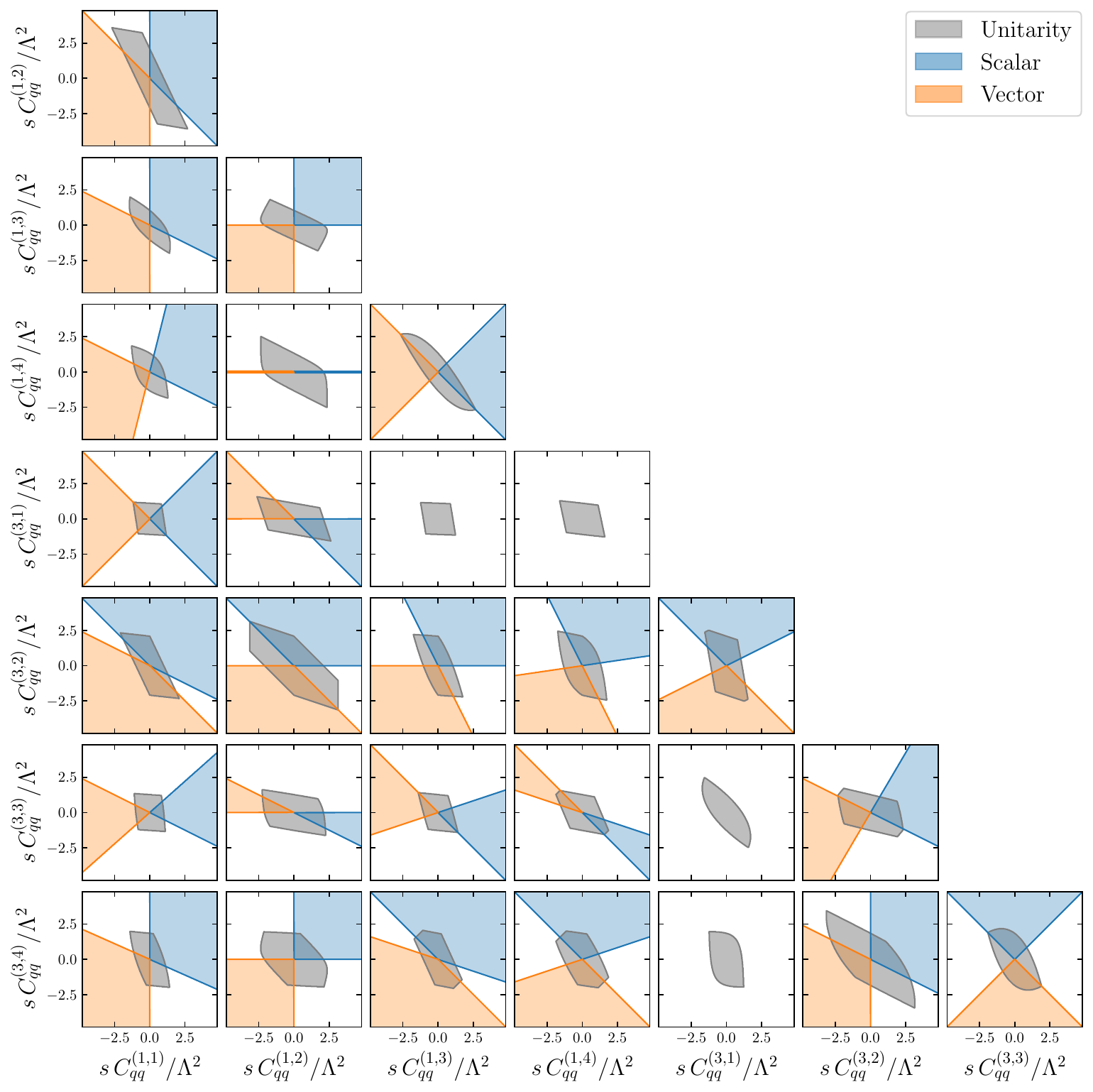}
    \caption{Partial-wave unitarity bounds and sum-rule constraints (under scalar and vector dominance) for the operators $O_{qq}^{(1,3)}$ under MFV, where $[C_{qq}^{(i)}]_{prst} =
            C_{qq}^{(i,1)}\delta_{pr}\delta_{st}
            + C_{qq}^{(i,2)}\delta_{pt}\delta_{sr}
            + C_{qq}^{(i,3)}((Y_u Y_u^\dagger)_{st}\delta_{pr} + (Y_u Y_u^\dagger)_{pr}\delta_{st}) + C_{qq}^{(i,4)}((Y_u Y_u^\dagger)_{sr}\delta_{pt} + (Y_u Y_u^\dagger)_{pt}\delta_{sr})$, $i=1,3$. The regions shown are two-dimensional slices of the full allowed region, obtained by setting to zero all the independent coefficients other than the two displayed.}
    \label{fig:Oqq_mfv}
\end{figure}

\begin{figure}[p]
    \centering
    \includegraphics[width=1\linewidth]{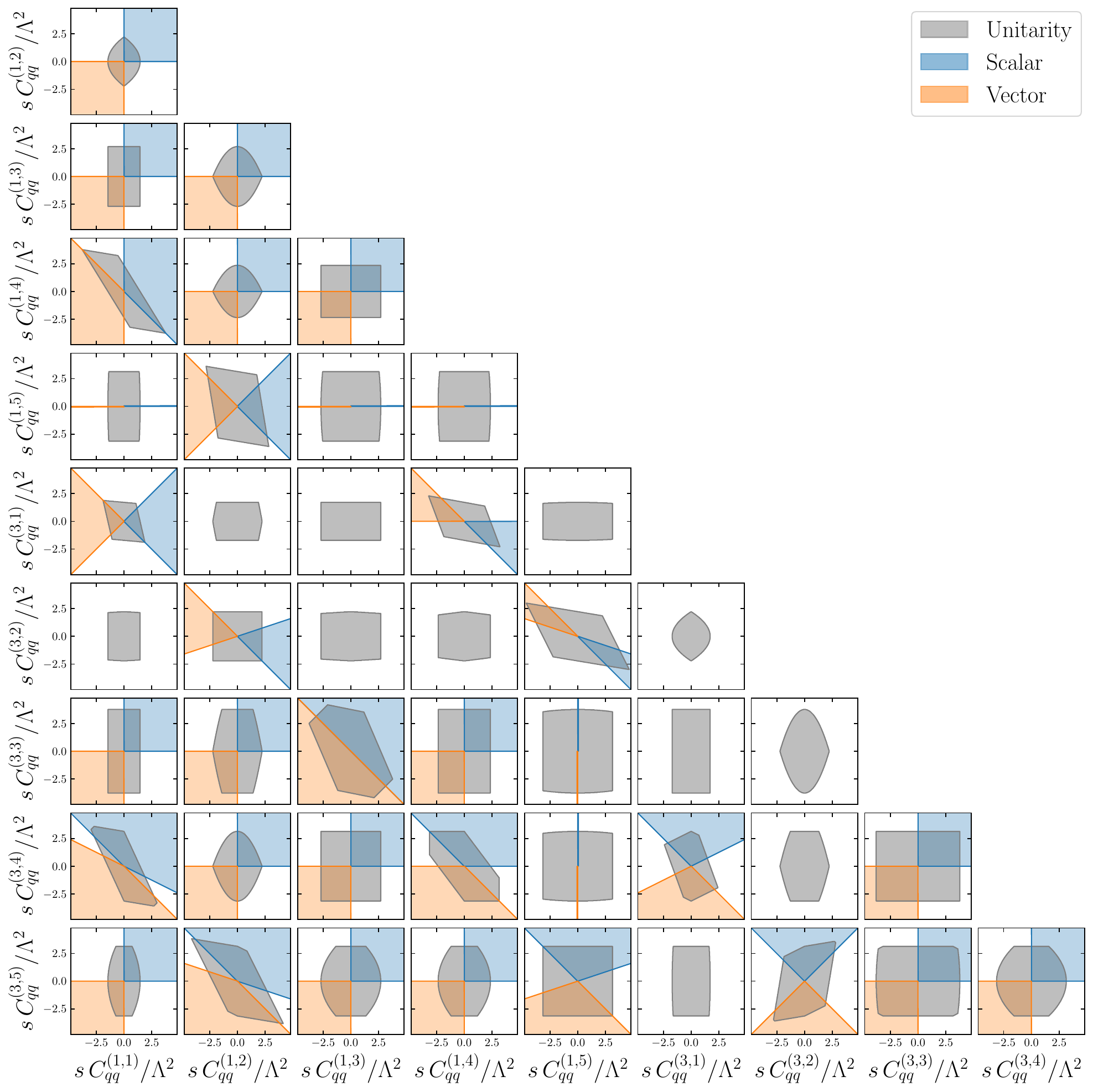}
    \caption{Partial-wave unitarity bounds and sum-rule constraints (under scalar and vector dominance) for the operators $O_{qq}^{(1,3)}$ under $\mathrm{U}(2)^5$ flavor symmetry, where $[C_{qq}^{(i)}]_{prst} =
            C_{qq}^{(i,1)}\Pij{p}{r}{a}\Pij{s}{t}{b}
            + C_{qq}^{(i,2)}(\Pij{s}{t}{a}\delta_{p3}\delta_{3r} + \Pij{p}{r}{a}\delta_{s3}\delta_{3t})
            + C_{qq}^{(i,3)}\delta_{p3}\delta_{3r}\delta_{s3}\delta_{3t}
            + C_{qq}^{(i,4)}\Pij{p}{t}{a}\Pij{s}{r}{b}
            + C_{qq}^{(i,5)}(\Pij{s}{r}{a}\delta_{p3}\delta_{3t} + \Pij{p}{t}{a}\delta_{s3}\delta_{3r})$, $i=1,3$. The regions shown are two-dimensional slices of the full allowed region, obtained by setting to zero all the independent coefficients other than the two displayed.}
    \label{fig:Oqq_u2}
\end{figure}

\begin{comment}
\begin{figure}[htbp]
    \centering
    \includegraphics[scale = 0.75]{long_paper_images/flavor/Olq_mfv.pdf}
    \caption{Partial-wave unitarity bounds and sum-rule constraints (under scalar and vector dominance) for the operators $O_{\ell q}^{(1,3)}$ under MFV, where $[C_{\ell q}^{(i)}]_{prst} =
            C_{\ell q}^{(i,1)}\delta_{pr}\delta_{st}
            + C_{\ell q}^{(i,2)}\delta_{pr}(Y_u Y_u^\dagger)_{st}$, $i=1,3$. The regions shown are two-dimensional slices of the full allowed region, obtained by setting to zero all the independent coefficients other than the two displayed.}
    \label{fig:Olq_mfv}
\end{figure}

\begin{figure}[htbp]
    \centering
    \includegraphics[width=1\linewidth]{long_paper_images/flavor/Olq_u2.pdf}
    \caption{Partial-wave unitarity bounds and sum-rule constraints (under scalar and vector dominance) for the operators $O_{\ell q}^{(1,3)}$ under $\mathrm{U}(2)^5$ flavor symmetry, where $[C_{\ell q}^{(i)}]_{prst} =
            C_{\ell q}^{(i,1)}\Pij{p}{r}{a}\Pij{s}{t}{b}
            + C_{\ell q}^{(i,2)}\delta_{p3}\delta_{3r}\Pij{s}{t}{a}
            + C_{\ell q}^{(i,3)}\Pij{p}{r}{a}\delta_{s3}\delta_{3t}
            + C_{\ell q}^{(i,4)}\delta_{p3}\delta_{3r}\delta_{s3}\delta_{3t}$, $i=1,3$. The regions shown are two-dimensional slices of the full allowed region, obtained by setting to zero all the independent coefficients other than the two displayed.}
    \label{fig:Olq_u2}
\end{figure}
\end{comment}

\begin{figure}[p]
    \centering
    \includegraphics[scale = 0.75]{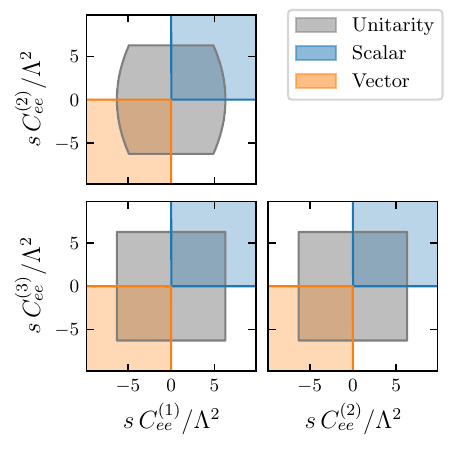}
    \caption{Partial-wave unitarity bounds and sum-rule constraints (under scalar and vector dominance) for the operator $O_{ee}$ under $\mathrm{U}(2)^5$ flavor symmetry, where $[C_{ee}]_{prst} =
            C_{ee}^{(1)}\Pij{p}{r}{a}\Pij{s}{t}{b}
            + C_{ee}^{(2)}(\Pij{s}{t}{a}\delta_{p3}\delta_{3r} + \Pij{p}{r}{a}\delta_{s3}\delta_{3t}) + C_{ee}^{(3)}\delta_{p3}\delta_{3r}\delta_{s3}\delta_{3t}$. The regions shown are two-dimensional slices of the full allowed region, obtained by setting to zero all the independent coefficients other than the two displayed.}
    \label{fig:Oee_u2}
\end{figure}

\begin{figure}[p]
    \centering
    \includegraphics[scale = 0.75]{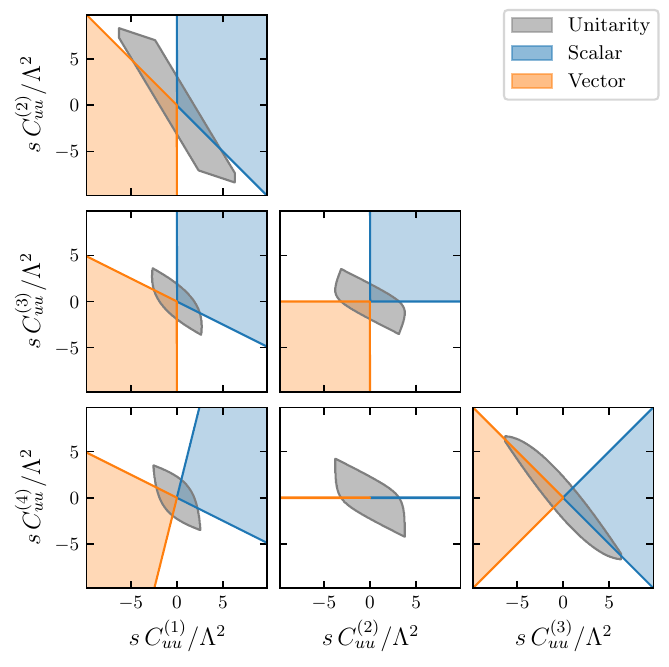}
    \caption{Partial-wave unitarity bounds and sum-rule constraints (under scalar and vector dominance) for the operator $O_{uu}$ under MFV, where $[C_{uu}]_{prst} =
            C_{uu}^{(1)}\delta_{pr}\delta_{st}
            + C_{uu}^{(2)}\delta_{pt}\delta_{sr}
            + C_{uu}^{(3)}((Y_u^\dagger Y_u)_{st}\delta_{pr} + (Y_u^\dagger Y_u)_{pr}\delta_{st})+ C_{uu}^{(4)}((Y_u^\dagger Y_u)_{sr}\delta_{pt} + (Y_u^\dagger Y_u)_{pt}\delta_{sr})$. The regions shown are two-dimensional slices of the full allowed region, obtained by setting to zero all the independent coefficients other than the two displayed.}
    \label{fig:Ouu_mfv}
\end{figure}

\begin{figure}[p]
    \centering
    \includegraphics[scale = 0.75]{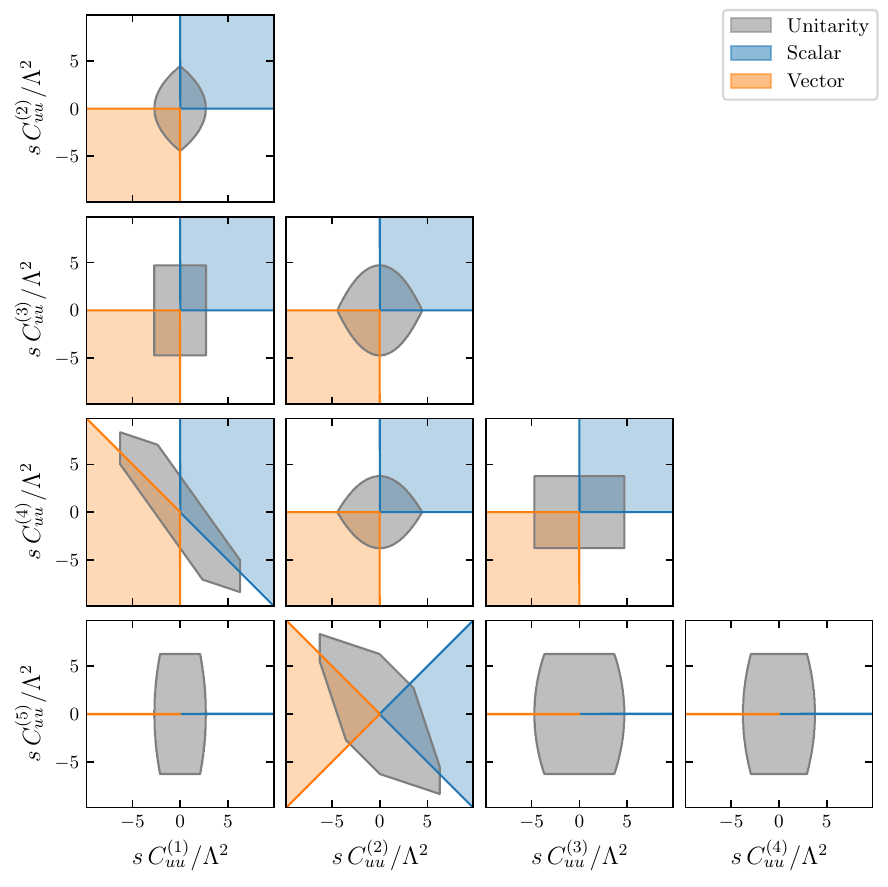}
    \caption{Partial-wave unitarity bounds and sum-rule constraints (under scalar and vector dominance) for the operator $O_{uu}$ under $\mathrm{U}(2)^5$ flavor symmetry, where $[C_{uu}]_{prst} =
            C_{uu}^{(1)}\Pij{p}{r}{a}\Pij{s}{t}{b}
            + C_{uu}^{(2)}(\Pij{s}{t}{a}\delta_{p3}\delta_{3r} + \Pij{p}{r}{a}\delta_{s3}\delta_{3t})
            + C_{uu}^{(3)}\delta_{p3}\delta_{3r}\delta_{s3}\delta_{3t}
            + C_{uu}^{(4)}\Pij{p}{t}{a}\Pij{s}{r}{b}
            + C_{uu}^{(5)}(\Pij{s}{r}{a}\delta_{p3}\delta_{3t} + \Pij{p}{t}{a}\delta_{s3}\delta_{3r})$. Same decomposition and plots are valid also for $O_{dd}$. The regions shown are two-dimensional slices of the full allowed region, obtained by setting to zero all the independent coefficients other than the two displayed.}
    \label{fig:Ouu_u2}
\end{figure}

\begin{figure}[p]
    \centering
    \includegraphics[scale = 0.75]{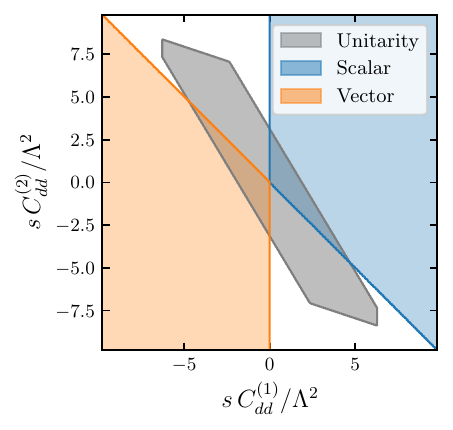}
    \caption{Partial-wave unitarity bounds and sum-rule constraints (under scalar and vector dominance) for the operator $O_{dd}$ under MFV, where $[C_{dd}]_{prst} =
            C_{dd}^{(1)}\delta_{pr}\delta_{st}
            + C_{dd}^{(2)}\delta_{pt}\delta_{sr}
            $.}
    \label{fig:Odd_mfv}
\end{figure}

\begin{figure}[p]
    \centering
    \includegraphics[scale = 0.75]{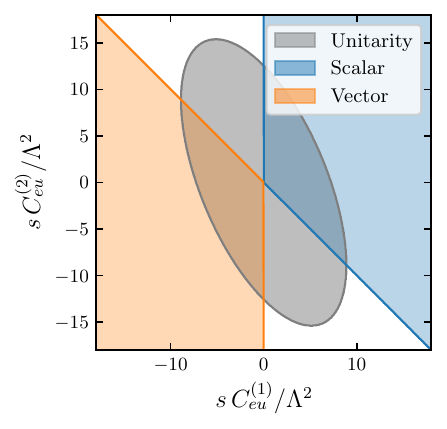}
    \caption{Partial-wave unitarity bounds and sum-rule constraints (under scalar and vector dominance) for the operator $O_{eu}$ under MFV, where $[C_{eu}]_{prst} =
            C_{eu}^{(1)}\delta_{pr}\delta_{st}
            + C_{eu}^{(2)}\delta_{pr}(Y_u^\dagger Y_u)_{st}
            $.}
    \label{fig:Oeu_mfv}
\end{figure}

\begin{figure}[p]
    \centering
    \includegraphics[scale = 0.75]{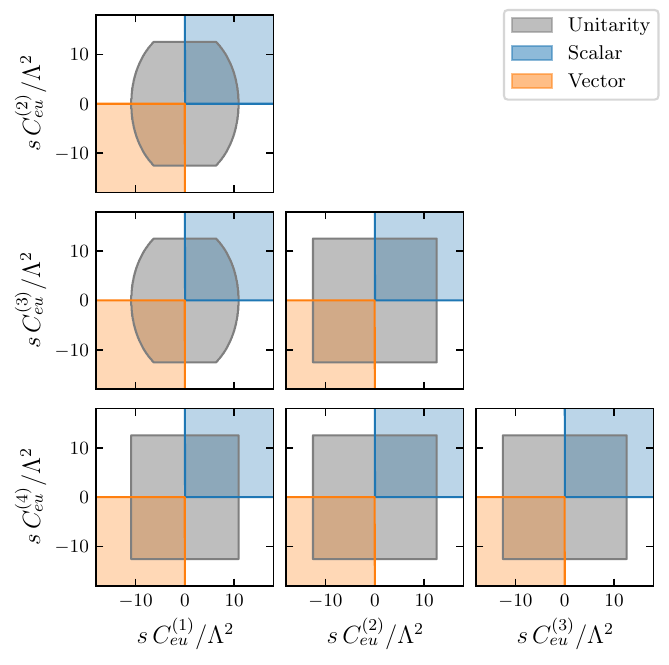}
    \caption{Partial-wave unitarity bounds and sum-rule constraints (under scalar and vector dominance) for the operator $O_{eu}$ under $\mathrm{U}(2)^5$ flavor symmetry, where $[C_{eu}]_{prst} =
            C_{eu}^{(1)}\Pij{p}{r}{a}\Pij{s}{t}{b}
            + C_{eu}^{(2)}\delta_{p3}\delta_{3r}\Pij{s}{t}{a} + C_{eu}^{(3)}\delta_{s3}\delta_{3t}\Pij{p}{r}{a}
            + C_{eu}^{(4)}\delta_{p3}\delta_{3r}\delta_{s3}\delta_{3t}$. Same decomposition and plots are valid also for $O_{ed}$. The regions shown are two-dimensional slices of the full allowed region, obtained by setting to zero all the independent coefficients other than the two displayed.}
    \label{fig:Oeu_u2}
\end{figure}

\begin{figure}[p]
    \centering
    \includegraphics[scale = 0.75]{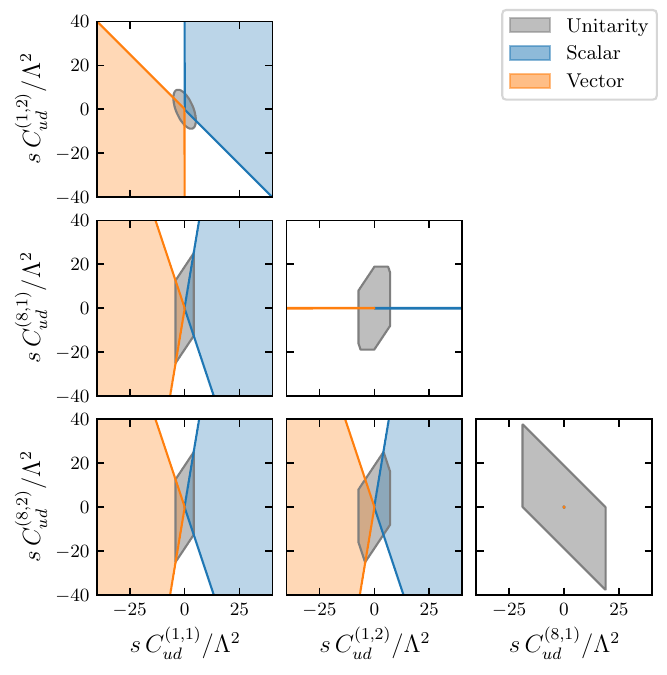}
    \caption{Partial-wave unitarity bounds and sum-rule constraints (under scalar and vector dominance) for the operators $O_{ud}^{(1,8)}$ under MFV, where $[C_{ud}^{(i)}]_{prst} =
            C_{ud}^{(i,1)}\delta_{pr}\delta_{st}
            + C_{ud}^{(i,2)}(Y_u^\dagger Y_u)_{pr}\delta_{st}$, $i=1,8$. The regions shown are two-dimensional slices of the full allowed region, obtained by setting to zero all the independent coefficients other than the two displayed.}
    \label{fig:Oud_mfv}
\end{figure}

\begin{figure}[p]
    \centering
    \includegraphics[width=\linewidth]{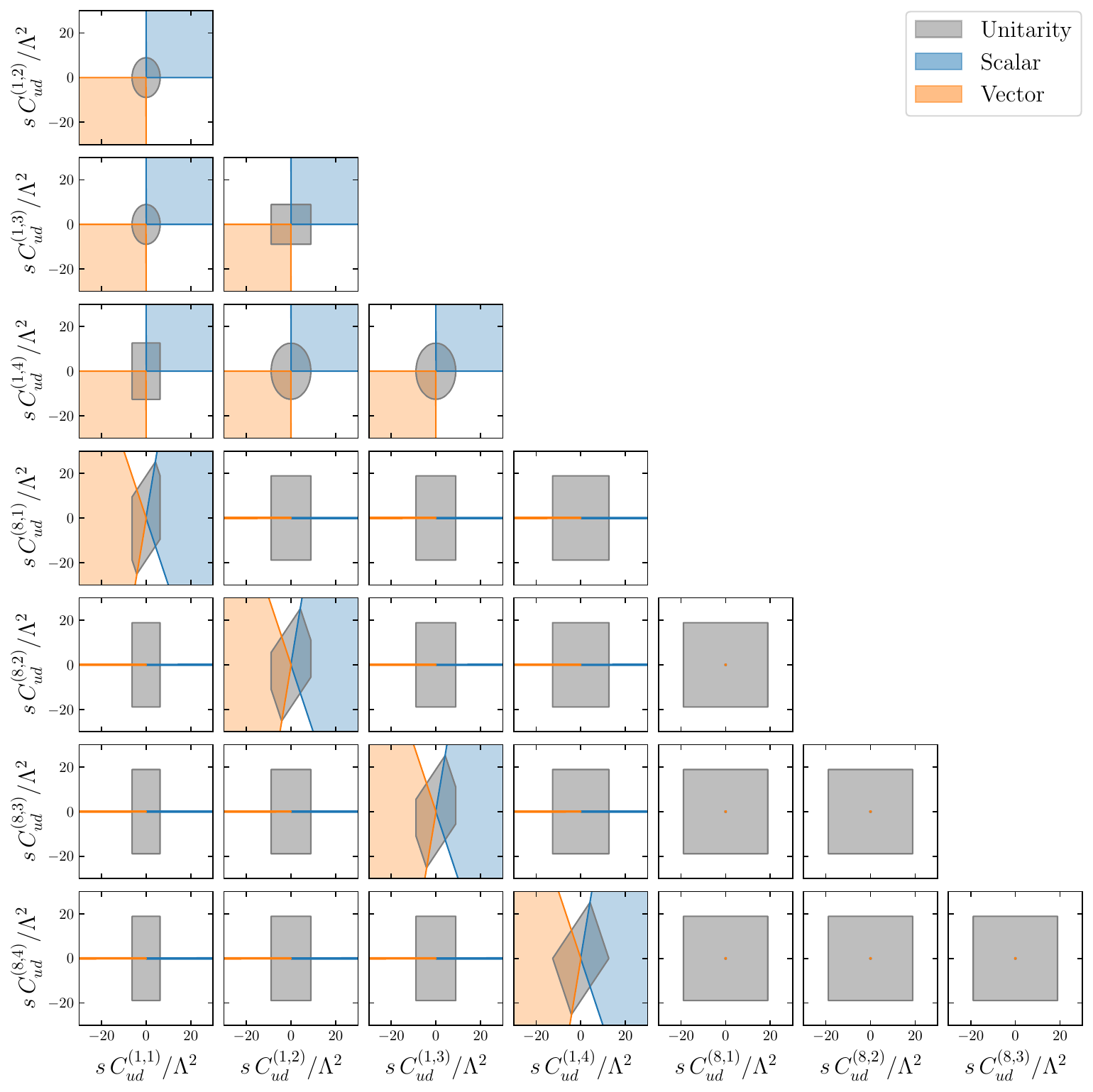}
    \caption{Partial-wave unitarity bounds and sum-rule constraints (under scalar and vector dominance) for the operators $O_{ud}^{(1,8)}$ under $\mathrm{U}(2)^5$ flavor symmetry, where $[C_{ud}^{(i)}]_{prst} =
            C_{ud}^{(i,1)}\Pij{p}{r}{a}\Pij{s}{t}{b}
            + C_{ud}^{(i,2)}\Pij{s}{t}{a}\delta_{p3}\delta_{3r}+ C_{ud}^{(i,3)}\Pij{p}{r}{a}\delta_{s3}\delta_{3t}
            + C_{ud}^{(i,4)}\delta_{p3}\delta_{3r}\delta_{s3}\delta_{3t}$, $i=1,8$. The regions shown are two-dimensional slices of the full allowed region, obtained by setting to zero all the independent coefficients other than the two displayed.}
    \label{fig:Oud_u2}
\end{figure}

\begin{figure}[p]
    \centering
    \includegraphics[scale = 0.75]{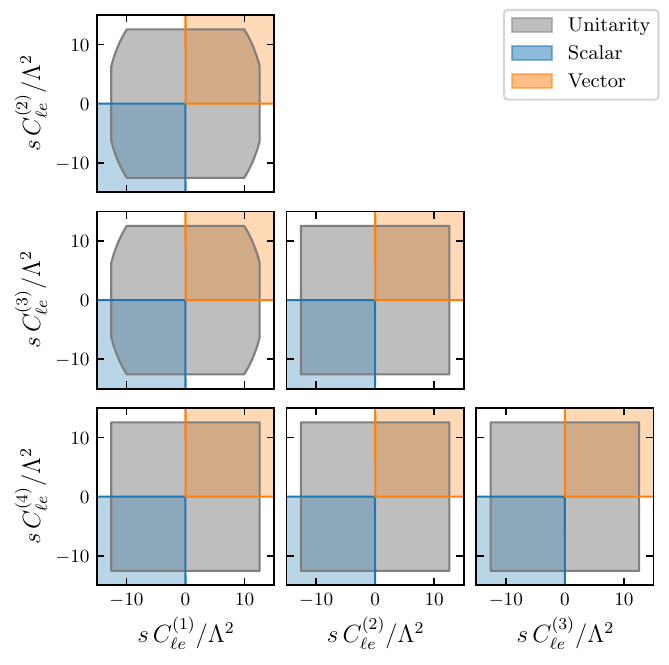}
    \caption{Partial-wave unitarity bounds and sum-rule constraints (under scalar and vector dominance) for the operator $O_{\ell e}$ under $\mathrm{U}(2)^5$ flavor symmetry, where $[C_{\ell e}]_{prst} =
            C_{\ell e}^{(1)}\Pij{p}{r}{a}\Pij{s}{t}{b}
            + C_{\ell e}^{(2)}\Pij{s}{t}{a}\delta_{p3}\delta_{3r}+ C_{\ell e}^{(3)}\Pij{p}{r}{a}\delta_{s3}\delta_{3t}
            + C_{\ell e}^{(4)}\delta_{p3}\delta_{3r}\delta_{s3}\delta_{3t}$. The regions shown are two-dimensional slices of the full allowed region, obtained by setting to zero all the independent coefficients other than the two displayed.}
    \label{fig:Ole_u2}
\end{figure}

\begin{figure}[p]
    \centering
    \includegraphics[scale = 0.75]{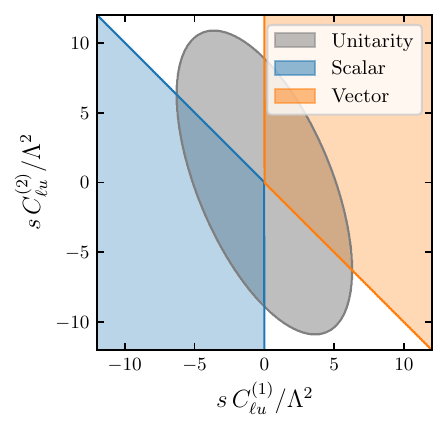}
    \caption{Partial-wave unitarity bounds and sum-rule constraints (under scalar and vector dominance) for the operator $O_{\ell u}$ under MFV, where $[C_{\ell u}]_{prst} = C_{\ell u}^{(1)}\delta_{pr}\delta_{st} + C_{\ell u}^{(2)}\delta_{pr}(Y_u^\dagger Y_u)_{st}$.}
    \label{fig:Olu_mfv}
\end{figure}

\begin{figure}[p]
    \centering
    \includegraphics[scale = 0.75]{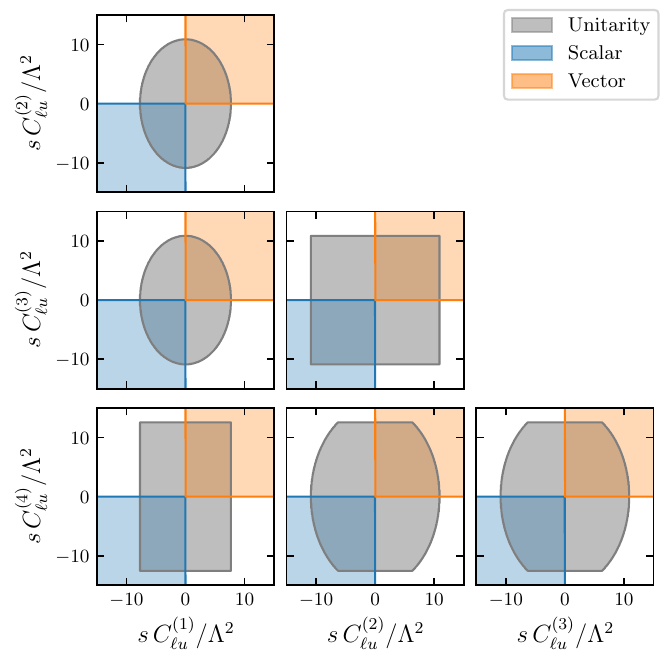}
    \caption{Partial-wave unitarity bounds and sum-rule constraints (under scalar and vector dominance) for the operator $O_{\ell u}$ under $\mathrm{U}(2)^5$ flavor symmetry, where $[C_{\ell u}]_{prst} =
            C_{\ell u}^{(1)}\Pij{p}{r}{a}\Pij{s}{t}{b}
            + C_{\ell u}^{(2)}\Pij{s}{t}{a}\delta_{p3}\delta_{3r}+ C_{\ell u}^{(3)}\Pij{p}{r}{a}\delta_{s3}\delta_{3t}
            + C_{\ell u}^{(4)}\delta_{p3}\delta_{3r}\delta_{s3}\delta_{3t}$. Same decomposition and plots are valid also for $O_{\ell d}$. The regions shown are two-dimensional slices of the full allowed region, obtained by setting to zero all the independent coefficients other than the two displayed.}
    \label{fig:Olu_u2}
\end{figure}

\begin{figure}[p]
    \centering
    \includegraphics[scale = 0.75]{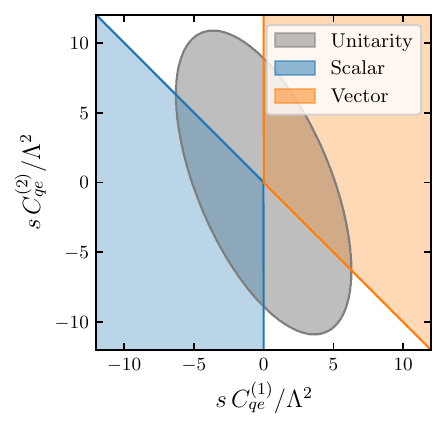}
    \caption{Partial-wave unitarity bounds and sum-rule constraints (under scalar and vector dominance) for the operator $O_{qe}$ under MFV, where $[C_{qe}]_{prst} = C_{qe}^{(1)}\delta_{pr}\delta_{st} + C_{qe}^{(2)}(Y_u Y_u^\dagger)_{pr}\delta_{st}$.}
    \label{fig:Oqe_mfv}
\end{figure}

\begin{figure}[p]
    \centering
    \includegraphics[scale = 0.75]{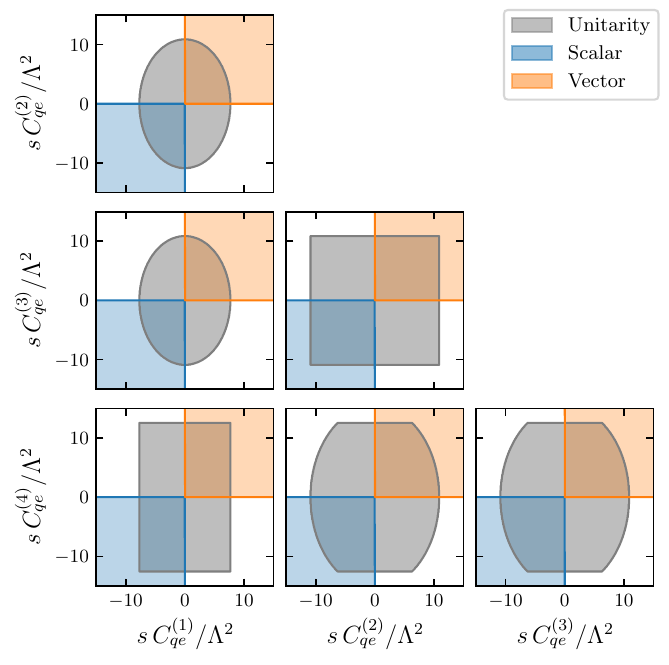}
    \caption{Partial-wave unitarity bounds and sum-rule constraints (under scalar and vector dominance) for the operator $O_{qe}$ under $\mathrm{U}(2)^5$ flavor symmetry, where $[C_{qe}]_{prst} =
            C_{qe}^{(1)}\Pij{p}{r}{a}\Pij{s}{t}{b}
            + C_{qe}^{(2)}\Pij{s}{t}{a}\delta_{p3}\delta_{3r}+ C_{qe}^{(3)}\Pij{p}{r}{a}\delta_{s3}\delta_{3t}
            + C_{qe}^{(4)}\delta_{p3}\delta_{3r}\delta_{s3}\delta_{3t}$. The regions shown are two-dimensional slices of the full allowed region, obtained by setting to zero all the independent coefficients other than the two displayed.}
    \label{fig:Oqe_u2}
\end{figure}

\begin{figure}[p]
    \centering
    \includegraphics[width=\linewidth]{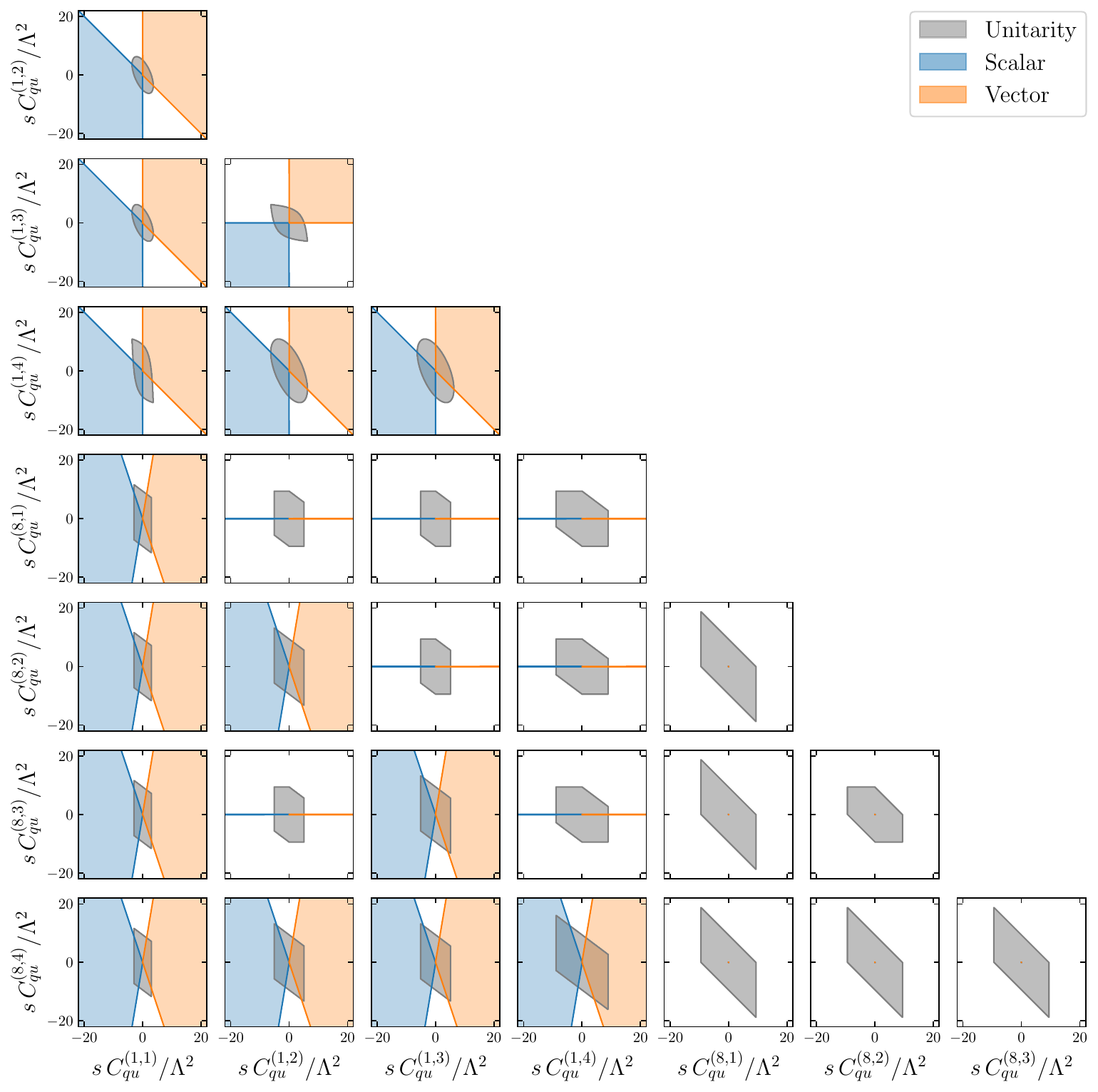}
    \caption{Partial-wave unitarity bounds and sum-rule constraints (under scalar and vector dominance) for the operators $O_{qu}^{(1,8)}$ under MFV, where $[C_{qu}^{(i)}]_{prst} =
            C_{qu}^{(i,1)}\delta_{pr}\delta_{st}
            + C_{qu}^{(i,2)}(Y_u Y_u^\dagger)_{pr}\delta_{st}+ C_{qu}^{(i,3)}\delta_{pr}(Y_u^\dagger Y_u)_{st}
            + C_{qu}^{(i,4)}(Y_u)_{pt}(Y_u^\dagger)_{sr}$, $i=1,8$. The regions shown are two-dimensional slices of the full allowed region, obtained by setting to zero all the independent coefficients other than the two displayed.}
    \label{fig:Oqu_mfv}
\end{figure}

\begin{figure}[p]
    \centering
    \includegraphics[width=\linewidth]{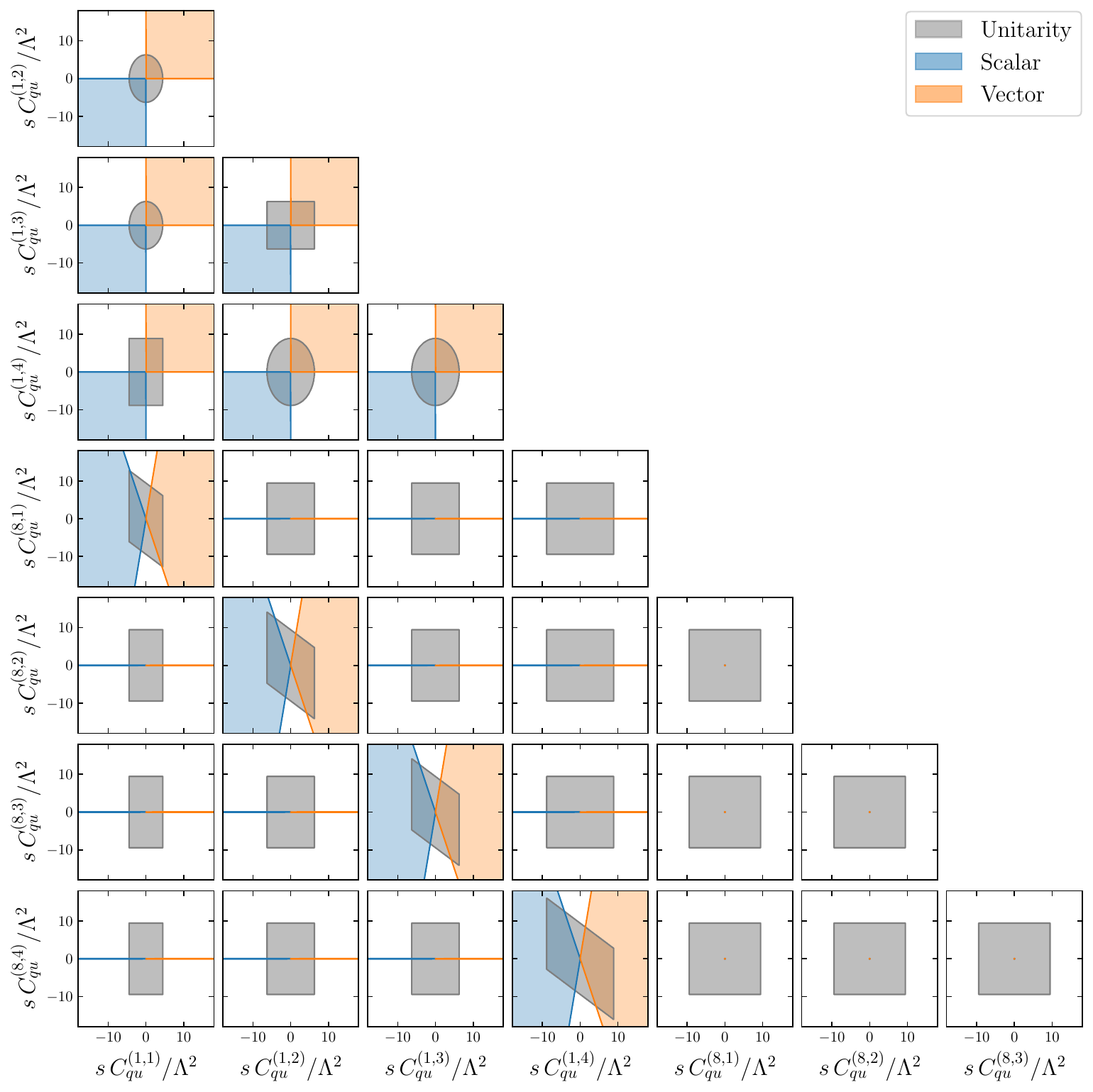}
    \caption{Partial-wave unitarity bounds and sum-rule constraints (under scalar and vector dominance) for the operators $O_{qu}^{(1,8)}$ under $\mathrm{U}(2)^5$ flavor symmetry, where $[C_{qu}^{(i)}]_{prst} =
            C_{qu}^{(i,1)}\Pij{p}{r}{a}\Pij{s}{t}{b}
            + C_{qu}^{(i,2)}\Pij{s}{t}{a}\delta_{p3}\delta_{3r}+ C_{qu}^{(i,3)}\Pij{p}{r}{a}\delta_{s3}\delta_{3t}
            + C_{qu}^{(i,4)}\delta_{p3}\delta_{3r}\delta_{s3}\delta_{3t}$, $i=1,8$. Same decomposition and plots are valid also for $O_{qd}^{(1,8)}$. The regions shown are two-dimensional slices of the full allowed region, obtained by setting to zero all the independent coefficients other than the two displayed.}
    \label{fig:Oqu_u2}
\end{figure}

\begin{figure}[p]
    \centering
    \includegraphics[scale = 0.75]{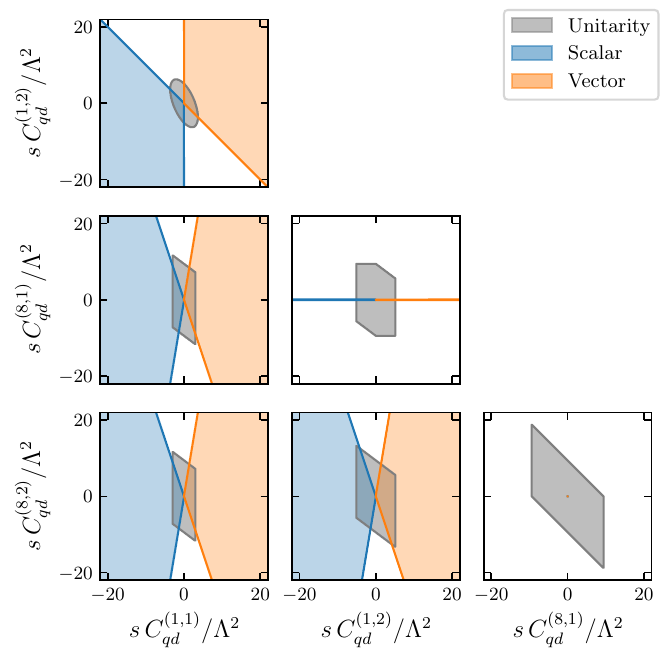}
    \caption{Partial-wave unitarity bounds and sum-rule constraints (under scalar and vector dominance) for the operators $O_{qd}^{(1,8)}$ under MFV, where $[C_{qd}^{(i)}]_{prst} =
            C_{qd}^{(i,1)}\delta_{pr}\delta_{st}
            + C_{qd}^{(i,2)}(Y_u Y_u^\dagger)_{pr}\delta_{st}$, $i=1,8$. The regions shown are two-dimensional slices of the full allowed region, obtained by setting to zero all the independent coefficients other than the two displayed.}
    \label{fig:Oqd_mfv}
\end{figure}

% Bibliography
\clearpage
\bibliographystyle{JHEP}
\bibliography{long_paper_bib}

@article{Bresciani:2026acy,
    author = "Bresciani, Luigi C. and Paradisi, Paride and Sainaghi, Andrea",
    title = "{Unitarity Bounds and Sum Rules in the Standard Model Effective Field Theory}",
    eprint = "2603.03423",
    archivePrefix = "arXiv",
    primaryClass = "hep-ph",
    doi = "10.1103/v5xg-dmss",
    journal = "Phys. Rev. Lett.",
    volume = "137",
    number = "15",
    pages = "151801",
    year = "2026"
}

@article{Low:2009di,
    author = "Low, Ian and Rattazzi, Riccardo and Vichi, Alessandro",
    title = "{Theoretical Constraints on the Higgs Effective Couplings}",
    eprint = "0907.5413",
    archivePrefix = "arXiv",
    primaryClass = "hep-ph",
    doi = "10.1007/JHEP04(2010)126",
    journal = "JHEP",
    volume = "04",
    pages = "126",
    year = "2010"
}

@article{Remmen:2020uze,
    author = "Remmen, Grant N. and Rodd, Nicholas L.",
    title = "{Signs, spin, SMEFT: Sum rules at dimension six}",
    eprint = "2010.04723",
    archivePrefix = "arXiv",
    primaryClass = "hep-ph",
    doi = "10.1103/PhysRevD.105.036006",
    journal = "Phys. Rev. D",
    volume = "105",
    number = "3",
    pages = "036006",
    year = "2022"
}

@article{Remmen:2022orj,
    author = "Remmen, Grant N. and Rodd, Nicholas L.",
    title = "{Spinning sum rules for the dimension-six SMEFT}",
    eprint = "2206.13524",
    archivePrefix = "arXiv",
    primaryClass = "hep-ph",
    reportNumber = "CERN-TH-2022-105",
    doi = "10.1007/JHEP09(2022)030",
    journal = "JHEP",
    volume = "09",
    pages = "030",
    year = "2022"
}

@article{Gu:2020thj,
    author = "Gu, Jiayin and Wang, Lian-Tao",
    title = "{Sum Rules in the Standard Model Effective Field Theory from Helicity Amplitudes}",
    eprint = "2008.07551",
    archivePrefix = "arXiv",
    primaryClass = "hep-ph",
    reportNumber = "MITP/20-040",
    doi = "10.1007/JHEP03(2021)149",
    journal = "JHEP",
    volume = "03",
    pages = "149",
    year = "2021"
}

@article{Altmannshofer:2025lun,
    author = "Altmannshofer, Wolfgang and Balme, Zev and Donohue, Christopher M. and Gori, Stefania and Mukundhan, Siddharth Vignesh",
    title = "{Targets for flavor-violating top decays}",
    eprint = "2504.18664",
    archivePrefix = "arXiv",
    primaryClass = "hep-ph",
    doi = "10.1007/JHEP08(2025)191",
    journal = "JHEP",
    volume = "08",
    pages = "191",
    year = "2025"
}

@article{Altmannshofer:2023bfk,
    author = "Altmannshofer, Wolfgang and Gori, Stefania and Lehmann, Benjamin V. and Zuo, Jianhong",
    title = "{UV physics from IR features: New prospects from top flavor violation}",
    eprint = "2303.00781",
    archivePrefix = "arXiv",
    primaryClass = "hep-ph",
    reportNumber = "MIT-CTP/5534",
    doi = "10.1103/PhysRevD.107.095025",
    journal = "Phys. Rev. D",
    volume = "107",
    number = "9",
    pages = "095025",
    year = "2023"
}

@article{Azatov:2021ygj,
    author = "Azatov, Aleksandr and Ghosh, Diptimoy and Singh, Amartya Harsh",
    title = "{Four-fermion operators at dimension 6: Dispersion relations and UV completions}",
    eprint = "2112.02302",
    archivePrefix = "arXiv",
    primaryClass = "hep-ph",
    reportNumber = "SISSA 21/2021/FISI",
    doi = "10.1103/PhysRevD.105.115019",
    journal = "Phys. Rev. D",
    volume = "105",
    number = "11",
    pages = "115019",
    year = "2022"
}

@article{Anders:1993dk,
    author = "Anders, T. B. and von Mellenthin, R. and Pfeil, B. and Selecker, H.",
    title = "{Unitarity bounds for four fermion contact interactions}",
    doi = "10.1007/BF01883720",
    journal = "Found. Phys.",
    volume = "23",
    pages = "399--410",
    year = "1993"
}

@article{Chang:2019vez,
    author = "Chang, Spencer and Luty, Markus A.",
    title = "{The Higgs Trilinear Coupling and the Scale of New Physics}",
    eprint = "1902.05556",
    archivePrefix = "arXiv",
    primaryClass = "hep-ph",
    doi = "10.1007/JHEP03(2020)140",
    journal = "JHEP",
    volume = "03",
    pages = "140",
    year = "2020"
}

@article{Jiang:2020rwz,
    author = "Jiang, Minyuan and Shu, Jing and Xiao, Ming-Lei and Zheng, Yu-Hui",
    title = "{Partial Wave Amplitude Basis and Selection Rules in Effective Field Theories}",
    eprint = "2001.04481",
    archivePrefix = "arXiv",
    primaryClass = "hep-ph",
    doi = "10.1103/PhysRevLett.126.011601",
    journal = "Phys. Rev. Lett.",
    volume = "126",
    number = "1",
    pages = "011601",
    year = "2021"
}

@article{Shu:2021qlr,
    author = "Shu, Jing and Xiao, Ming-Lei and Zheng, Yu-Hui",
    title = "{Constructing the general partial wave and renormalization in effective field theory}",
    eprint = "2111.08019",
    archivePrefix = "arXiv",
    primaryClass = "hep-th",
    doi = "10.1103/PhysRevD.107.095040",
    journal = "Phys. Rev. D",
    volume = "107",
    number = "9",
    pages = "095040",
    year = "2023"
}

@article{Falkowski:2019tft,
    author = "Falkowski, Adam and Rattazzi, Riccardo",
    title = "{Which EFT}",
    eprint = "1902.05936",
    archivePrefix = "arXiv",
    primaryClass = "hep-ph",
    reportNumber = "LPT Orsay 19-05",
    doi = "10.1007/JHEP10(2019)255",
    journal = "JHEP",
    volume = "10",
    pages = "255",
    year = "2019"
}

@article{Allwicher:2023shc,
    author = "Allwicher, Lukas and Cornella, Claudia and Isidori, Gino and Stefanek, Ben A.",
    title = "{New physics in the third generation. A comprehensive SMEFT analysis and future prospects}",
    eprint = "2311.00020",
    archivePrefix = "arXiv",
    primaryClass = "hep-ph",
    reportNumber = "ZU-TH 71/23, MITP-23-060, KCL-PH-TH/2023-59",
    doi = "10.1007/JHEP03(2024)049",
    journal = "JHEP",
    volume = "03",
    pages = "049",
    year = "2024"
}

@article{Corbett:2014ora,
    author = "Corbett, Tyler and {\'E}boli, O. J. P. and Gonzalez-Garcia, M. C.",
    title = "{Unitarity Constraints on Dimension-Six Operators}",
    eprint = "1411.5026",
    archivePrefix = "arXiv",
    primaryClass = "hep-ph",
    reportNumber = "YITP-SB-14-46",
    doi = "10.1103/PhysRevD.91.035014",
    journal = "Phys. Rev. D",
    volume = "91",
    number = "3",
    pages = "035014",
    year = "2015"
}

@article{Corbett:2017qgl,
    author = "Corbett, Tyler and {\'E}boli, O. J. P. and Gonzalez-Garcia, M. C.",
    title = "{Unitarity Constraints on Dimension-six Operators II: Including Fermionic Operators}",
    eprint = "1705.09294",
    archivePrefix = "arXiv",
    primaryClass = "hep-ph",
    doi = "10.1103/PhysRevD.96.035006",
    journal = "Phys. Rev. D",
    volume = "96",
    number = "3",
    pages = "035006",
    year = "2017"
}

@article{Cao:2023qks,
    author = "Cao, Qing-Hong and Liu, Yandong and Yuan, Shu-Run",
    title = "{Unitarity bounds and basis transformations in SMEFT: An analysis of Warsaw and SILH bases}",
    doi = "10.1016/j.nuclphysb.2024.116781",
    journal = "Nucl. Phys. B",
    volume = "1010",
    pages = "116781",
    year = "2025",
    eprint = "2309.14079",
    archivePrefix = "arXiv",
    primaryClass = "hep-ph"
}

@article{deBlas:2017xtg,
    author = "de Blas, J. and Criado, J. C. and Perez-Victoria, M. and Santiago, J.",
    title = "{Effective description of general extensions of the Standard Model: the complete tree-level dictionary}",
    eprint = "1711.10391",
    archivePrefix = "arXiv",
    primaryClass = "hep-ph",
    reportNumber = "CERN-TH-2017-251",
    doi = "10.1007/JHEP03(2018)109",
    journal = "JHEP",
    volume = "03",
    pages = "109",
    year = "2018"
}

@article{Grzadkowski:2010es,
    author = "Grzadkowski, B. and Iskrzynski, M. and Misiak, M. and Rosiek, J.",
    title = "{Dimension-Six Terms in the Standard Model Lagrangian}",
    eprint = "1008.4884",
    archivePrefix = "arXiv",
    primaryClass = "hep-ph",
    reportNumber = "IFT-9-2010, TTP10-35",
    doi = "10.1007/JHEP10(2010)085",
    journal = "JHEP",
    volume = "10",
    pages = "085",
    year = "2010"
}

@article{Bresciani:2025ojh,
    author = "Bresciani, Luigi C. and Levati, Gabriele and Paradisi, Paride",
    title = "{Positivity and partial wave unitarity bounds on ALP theories via amplitude methods}",
    eprint = "2510.13953",
    archivePrefix = "arXiv",
    primaryClass = "hep-ph",
    doi = "10.1007/JHEP07(2026)172",
    journal = "JHEP",
    volume = "07",
    pages = "172",
    year = "2026"
}

@article{Bresciani:2025toe,
    author = "Bresciani, Luigi C. and Levati, Gabriele and Paradisi, Paride",
    title = "{Amplitudes and partial wave unitarity bounds}",
    eprint = "2504.12855",
    archivePrefix = "arXiv",
    primaryClass = "hep-ph",
    doi = "10.1103/tp6m-mcyn",
    journal = "Phys. Rev. D",
    volume = "113",
    number = "7",
    pages = "L071702",
    year = "2026"
}

@article{Ghosh:2022qqq,
    author = "Ghosh, Diptimoy and Sharma, Rajat and Ullah, Farman",
    title = "{Amplitude{\textquoteright}s positivity vs. subluminality: causality and unitarity constraints on dimension 6 {\&} 8 gluonic operators in the SMEFT}",
    eprint = "2211.01322",
    archivePrefix = "arXiv",
    primaryClass = "hep-ph",
    doi = "10.1007/JHEP02(2023)199",
    journal = "JHEP",
    volume = "02",
    pages = "199",
    year = "2023"
}

@article{Jacob:1959at,
    author = "Jacob, M. and Wick, G. C.",
    title = "{On the General Theory of Collisions for Particles with Spin}",
    doi = "10.1006/aphy.2000.6022",
    journal = "Annals Phys.",
    volume = "7",
    pages = "404--428",
    year = "1959"
}

@article{Lee:1977yc,
    author = "Lee, Benjamin W. and Quigg, C. and Thacker, H. B.",
    title = "{The Strength of Weak Interactions at Very High-Energies and the Higgs Boson Mass}",
    reportNumber = "FERMILAB-PUB-77-022-T",
    doi = "10.1103/PhysRevLett.38.883",
    journal = "Phys. Rev. Lett.",
    volume = "38",
    pages = "883--885",
    year = "1977"
}

@article{Lee:1977eg,
    author = "Lee, Benjamin W. and Quigg, C. and Thacker, H. B.",
    title = "{Weak Interactions at Very High-Energies: The Role of the Higgs Boson Mass}",
    reportNumber = "FERMILAB-PUB-77-030-T",
    doi = "10.1103/PhysRevD.16.1519",
    journal = "Phys. Rev. D",
    volume = "16",
    pages = "1519",
    year = "1977"
}

@article{Gounaris:1994cm,
    author = "Gounaris, G. J. and Layssac, J. and Paschalis, J. E. and Renard, F. M.",
    title = "{Unitarity constraints for new physics induced by dim-6 operators}",
    eprint = "hep-ph/9409260",
    archivePrefix = "arXiv",
    reportNumber = "PM-94-28, THES-TP-94-08",
    doi = "10.1007/BF01579637",
    journal = "Z. Phys. C",
    volume = "66",
    pages = "619--632",
    year = "1995"
}

@article{DiLuzio:2017chi,
    author = "Di Luzio, Luca and Nardecchia, Marco",
    title = "{What is the scale of new physics behind the $B$-flavour anomalies?}",
    eprint = "1706.01868",
    archivePrefix = "arXiv",
    primaryClass = "hep-ph",
    reportNumber = "IPPP-17-50, CERN-TH-2017-126",
    doi = "10.1140/epjc/s10052-017-5118-9",
    journal = "Eur. Phys. J. C",
    volume = "77",
    number = "8",
    pages = "536",
    year = "2017"
}

@article{Eboli:2026ccq,
    author = "{\'E}boli, O. J. P. and Gonzalez-Garcia, M. C.",
    title = "{Unitarity limits on triple gauge boson production}",
    eprint = "2601.07920",
    archivePrefix = "arXiv",
    primaryClass = "hep-ph",
    reportNumber = "YITP-SB-2026-01",
    doi = "10.1103/n529-w187",
    journal = "Phys. Rev. D",
    volume = "113",
    number = "7",
    pages = "073006",
    year = "2026"
}

@article{Degrande:2025uil,
    author = "Degrande, C{\'e}line and Li, Hao-Lin and Xu, Ling-Xiao",
    title = "{Partial-Wave Unitarity Bounds on Higher-Dimensional Operators from 2-to-$N$ Scattering}",
    eprint = "2511.15524",
    archivePrefix = "arXiv",
    primaryClass = "hep-ph",
    month = "11",
    year = "2025"
}

@article{DiLuzio:2016sur,
    author = "Di Luzio, Luca and Kamenik, Jernej F. and Nardecchia, Marco",
    title = "{Implications of perturbative unitarity for scalar di-boson resonance searches at LHC}",
    eprint = "1604.05746",
    archivePrefix = "arXiv",
    primaryClass = "hep-ph",
    reportNumber = "IPPP-17-2, IPPP/17/2",
    doi = "10.1140/epjc/s10052-017-4594-2",
    journal = "Eur. Phys. J. C",
    volume = "77",
    number = "1",
    pages = "30",
    year = "2017"
}

@article{Mahmud:2024iyn,
    author = "Mahmud, Shameran and Tobioka, Kohsaku",
    title = "{Energy growth in V$_{L}$V$_{L}${\textrightarrow} V$_{L}$V$_{L}$, V$_{L}$V$_{L}$h scattering to probe Higgs cubic and HEFT interactions}",
    eprint = "2406.03522",
    archivePrefix = "arXiv",
    primaryClass = "hep-ph",
    reportNumber = "KEK-TH-2626",
    doi = "10.1007/JHEP09(2024)073",
    journal = "JHEP",
    volume = "09",
    pages = "073",
    year = "2024"
}

@article{Allwicher:2021jkr,
    author = "Allwicher, Lukas and Di Luzio, Luca and Fedele, Marco and Mescia, Federico and Nardecchia, Marco",
    title = "{What is the scale of new physics behind the muon g-2?}",
    eprint = "2105.13981",
    archivePrefix = "arXiv",
    primaryClass = "hep-ph",
    reportNumber = "TTP21-014, P3H-21-037, ZU-TH 24/21",
    doi = "10.1103/PhysRevD.104.055035",
    journal = "Phys. Rev. D",
    volume = "104",
    number = "5",
    pages = "055035",
    year = "2021"
}

@article{Cohen:2021ucp,
    author = "Cohen, Timothy and Craig, Nathaniel and Lu, Xiaochuan and Sutherland, Dave",
    title = "{Unitarity violation and the geometry of Higgs EFTs}",
    eprint = "2108.03240",
    archivePrefix = "arXiv",
    primaryClass = "hep-ph",
    doi = "10.1007/JHEP12(2021)003",
    journal = "JHEP",
    volume = "12",
    pages = "003",
    year = "2021"
}

@article{Almeida:2020ylr,
    author = "Almeida, Eduardo da Silva and {\'E}boli, O. J. P. and Gonzalez{\textendash}Garcia, M. C.",
    title = "{Unitarity constraints on anomalous quartic couplings}",
    eprint = "2004.05174",
    archivePrefix = "arXiv",
    primaryClass = "hep-ph",
    reportNumber = "YITP-SB-2020-8",
    doi = "10.1103/PhysRevD.101.113003",
    journal = "Phys. Rev. D",
    volume = "101",
    number = "11",
    pages = "113003",
    year = "2020"
}

@article{Cohen:2021gdw,
    author = "Cohen, Timothy and Doss, Joel and Lu, Xiaochuan",
    title = "{Unitarity bounds on effective field theories at the LHC}",
    eprint = "2111.09895",
    archivePrefix = "arXiv",
    primaryClass = "hep-ph",
    doi = "10.1007/JHEP04(2022)155",
    journal = "JHEP",
    volume = "04",
    pages = "155",
    year = "2022"
}

@article{Abu-Ajamieh:2020yqi,
    author = "Abu-Ajamieh, Fayez and Chang, Spencer and Chen, Miranda and Luty, Markus A.",
    title = "{Higgs coupling measurements and the scale of new physics}",
    eprint = "2009.11293",
    archivePrefix = "arXiv",
    primaryClass = "hep-ph",
    doi = "10.1007/JHEP07(2021)056",
    journal = "JHEP",
    volume = "07",
    pages = "056",
    year = "2021"
}

@article{Brivio:2021fog,
    author = "Brivio, I. and {\'E}boli, O. J. P. and Gonzalez-Garcia, M. C.",
    title = "{Unitarity constraints on ALP interactions}",
    eprint = "2106.05977",
    archivePrefix = "arXiv",
    primaryClass = "hep-ph",
    reportNumber = "YITP-SB-2021-8",
    doi = "10.1103/PhysRevD.104.035027",
    journal = "Phys. Rev. D",
    volume = "104",
    number = "3",
    pages = "035027",
    year = "2021"
}

@article{Alonso:2025ksv,
    author = "Alonso, Rodrigo and Chattopadhyay, Susobhan and Ingoldby, James",
    title = "{The potential of HEFT and the scale of new physics}",
    eprint = "2512.13612",
    archivePrefix = "arXiv",
    primaryClass = "hep-ph",
    reportNumber = "IPPP/25/86, TIFR/TH/25-24",
    doi = "10.1007/JHEP07(2026)231",
    journal = "JHEP",
    volume = "07",
    pages = "231",
    year = "2026"
}

@article{Allwicher:2020esa,
    author = "Allwicher, Lukas and Isidori, Gino and Thomsen, Anders Eller",
    title = "{Stability of the Higgs Sector in a Flavor-Inspired Multi-Scale Model}",
    eprint = "2011.01946",
    archivePrefix = "arXiv",
    primaryClass = "hep-ph",
    reportNumber = "ZU-TH-41/20",
    doi = "10.1007/JHEP01(2021)191",
    journal = "JHEP",
    volume = "01",
    pages = "191",
    year = "2021"
}

@article{Covone:2024elw,
    author = "Covone, Sebastiano and Davighi, Joe and Isidori, Gino and Pesut, Marko",
    title = "{Flavour deconstructing the composite Higgs}",
    eprint = "2407.10950",
    archivePrefix = "arXiv",
    primaryClass = "hep-ph",
    reportNumber = "CERN-TH-2024-112",
    doi = "10.1007/JHEP01(2025)041",
    journal = "JHEP",
    volume = "01",
    pages = "041",
    year = "2025"
}

@article{FernandezNavarro:2023rhv,
    author = "Fern{\'a}ndez Navarro, Mario and King, Stephen F.",
    title = "{Tri-hypercharge: a separate gauged weak hypercharge for each fermion family as the origin of flavour}",
    eprint = "2305.07690",
    archivePrefix = "arXiv",
    primaryClass = "hep-ph",
    doi = "10.1007/JHEP08(2023)020",
    journal = "JHEP",
    volume = "08",
    pages = "020",
    year = "2023"
}

@article{FernandezNavarro:2024hnv,
    author = "Fern{\'a}ndez Navarro, Mario and King, Stephen F. and Vicente, Avelino",
    title = "{Minimal complete tri-hypercharge theories of flavour}",
    eprint = "2404.12442",
    archivePrefix = "arXiv",
    primaryClass = "hep-ph",
    doi = "10.1007/JHEP07(2024)147",
    journal = "JHEP",
    volume = "07",
    pages = "147",
    year = "2024"
}

@article{FernandezNavarro:2025zmb,
    author = "Fern{\'a}ndez Navarro, Mario and King, Stephen F. and Vicente, Avelino",
    title = "{Natural neutrino mass hierarchy in a theory of gauge flavour deconstruction}",
    eprint = "2506.21687",
    archivePrefix = "arXiv",
    primaryClass = "hep-ph",
    doi = "10.1007/JHEP02(2026)046",
    journal = "JHEP",
    volume = "02",
    pages = "046",
    year = "2026"
}

@article{Beneito:2025fzf,
    author = "{Bas i Beneito}, Arnau and Palavri{\'c}, Ajdin and Sainaghi, Andrea",
    title = "{Implications of flavor symmetries for baryon number violation}",
    eprint = "2512.14813",
    archivePrefix = "arXiv",
    primaryClass = "hep-ph",
    doi = "10.1103/dlq2-9yl8",
    journal = "Phys. Rev. D",
    volume = "113",
    number = "5",
    pages = "055021",
    year = "2026"
}

@article{Fuentes-Martin:2024fpx,
    author = "Fuentes-Mart{\'\i}n, Javier and Lizana, Javier M.",
    title = "{Deconstructing flavor anomalously}",
    eprint = "2402.09507",
    archivePrefix = "arXiv",
    primaryClass = "hep-ph",
    reportNumber = "IFT-UAM/CSIC-24-21",
    doi = "10.1007/JHEP07(2024)117",
    journal = "JHEP",
    volume = "07",
    pages = "117",
    year = "2024"
}

@article{Davighi:2023evx,
    author = "Davighi, Joe and Stefanek, Ben A.",
    title = "{Deconstructed hypercharge: a natural model of flavour}",
    eprint = "2305.16280",
    archivePrefix = "arXiv",
    primaryClass = "hep-ph",
    reportNumber = "ZU-TH 24/23",
    doi = "10.1007/JHEP11(2023)100",
    journal = "JHEP",
    volume = "11",
    pages = "100",
    year = "2023"
}

@article{Fuentes-Martin:2020bnh,
    author = "Fuentes-Mart{\'\i}n, Javier and Stangl, Peter",
    title = "{Third-family quark-lepton unification with a fundamental composite Higgs}",
    eprint = "2004.11376",
    archivePrefix = "arXiv",
    primaryClass = "hep-ph",
    reportNumber = "ZU-TH-08/20, LAPTH-014/20",
    doi = "10.1016/j.physletb.2020.135953",
    journal = "Phys. Lett. B",
    volume = "811",
    pages = "135953",
    year = "2020"
}

@article{Lizana:2024jby,
    author = "Lizana, Javier M.",
    title = "{A common origin of the Higgs boson and the flavor hierarchies}",
    eprint = "2412.14243",
    archivePrefix = "arXiv",
    primaryClass = "hep-ph",
    reportNumber = "IFT-UAM/CSIC-24-185",
    doi = "10.1007/JHEP05(2025)176",
    journal = "JHEP",
    volume = "05",
    pages = "176",
    year = "2025"
}

@article{Isidori:2025rci,
    author = "Isidori, Gino and Paradisi, Paride and Sainaghi, Andrea and Selimovic, Nudzeim",
    title = "{Anarchic neutrinos from flavor deconstruction: phenomenology of the lepton sector}",
    eprint = "2510.23703",
    archivePrefix = "arXiv",
    primaryClass = "hep-ph",
    doi = "10.1007/JHEP02(2026)146",
    journal = "JHEP",
    volume = "02",
    pages = "146",
    year = "2026"
}

@article{Greljo:2018tuh,
    author = "Greljo, Admir and Stefanek, Ben A.",
    title = "{Third family quark\textendash{}lepton unification at the TeV scale}",
    eprint = "1802.04274",
    archivePrefix = "arXiv",
    primaryClass = "hep-ph",
    reportNumber = "MITP-18-012",
    doi = "10.1016/j.physletb.2018.05.033",
    journal = "Phys. Lett. B",
    volume = "782",
    pages = "131--138",
    year = "2018"
}

@article{Barbieri:2023qpf,
    author = "Barbieri, Riccardo and Isidori, Gino",
    title = "{Minimal flavour deconstruction}",
    eprint = "2312.14004",
    archivePrefix = "arXiv",
    primaryClass = "hep-ph",
    doi = "10.1007/JHEP05(2024)033",
    journal = "JHEP",
    volume = "05",
    pages = "033",
    year = "2024"
}

@article{Barbieri:2024zkh,
    author = "Barbieri, Riccardo",
    title = "{Phenomenology of Minimal Flavour Deconstruction at the lowest new scale}",
    eprint = "2409.08657",
    archivePrefix = "arXiv",
    primaryClass = "hep-ph",
    month = "9",
    year = "2024"
}

@article{Davighi:2023iks,
    author = "Davighi, Joe and Isidori, Gino",
    title = "{Non-universal gauge interactions addressing the inescapable link between Higgs and flavour}",
    eprint = "2303.01520",
    archivePrefix = "arXiv",
    primaryClass = "hep-ph",
    doi = "10.1007/JHEP07(2023)147",
    journal = "JHEP",
    volume = "07",
    pages = "147",
    year = "2023"
}

@article{Bordone:2017bld,
    author = "Bordone, Marzia and Cornella, Claudia and Fuentes-Martin, Javier and Isidori, Gino",
    title = "{A three-site gauge model for flavor hierarchies and flavor anomalies}",
    eprint = "1712.01368",
    archivePrefix = "arXiv",
    primaryClass = "hep-ph",
    reportNumber = "ZU-TH-36-17",
    doi = "10.1016/j.physletb.2018.02.011",
    journal = "Phys. Lett. B",
    volume = "779",
    pages = "317--323",
    year = "2018"
}

@article{Fuentes-Martin:2022xnb,
    author = "Fuentes-Martin, Javier and Isidori, Gino and Lizana, Javier M. and Selimovic, Nudzeim and Stefanek, Ben A.",
    title = "{Flavor hierarchies, flavor anomalies, and Higgs mass from a warped extra dimension}",
    eprint = "2203.01952",
    archivePrefix = "arXiv",
    primaryClass = "hep-ph",
    reportNumber = "ZU-TH-08/22",
    doi = "10.1016/j.physletb.2022.137382",
    journal = "Phys. Lett. B",
    volume = "834",
    pages = "137382",
    year = "2022"
}

@article{Greljo:2025mwj,
    author = "Greljo, Admir and Palavri{\'c}, Ajdin and Stefanek, Ben A.",
    title = "{Minimal Flavor Protection for TeV-scale New Physics}",
    eprint = "2512.04159",
    archivePrefix = "arXiv",
    primaryClass = "hep-ph",
    month = "12",
    year = "2025"
}

@article{Isidori:2023pyp,
    author = "Isidori, Gino and Wilsch, Felix and Wyler, Daniel",
    title = "{The standard model effective field theory at work}",
    eprint = "2303.16922",
    archivePrefix = "arXiv",
    primaryClass = "hep-ph",
    reportNumber = "ZU-TH 14/23",
    doi = "10.1103/RevModPhys.96.015006",
    journal = "Rev. Mod. Phys.",
    volume = "96",
    number = "1",
    pages = "015006",
    year = "2024"
}

@article{Brivio:2017vri,
    author = "Brivio, Ilaria and Trott, Michael",
    title = "{The Standard Model as an Effective Field Theory}",
    eprint = "1706.08945",
    archivePrefix = "arXiv",
    primaryClass = "hep-ph",
    doi = "10.1016/j.physrep.2018.11.002",
    journal = "Phys. Rept.",
    volume = "793",
    pages = "1--98",
    year = "2019"
}

@article{Buchmuller:1985jz,
    author = "Buchmuller, W. and Wyler, D.",
    title = "{Effective Lagrangian Analysis of New Interactions and Flavor Conservation}",
    reportNumber = "CERN-TH-4254/85",
    doi = "10.1016/0550-3213(86)90262-2",
    journal = "Nucl. Phys. B",
    volume = "268",
    pages = "621--653",
    year = "1986"
}

@article{Aebischer:2025qhh,
    author = "Aebischer, Jason and Buras, Andrzej J. and Kumar, Jacky",
    title = "{SMEFT ATLAS: The landscape beyond the Standard Model}",
    eprint = "2507.05926",
    archivePrefix = "arXiv",
    primaryClass = "hep-ph",
    reportNumber = "AJB-25-1, CERN-TH-2025-129, LA-UR-24-24665",
    doi = "10.1016/j.physrep.2026.07.003",
    journal = "Phys. Rept.",
    volume = "1198",
    pages = "1--181",
    year = "2026"
}

@article{Faroughy:2020ina,
    author = "Faroughy, Darius A. and Isidori, Gino and Wilsch, Felix and Yamamoto, Kei",
    title = "{Flavour symmetries in the SMEFT}",
    eprint = "2005.05366",
    archivePrefix = "arXiv",
    primaryClass = "hep-ph",
    doi = "10.1007/JHEP08(2020)166",
    journal = "JHEP",
    volume = "08",
    pages = "166",
    year = "2020"
}

@article{Greljo:2022cah,
    author = "Greljo, Admir and Palavri{\'c}, Ajdin and Thomsen, Anders Eller",
    title = "{Adding Flavor to the SMEFT}",
    eprint = "2203.09561",
    archivePrefix = "arXiv",
    primaryClass = "hep-ph",
    doi = "10.1007/JHEP10(2022)010",
    journal = "JHEP",
    volume = "10",
    pages = "010",
    year = "2022"
}

@article{DAmbrosio:2002vsn,
    author = "D'Ambrosio, G. and Giudice, G. F. and Isidori, G. and Strumia, A.",
    title = "{Minimal flavor violation: An Effective field theory approach}",
    eprint = "hep-ph/0207036",
    archivePrefix = "arXiv",
    reportNumber = "CERN-TH-2002-147, IFUP-TH-2002-17",
    doi = "10.1016/S0550-3213(02)00836-2",
    journal = "Nucl. Phys. B",
    volume = "645",
    pages = "155--187",
    year = "2002"
}

@article{Cirigliano:2005ck,
    author = "Cirigliano, Vincenzo and Grinstein, Benjamin and Isidori, Gino and Wise, Mark B.",
    title = "{Minimal flavor violation in the lepton sector}",
    eprint = "hep-ph/0507001",
    archivePrefix = "arXiv",
    reportNumber = "UCSD-PTH-05-11, CALT-68-2566",
    doi = "10.1016/j.nuclphysb.2005.08.037",
    journal = "Nucl. Phys. B",
    volume = "728",
    pages = "121--134",
    year = "2005"
}

@article{Masiero:2026qon,
    author = "Masiero, Antonio and Paradisi, Paride and Queiroz, Daniel and Sainaghi, Andrea and Valori, Nicola and Vives, Oscar",
    title = "{Anatomy and Phenomenology of Minimal Flavor Deconstruction in the Lepton Sector}",
    eprint = "2606.02709",
    archivePrefix = "arXiv",
    primaryClass = "hep-ph",
    month = "6",
    year = "2026"
}

@article{FuentesZamoro:2025exp,
    author = "Fuentes Zamoro, Marta and Grinstein, Benjam{\'\i}n and Qu{\'\i}lez, Pablo",
    title = "{Taming forward scattering singularities in partial waves}",
    eprint = "2510.08784",
    archivePrefix = "arXiv",
    primaryClass = "hep-ph",
    month = "10",
    year = "2025"
}

@article{Plestid:2026pzn,
    author = "Plestid, Ryan and Lasanta, Pablo Qu{\'\i}lez",
    title = "{Partial-wave unitarity and long-range interactions}",
    eprint = "2606.19432",
    archivePrefix = "arXiv",
    primaryClass = "hep-ph",
    reportNumber = "CERN-TH/2026-119",
    month = "6",
    year = "2026"
}

@article{Hannesdottir:2019opa,
    author = "Hannesdottir, Holmfridur and Schwartz, Matthew D.",
    title = "{$S$-Matrix for massless particles}",
    eprint = "1911.06821",
    archivePrefix = "arXiv",
    primaryClass = "hep-th",
    doi = "10.1103/PhysRevD.101.105001",
    journal = "Phys. Rev. D",
    volume = "101",
    number = "10",
    pages = "105001",
    year = "2020"
}

@article{Hannesdottir:2019umk,
    author = "Hannesdottir, Holmfridur and Schwartz, Matthew D.",
    title = "{Finite $S$ matrix}",
    eprint = "1906.03271",
    archivePrefix = "arXiv",
    primaryClass = "hep-th",
    doi = "10.1103/PhysRevD.107.L021701",
    journal = "Phys. Rev. D",
    volume = "107",
    number = "2",
    pages = "L021701",
    year = "2023"
}

@article{Carney:2017oxp,
    author = "Carney, Daniel and Chaurette, Laurent and Neuenfeld, Dominik and Semenoff, Gordon Walter",
    title = "{Dressed infrared quantum information}",
    eprint = "1710.02531",
    archivePrefix = "arXiv",
    primaryClass = "hep-th",
    doi = "10.1103/PhysRevD.97.025007",
    journal = "Phys. Rev. D",
    volume = "97",
    number = "2",
    pages = "025007",
    year = "2018"
}

@article{Carney:2018ygh,
    author = "Carney, Daniel and Chaurette, Laurent and Neuenfeld, Dominik and Semenoff, Gordon",
    title = "{On the need for soft dressing}",
    eprint = "1803.02370",
    archivePrefix = "arXiv",
    primaryClass = "hep-th",
    doi = "10.1007/JHEP09(2018)121",
    journal = "JHEP",
    volume = "09",
    pages = "121",
    year = "2018"
}

@article{Lippstreu:2023vvg,
    author = "Lippstreu, Luke",
    title = "{A perturbation theory for the Coulomb-phase infrared divergence}",
    eprint = "2312.08455",
    archivePrefix = "arXiv",
    primaryClass = "hep-th",
    month = "12",
    year = "2023"
}

@article{Lippstreu:2025jit,
    author = "Lippstreu, Luke",
    title = "{Analytic Properties of Infrared-Finite Amplitudes in Theories with Long-Range Forces}",
    eprint = "2505.04702",
    archivePrefix = "arXiv",
    primaryClass = "hep-th",
    month = "5",
    year = "2025"
}

@article{Bellazzini:2021oaj,
    author = "Bellazzini, Brando and Riembau, Marc and Riva, Francesco",
    title = "{IR side of positivity bounds}",
    eprint = "2112.12561",
    archivePrefix = "arXiv",
    primaryClass = "hep-th",
    doi = "10.1103/PhysRevD.106.105008",
    journal = "Phys. Rev. D",
    volume = "106",
    number = "10",
    pages = "105008",
    year = "2022"
}

@article{Caron-Huot:2022ugt,
    author = "Caron-Huot, Simon and Li, Yue-Zhou and Parra-Martinez, Julio and Simmons-Duffin, David",
    title = "{Causality constraints on corrections to Einstein gravity}",
    eprint = "2201.06602",
    archivePrefix = "arXiv",
    primaryClass = "hep-th",
    reportNumber = "CALT-TH 2021-003",
    doi = "10.1007/JHEP05(2023)122",
    journal = "JHEP",
    volume = "05",
    pages = "122",
    year = "2023"
}

@article{Caron-Huot:2022jli,
    author = "Caron-Huot, Simon and Li, Yue-Zhou and Parra-Martinez, Julio and Simmons-Duffin, David",
    title = "{Graviton partial waves and causality in higher dimensions}",
    eprint = "2205.01495",
    archivePrefix = "arXiv",
    primaryClass = "hep-th",
    reportNumber = "CALT-TH 2022-17",
    doi = "10.1103/PhysRevD.108.026007",
    journal = "Phys. Rev. D",
    volume = "108",
    number = "2",
    pages = "026007",
    year = "2023"
}

@article{Chang:2025cxc,
    author = "Chang, Cyuan-Han and Parra-Martinez, Julio",
    title = "{Graviton loops and negativity}",
    eprint = "2501.17949",
    archivePrefix = "arXiv",
    primaryClass = "hep-th",
    doi = "10.1007/JHEP08(2025)175",
    journal = "JHEP",
    volume = "08",
    pages = "175",
    year = "2025"
}

@article{Bellazzini:2025bay,
    author = "Bellazzini, B. and Berman, J. and Isabella, G. and Riva, F. and Romano, M. and Sciotti, F.",
    title = "{Positivity with long-range interactions}",
    eprint = "2512.13780",
    archivePrefix = "arXiv",
    primaryClass = "hep-th",
    doi = "10.1007/JHEP07(2026)235",
    journal = "JHEP",
    volume = "07",
    pages = "235",
    year = "2026"
}

@book{Varshalovich:1988ifq,
    author = "Varshalovich, D. A. and Moskalev, A. N. and Khersonskii, V. K.",
    title = "{Quantum Theory of Angular Momentum}: {Irreducible Tensors, Spherical Harmonics, Vector Coupling Coefficients, 3nj Symbols}",
    doi = "10.1142/0270",
    isbn = "978-981-4415-49-1, 978-9971-5-0107-5",
    publisher = "World Scientific Publishing Company",
    year = "1988"
}

@article{Singh:2026xak,
    author = "Singh, Aadarsh and D'Ambrosio, Giancarlo and Vempati, Sudhir K.",
    title = "{Theoretical and experimental constraints in the {\ensuremath{\mu}}-{\ensuremath{\tau}} four-lepton sector of the SMEFT: Implications to neutrino self-interactions}",
    eprint = "2604.27716",
    archivePrefix = "arXiv",
    primaryClass = "hep-ph",
    doi = "10.1103/8z45-8cw5",
    journal = "Phys. Rev. D",
    volume = "114",
    number = "3",
    pages = "035038",
    year = "2026"
}

@article{Dobado:2026kll,
    author = "Dobado, A. and Domenech, D. and Herrero, M. and Morales, R. A.",
    title = "{Unitarity limits on HH production from VBF mediated channels}",
    eprint = "2610.07194",
    archivePrefix = "arXiv",
    primaryClass = "hep-ph",
    reportNumber = "IFT-UAM/CSIC-26-104",
    month = "10",
    year = "2026"
}

\end{document}